\documentclass{mn2e}
\newcommand{\gtsim}{\mbox{{\raisebox{-0.4ex}{$\stackrel{>}{{\scriptstyle\sim
}}
$}}}}
\newcommand{\ltsim}{\mbox{{\raisebox{-0.4ex}{$\stackrel{<}{{\scriptstyle\sim
}}
$}}}}

\usepackage{times}
\usepackage{amssymb}
\usepackage{epsfig}
\usepackage{lscape}
\usepackage{graphicx}
          
\title[Galaxies at z = 6 -- 9 from the WFC3/IR imaging of the HUDF]{Galaxies at z = 6 -- 9 from the WFC3/IR imaging of the HUDF}

\author[R.J..~McLure, et al.]
{R.\,J.~McLure$^{1}$\thanks{Email: rjm@roe.ac.uk}, J.\,S.~Dunlop$^{1}$, M.~Cirasuolo$^{1,2}$, A.\,M.~Koekemoer$^3$, E.~Sabbi$^3$, \and D.\,P.~Stark$^4$, T.\,A.~Targett$^5$, R.S. Ellis$^6$\\
\footnotesize\\
$^{1}$ SUPA\thanks{Scottish Universities Physics Alliance}, 
Institute for Astronomy, University of Edinburgh, 
Royal Observatory, Edinburgh, EH9 3HJ, UK\\
$^{2}$UK Astronomical Technology Centre, 
Royal Observatory, Edinburgh, EH9 3HJ, UK\\
$^{3}$Space Telescope Science Institute, 3700 San Martin Drive, Baltimore,
MD 21218, USA\\
$^{4}$Kavli Institute of Cosmology, University of Cambridge, Madingley Road, 
Cambridge, CB3 0HA, UK\\
$^{5}$Department of Physics \& Astronomy, University of British Columbia, 6224 Agricultural Road, Vancouver, BC V6T 1Z1, Canada\\
$^{6}$Department of Astronomy, California Institute of Technology, Pasadena, CA 91125, USA}
\date{Accepted ... ; Received ... ; in original form ...}

\pagerange{000--000}

\begin{document}
\hsize=6truein

\maketitle

\begin{abstract}
We present the results of a systematic search for galaxies in the redshift range $z = 6 - 9$,
within the new, deep, near-infrared ($Y,J,H$) imaging of the 
Hubble Ultra Deep Field provided by the Wide Field Camera 3 (WFC3) on the Hubble
Space Telescope. We have performed full spectral energy distribution fitting
to the optical+infrared photometry of all high-redshift galaxy candidates 
detected at $\geq5 \sigma$ significance in at least one of the WFC3/IR broad-band 
filters. After careful rejection of contaminants, the result is 
a sample of 49 galaxies with primary photometric redshift 
solutions $z > 5.9$, within the 4.5 arcmin$^2$ field covered by the new near-infrared imaging.
Our sample, selected without recourse to specific 
colour cuts, re-selects all but the faintest one of the 16 $z_{850}$-drops selected by Oesch et al. (2009),
recovers all 5 of the $Y_{105}$-drops reported by Bouwens et al. (2009), and adds a further 29 
equally plausible galaxy candidates, of which 12 lie beyond $z \simeq 6.3$, and 4 lie beyond $z \simeq 7.0$. However, we also present 
confidence intervals on our photometric redshift estimates, including alternative secondary redshift
solutions. As a result of this analysis we caution that acceptable low-redshift ($z < 2$) solutions
exist for 28 out of the 37 galaxies at $z > 6.3$, and in particular for all 8 
of the galaxy candidates reported here at $z > 7.5$. Nevertheless, we note that the
very highest redshift candidates appear to be strongly clustered in
the field. Based on our photometric redshift analysis we derive new estimates of the ultraviolet galaxy luminosity function at $z \simeq 7$ and $z \simeq 8$.
Where our results are most robust, at a characteristic luminosity $M_{1500} \simeq -19.5 (AB)$, we find that the comoving 
number density of galaxies declines by a factor of $\simeq 2.5$ between $z \simeq 6$ and $z \simeq 7$, and by a further 
factor of $\simeq 2$ by $z \simeq 8$. These results suggest 
that it is difficult for the observed population of
high-redshift 
star-forming galaxies to achieve reionisation by $z \simeq 6$ without a significant
contribution from galaxies well below the detection limits, plus
alterations in the escape fraction of ionising photons and/or
continued vigorous star formation at $z>15$.
\end{abstract}


\section{INTRODUCTION}

A key goal of modern observational cosmology is to discover and study the first 
galaxies, and to quantify their role in the reionisation of the neutral universe
at redshifts $z > 7$ (Dunkley et al., 2009). Studies of quasars have now reached 
beyond $z \simeq 6$ (e.g. Fan et al. 2006) and indicate that the 
Gunn-Peterson (1965) optical depth due to neutral hydrogen in the intergalactic medium (IGM) increases significantly
at these redshifts. Whatever the precise implications for reionisation physics, 
the increased strength of the Lyman-break produced by this enhanced optical depth 
should make it relatively straightforward to select still higher-redshift
objects via the Lyman-break technique (e.g. Steidel et al. 1996), given sufficiently sensitive broad-band  
imaging.

Over recent years, galaxy evolution studies have indeed been successfully extended 
to redshifts $z > 6$, via both ground-based and {\it Hubble Space Telescope} (HST) searches for 
objects with either strong Lyman breaks and/or strong Lyman-$\alpha$ emission at 
observed wavelengths $\lambda_{obs} > 8500$\AA\ (e.g. Bouwens et al. 2004; Yan \& Windhorst 2004; Stark et al., 2007; Bouwens et al. 2008; Richard et al. 2008;
McLure et al. 2009; Oesch et al. 2009a; Zheng et al. 2009; Ouchi et al. 2009). However, it has proved very difficult 
to push the discovery space beyond $z \simeq 7$. The highest spectroscopically
confirmed redshift is $z = 6.96$ (Iye et al. 2006), and there currently exist at most 2 or 3 convincing 
candidates at redshifts just greater than $z = 7$ (Bouwens et al. 2008; Bradley et al. 2008; Richard et al. 2008; Oesch et al. 2009a; see Section 2). 
This current redshift ``limit''  is almost certainly in part due to continued (potentially steepening) decline 
in the number density of galaxies at higher redshifts, as anticipated from the 
negative evolution of the galaxy luminosity function found 
between $z = 4$ and $z = 6$ (Bouwens et al. 2006; McLure et al. 2009). However, action is clearly 
required at higher redshifts to commence and maintain reionisation, and 
extensive star-formation at $z > 7$ is implied by the stellar masses 
of galaxies at $z \simeq 6$ (Stark et al. 2009) and by reports of 
Balmer breaks in their spectral energy distributions (SEDs) (Eyles et al. 2007). Most recently, spectacular direct 
evidence of action at $z > 8$ has been provided by the discovery
of a gamma ray burst at $z \simeq  8.2$ (Tanvir et al. 2009; Salvaterra et al. 2009). Consequently the most likely 
reason for the current galaxy redshift limit of $z \simeq 7$ is a practical one, namely that the convincing 
discovery of galaxies at higher redshifts requires more sensitive near-infrared imaging
than has been available until now. Specifically, to uncover a galaxy
at $z \simeq 8$ on the basis of a strong Lyman break at $\lambda_{rest} = 1215$\AA\ requires 
deep imaging at $\lambda_{obs} > 1.1 {\rm \mu m}$ in at least 2 filters (to establish the existence of a blue continuum longward
of the break), {\it and} imaging at $\lambda_{obs} \simeq 0.9-1.1 {\rm \mu m}$ of sufficient depth 
for a non-detection to provide convincing evidence of a sharp Lyman break in the spectrum. 
The recent installation of Wide Field Camera 3 (WFC3) on the HST
offers the first realistic opportunity to achieve the required 
multi-filter near-infrared sensitivity levels over significant areas of sky. In this study we use the first 
deep imaging from this revolutionary new instrument to extend the 
study of galaxy evolution into the crucial reionisation 
epoch corresponding to $z = 7 - 10$ (Dunkley et al. 2009).
 
The specific aim of the work presented here is to use the first release of deep HST WFC3/IR imaging
in the Hubble Ultra Deep Field (HUDF; Beckwith et al. 2006), in tandem with the 
existing HST Advanced Camera for Surveys (ACS) optical imaging to select and study a new sample of 
galaxies at $z > 6$, with the expectation that this new sample could extend to $ z > 8$ (due to
the extreme depth of the new near-infrared imaging -- see Section 3).

The paper is structured as follows. In Section 2 we summarize what is currently known about high-redshift
galaxies in the HUDF, based on the handful of potential $z > 6.5$ galaxy candidates selected 
using the existing HST NICMOS+ACS dataset. The new WFC3/IR data are described in Section 3, where we summarize 
the key properties of the new WFC3/IR instrument, discuss the observing strategy of the HST HUDF09 Treasury programme,
and explain how we reduced the raw near-infrared imaging data after public release on September 9, 2009. Next, in Section 4, we describe how we
selected and refined our high-redshift galaxy catalogue from the combined WFC3/IR+ACS+Spitzer imaging data now available in
the field. We also explain how we used SED fitting to derive photometric redshifts 
(with confidence intervals) for all plausible high-redshift galaxy candidates, and used this information to 
clean the sample for a range of contaminants. In Section 5 we present and discuss the final sample of galaxies in the redshift
range $z = 6 - 8.5$, explore the robustness and implications of the new very high-redshift sub-sample, derive a first 
meaningful estimate of the evolution of the galaxy luminosity function out to $z \simeq 8$, and briefly explore the implications for our understanding
of cosmic reionisation. Finally, we summarize our conclusions in Section 6. 

We give all magnitudes in the AB system (Oke \& Gunn 1983), and for cosmological 
calculations assume ${\rm \Omega_0 = 0.3}$, ${\rm \Omega_{\Lambda} = 0.7}$, and
$H_0 = 70\,{\rm km\,s^{-1}\,Mpc^{-1}}$.

\section{Existing redshift 7 galaxy candidates in the HUDF} 

The first study to claim to have isolated galaxies at ``$z \simeq 7 - 8$'' was that of Bouwens
et al. (2004), who reported 6 $z_{850}$-drops derived 
from the combined NICMOS $J_{110}$, $H_{160}$ and ACS $z_{850}$ 
imaging of the HUDF. Bouwens et al. (2004) suspected that two of their candidates were possibly 
not real (UDF-491-880, and UDF-818-886), due to the fact they were not found in an independent reduction
of the NICMOS imaging undertaken by Massimo Robberto. This scepticism was justified, as neither of these 
objects is seen in the new, deeper WFC3/IR imaging. Subsequently, Bouwens et al. (2008) added a fifth plausible candidate.
This, plus one of the original 4 candidates, was also identified as a
high-redshift candidate in an independent study by Oesch et al. (2009a).

The positions of these 5 surviving NICMOS-selected $z \simeq 7$ galaxy candidates, along with their 
pre-existing alternative catalogue names, are summarized in Table 1. Our own analysis 
of the NICMOS+ACS photometry for these objects in fact indicates that only the first two objects listed
in Table 1 (i.e. the two galaxies confirmed by Oesch et al. 2009a) actually lie at $z > 7$, and SED fits and 
redshift constraints on these two galaxies are illustrated in Fig. 1 
(see Section 4.2 for details of the SED fitting procedure). 
 
\begin{table}
\begin{tabular}{ccll}
\hline
RA(J2000)& Dec(J2000) & Bouwens et al. 04/08 & Oesch et al. 09a\\
\hline
03 02 38.80 &$-$27 47 07.2 & UDF-983-964   & HUDF-480 \\
03 32 44.02 &$-$27 47 27.3 & UDF-3244-4727 & HUDF-708 \\
03 32 42.56 &$-$27 46 56.6 & UDF-640-1417  &          \\
03 32 42.56 &$-$27 47 31.4 & UDF-387-1125  &          \\
03 32 39.54 &$-$27 47 17.4 & UDF-825-950   &          \\
\hline
\end{tabular}
\caption{Existing galaxy candidates at $z > 6.5$ in the HUDF as derived from the NICMOS+ACS imaging
available in the field since 2004. Five candidates have survived, two of which plausibly lie 
at $z > 7$ (see Fig. 1). As well as positions, catalogue numbers as used by
Bouwens et al. (2004,2008) and by Oesch et al. (2009a) are given for ease of reference. All except the 
second of these candidates have been re-imaged in the current programme with WFC3, and survive in our  
final high-redshift galaxy sample presented in Table 2.}
\end{table}

\begin{figure*}
\begin{tabular}{llll}
\includegraphics[width=0.48\textwidth]{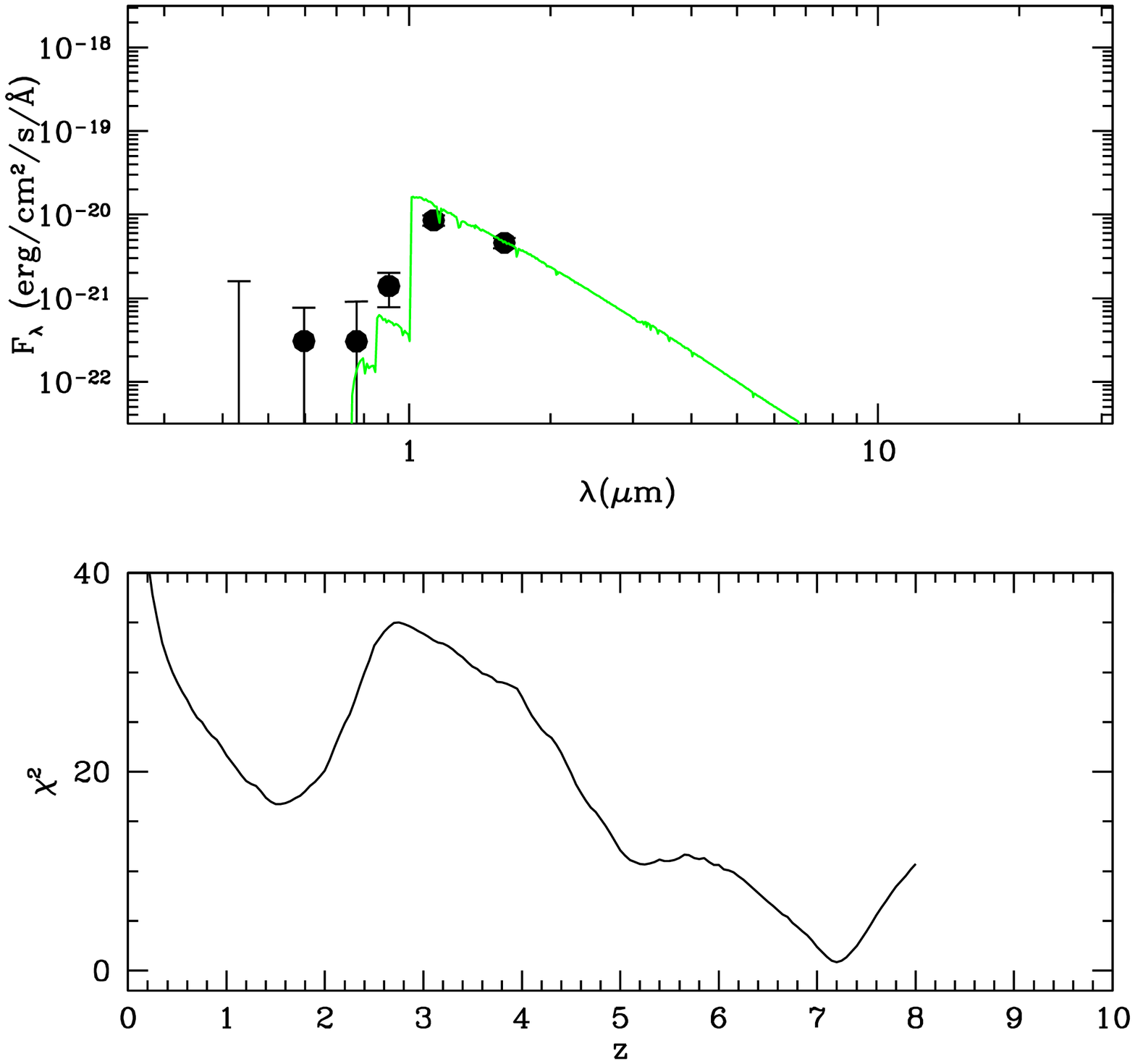}&
\includegraphics[width=0.48\textwidth]{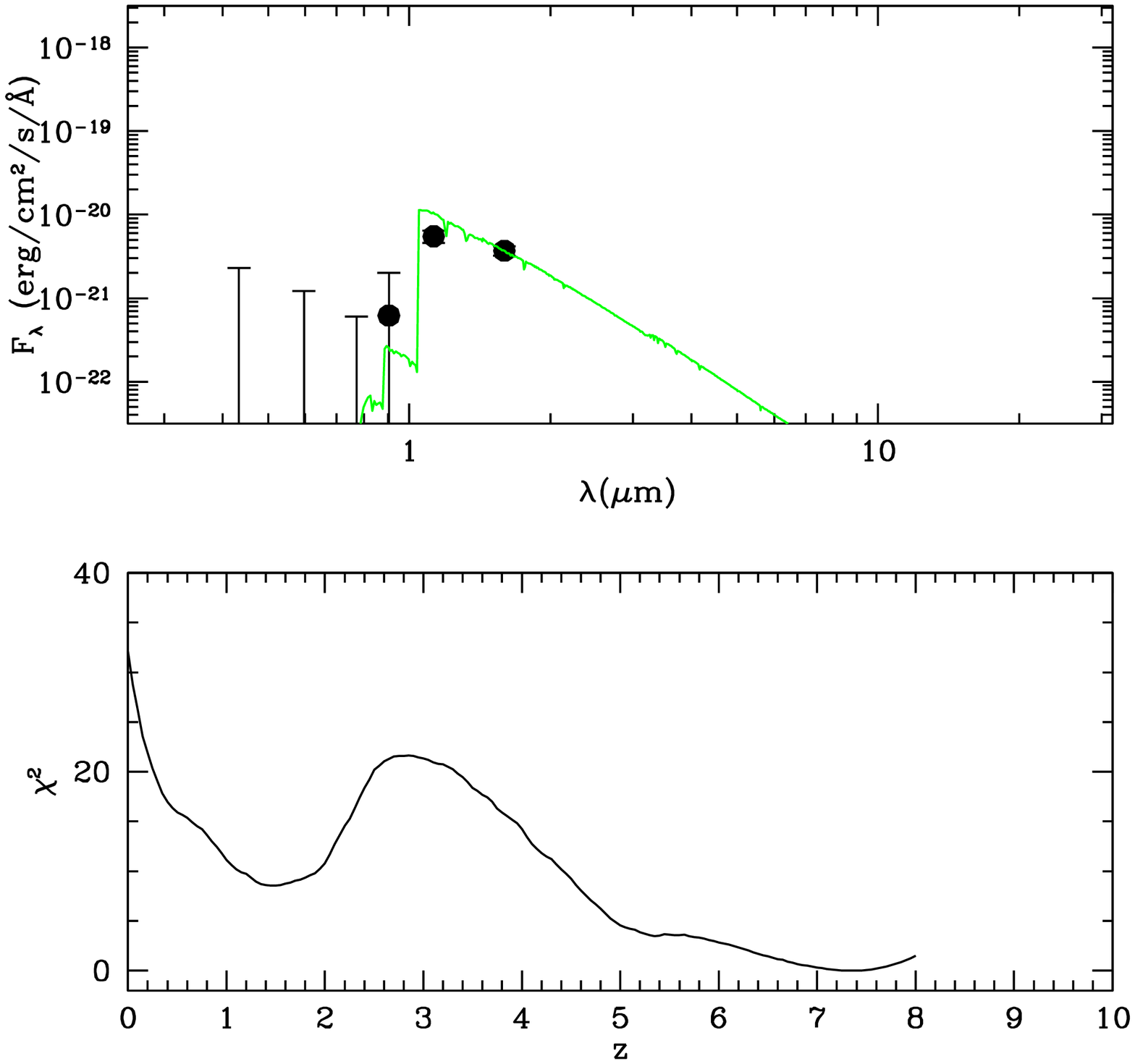}\\
\hspace*{1cm}UDF-983-964 & \hspace*{1cm}UDF-3244-4727\\
\end{tabular}
\caption{The results of our SED fitting to the optical (ACS) and near-infrared 
(NICMOS F110W \& F160W) photometry for the two most plausible galaxy 
candidates at $z > 7$ discovered within the HUDF prior to the new WFC3/IR 
imaging campaign. The upper panels show the best-fitting SED while the
lower panels show the dependence of $\chi^2$ on redshift (marginalised over all other fitted parameters). }
\end{figure*}

Unfortunately UDF-3244-4727 (= HUDF-708) has been missed by the pointing chosen for the new WFC3/IR image 
within the HUDF (see Section 3.3). However, the reality of the other 4 candidates listed in Table 1 has been confirmed by the new WFC3 
imaging, and we present revised redshift estimates for these objects (based on the vastly improved photometry) 
in Section 4 alongside the results for the new high-redshift galaxy candidates.

\section{New HST WFC3/IR data}

\subsection{Wide Field Camera 3}
Wide Field Camera 3 (WFC3) is the new near-infrared (WFC3/IR) and optical/ultra-violet (WFC3/UVIS) camera 
onboard the refurbished HST. The detector on the near-infrared channel 
is a 1024 $\times$ 1024 HgCdTe array, in which the semi-conductor band-gap has been enhanced to produce 
a sharp drop in sensitivity beyond $\lambda = 1.7\,{\rm \mu m}$. 1014 $\times$ 1014 pixels are exposed to the sky (the remainder 
providing reference pixels to track bias drifts), and the resulting field of view is
123 $\times$ 136\,arcsec, with a pixel scale of 0.13\,arcsec.

WFC3/IR is equipped with 17 filters. The filters used in the present study are the wide filters 
F105W ($Y_{105}$), F125W ($J_{125}$), and F160W ($H_{160}$), with these 3 bands being selected as most appropriate 
for providing sufficient depth {\it and} also sufficient colour information to successfully isolate high-redshift galaxies (i.e. 
although F105W and F125W do overlap somewhat, in combination they
offer much better redshift discrimination than F110W, the even wider $Y+J$ filter previously used in 
deep NICMOS imaging).

The WFC3/IR instrument represents a major advance over the NICMOS camera. Its field of view is 
$\simeq 6$ times greater, it is $2-4$ times as sensitive, and it
offers a $\simeq 2$ times improvement
in angular resolution. For a programme that aims to conduct a blank-field search for faint, compact, 
objects, these improvements add up to a factor of $\simeq 30-40$ improvement in survey efficiency.

\subsection{Observing Programme}

This study utilises new deep $Y_{105}$, $J_{125}$ and $H_{160}$ imaging taken 
with a single $\simeq 4.5$\,arcmin$^2$ pointing of WFC3/IR 
within the wider 11\,arcmin$^2$ covered by the existing 
HST ACS optical imaging of the HUDF. This is one of 3 ultra-deep 
WFC3/IR pointings which will be completed over the next year (the other two
will be located within the nearby HUDF05 parallel fields; Oesch et al. 2007) 
as part of the 192-orbit HUDF09 HST Treasury programme GO-11563 (P.I. G. Illingworth). 

The first 62 orbits of observations in the HUDF were obtained from 
August 26, 2009 to September 6, 2009, with the raw data being 
released to the public on September 9, 2009.

\subsection{Field selection and supporting data}

The WFC3/IR pointing in the HUDF is centred on $3^h 32^m 38.5^s$, $-27^\circ 47^{\prime} 0.0^{\prime \prime}$.
As shown in Fig. 2, this pointing lies somewhat to the north of the existing NICMOS F110W and F160W imaging within the HUDF
(Thompson et al. 2005), and the deep VLT ISAAC $K'$-band image referred to by Labbe et al. (2006) (which 
reaches a $5\sigma$ detection limit of $K'(AB) \simeq 25.5$).

The WFC3/IR pointing still lies within the ACS image, 
and thus benefits from the ultra-deep, high-resolution, optical data 
obtained by HST in 
4 different filters: $F435W\,(B_{435})$, $F606W\,(V_{606})$, $F775W\,(i_{775})$ and 
$F850LP\,(z_{850})$ (Beckwith et al. 2006). These deep optical data (in particular
the 150 orbits of $z_{850}$ imaging) are crucial for the effective exploitation of the 
new deep WFC3/IR data in the search for galaxies at $z > 7$, where the fundamental 
aim is to establish the existence of a strong Lyman break ($\lambda_{rest} = 1215$\AA) 
redshifted to $\lambda_{obs} > 9500$\AA. Within a 0.6\,arcsec diameter aperture, we find these
optical data reach $5\sigma$ detection limits of $B_{435} = 28.8$, $V_{606} = 29.2$, 
$i_{775} = 28.8$, $z_{850} = 28.1$ (consistent with the smaller aperture depths quoted 
by Bouwens et al. 2006).

Finally, the HUDF is also covered by ultra-deep {\it Spitzer} imaging with the Infrared Array Camera (IRAC: Fazio et al. 2004) 
taken as part of GOODS (proposal ID 194, Dickinson et al., in preparation), in 
all 4 IRAC channels (3.6, 4.5, 5.6, and 8.0${\rm \mu m}$).
The IRAC 5$\,\sigma$ (AB mag) detection limits are $S_{3.6} \simeq 25.9$, $S_{4.5} \simeq 25.5$, $S_{5.6} \simeq 23.3$, $S_{8.0} \simeq 22.9$.

\begin{figure}
\includegraphics[width=0.47\textwidth]{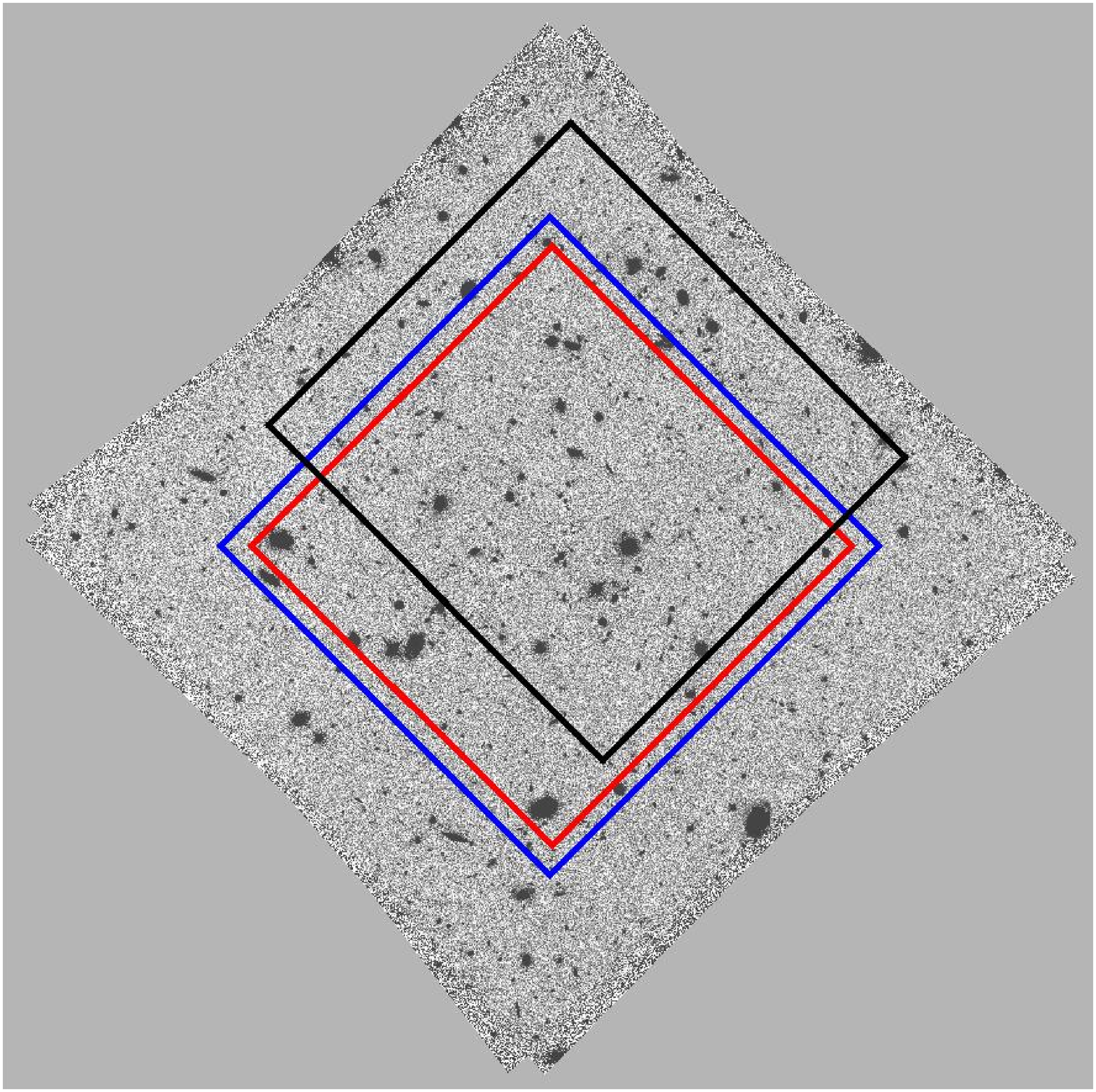}
\caption{The new WFC3 pointing is indicated by the black rectangle superimposed on the HST ACS
$z_{850}$ image of the HUDF. The blue box shows the location of the existing HST NICMOS F110W and F160W 
imaging of the field (Thompson et al. 2005), while the red box indicates the region covered by the deep 
$K'$-band image obtained with ISAAC on the VLT (Labb\'{e} et al. 2006).}
\end{figure}

\subsection{Data acquired}
The new HST near-infrared data were obtained over a total of 31 2-orbit visits 
with 2 WFC3/IR exposures being taken in each orbit. 
The total of 62 orbits was divided into 
28 orbits in $H_{160}$, 16 orbits in $J_{125}$, and 18 orbits in $Y_{105}$.
Unfortunately, the $Y_{105}-$band exposures in two orbits were
significantly affected by persistence from bright sources in an
unrelated HST observing programme, and were excluded from subsequent
image combination.

All exposures had the same
integration time of 1400 seconds and were obtained in MULTIACCUM mode
using the SPARS100 readout sequence, which involves a series of
non-destructive samples of the detector at equally spaced intervals
of 100 seconds. An additional pair of reads was obtained at the start
of each exposure, with the first read effectively serving as the bias
level for the remainder of the exposures. Small ($< 1$\,arcsec) offsets were made between each of the four 1400-second 
exposures in a given visit. Larger offsets, 
of up to 10\,arcsec, were made between different visits.

\subsection{Data reduction}

The exposures were calibrated using the latest version of the STSDAS
Pyraf task calwfc3 (released on September 9, 2009) which
applies bias level removal using the zero-th read of each exposure,
as well as dark current subtraction, flat-fielding and linearity correction.
Since the standard pipeline calibration was based on ground-based dark
current and flat-field files, a new "superdark" file was constructed
from on-orbit dark exposures obtained during approximately the same
time frame as the observations, and which became public at the same
time. Comparison tests were carried out between the calibration
results of the pipeline ground-based darks and the new on-orbit darks,
showing that the latter resulted in improved dark current subtraction
of the exposures.

Remaining low-level residuals in the images were removed
iteratively, after first combining the images using MultiDrizzle
(Koekemoer et al. 2002) which makes use of the drizzle software (Fruchter \& Hook 2002).
This allowed us to create a clean image of the field
which could then be used to create a high signal-to-noise source mask.
This source mask was then transformed back to the frame of each
detector, and a large-scale, low-resolution median filter was applied
to the remaining pixels in order to create an image characterizing
the low-order residuals for each filter. This process also involved
treating the four amplifier quadrants separately, in order to include
differences between the amplifiers in the low-level residual image.
When running MultiDrizzle, the images were combined using inverse
variance weighting, which involves calculating for each exposure
the total expected noise due to the background sky (modulated by
the flatfield structure), as well as the accumulated dark current
and read noise added in quadrature, giving the total noise
component due to the background. This is compared with the final
r.m.s. in the multidrizzled images, in order to assess the amount of
correlated noise in the final images.

After applying the scaled residual image correction to each exposure,
the exposures were verified to have no significant remaining residual
signature and were combined using MultiDrizzle (Koekemoer et al. 2002),
this time creating a new drizzled image for each of the 3 filters, with
no signifcant low-level residuals remaining. The output scale was set
to 0.03 arcsec per pixel in order to match the original ACS images; a
total of 850 - 900 sources were matched between the WFC3 and ACS images
in order to determine the astrometric solution and geometric transformation
from the WFC3 to the original UDF ACS images, which were taken as
the astrometric reference frame. The resulting astrometric solutions
were applied to the image header world coordinate system parameters and 
distortion coefficients for each exposure, and Multidrizzle was then
re-run to create images at both 0.03-arcsec and 0.06-arcsec per pixel, matched
to the ACS astrometry.

The resulting images are aligned to the ACS data
with an r.m.s. of $\simeq 0.1$ times the final WFC3 PSF and are on the same pixel
grid as the ACS images, enabling direct photometry to be carried out
between the datasets. A pixfrac of 0.6 was used to minimize additional
convolution; the small output pixel size also reduces the convolution
of the PSF, with the resulting PSF widths being 0.15 arcsec in $Y_{\rm 105}$,
0.16 arcsec in $J_{\rm 125}$, and 0.18 arcsec in $H_{\rm 160}$, matching the
theoretical values expected after convolution by the 0.13-arcsec WFC3/IR
pixels is taken into account.

The final WFC3/IR images have 
total exposure times of $\simeq 11$ hours in $Y_{105}$, $\simeq 12$
hours in $J_{125}$, and $\simeq 22$ hours in $H_{160}$.

\subsection{Photometric Zero Points}

We established our own independent zero points by comparison with catalogues
derived from deep ground-based imaging of the field in the $Y-$band
with VLT+HAWK-I (Programme ID: 60.A-9284(A)) and in the $J-$ and
$H-$bands with VLT+ISAAC (Programme ID: 168.A-0485). Moreover, we
also independently checked our $H-$band zero point determination using objects
in the overlap region between the new WFC3 imaging and the existing
deep NICMOS F160W imaging of the HUDF (Thompson et al. 2005). Finally,
we confirmed our zero point determination using large aperture 
(6-arcsec diameter) photometry of two bright stars in the HUDF which were found
to have approximately flat $Y-H$ colours from the existing ground-based imaging. This
process led to the determination of the following AB zero points:
$Y_{105}(zpt)=26.25$, $J_{125}(zpt)=26.25$ and $H_{160}(zpt)=26.00$, 
with a typical uncertainty of $\simeq 0.05$ magnitudes. We note here that our
independent determinations of the $Y_{105}$, $J_{125}$ and $H_{160}-$band zero points are
within $0.04$ magnitudes of the most recent, on-orbit values\footnote{http://www.stsci.edu/hst/wfc3}. 

Following the determination of the photometric zero points we derived
robust measurements of the image depths in the three near-infrared bands. To
achieve this we placed a grid of $\geq 10,000$ circular apertures on
blank sky regions within the final WFC3/IR mosaics. The resulting
distribution of blank sky aperture fluxes was then used to determine
the r.m.s. background noise of the WFC3/IR mosaics on the scale of our
chosen apertures. In order to determine what regions of the images were
largely uncontaminated by object flux we used a $\chi^{2}$
image (Szalay et al. 1999) constructed from the deep $BViz$ ACS
imaging of the HUDF. The results of this analysis
demonstrated that, within 0.4-arcsec diameter apertures, the $5\sigma$
depths of our reductions of the WFC3 images are $Y_{105}=29.01$,
$J_{125}=29.07$ and $H_{160}=29.08$ (we note that this is consistent with
the approximate $5\sigma$ depth of $\simeq 29$ in all bands derived from their
own independent reduction by Bouwens et al. 2009).

\begin{figure*}
\begin{tabular}{llll}
\includegraphics[width=0.48\textwidth]{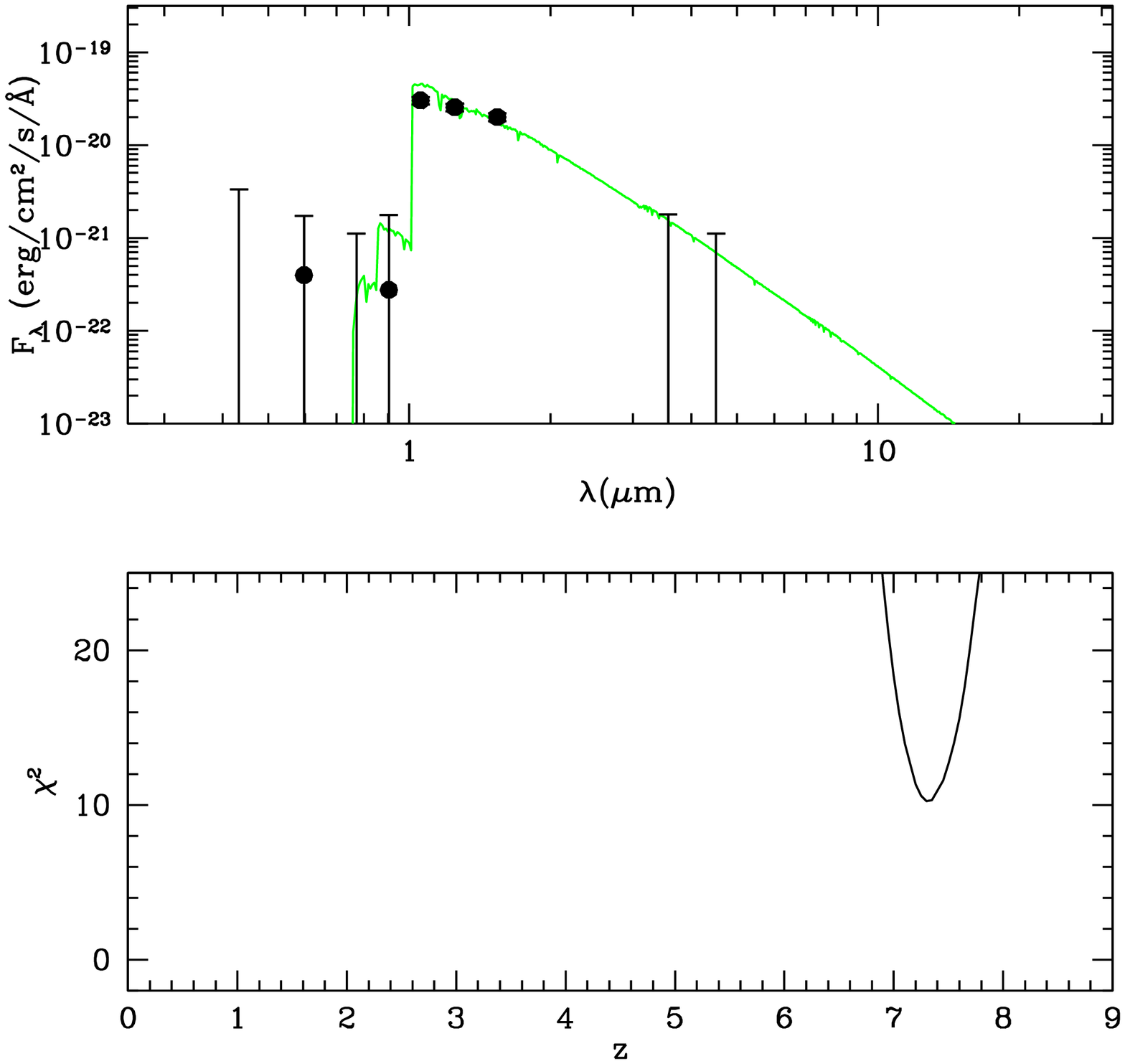}&
\includegraphics[width=0.48\textwidth]{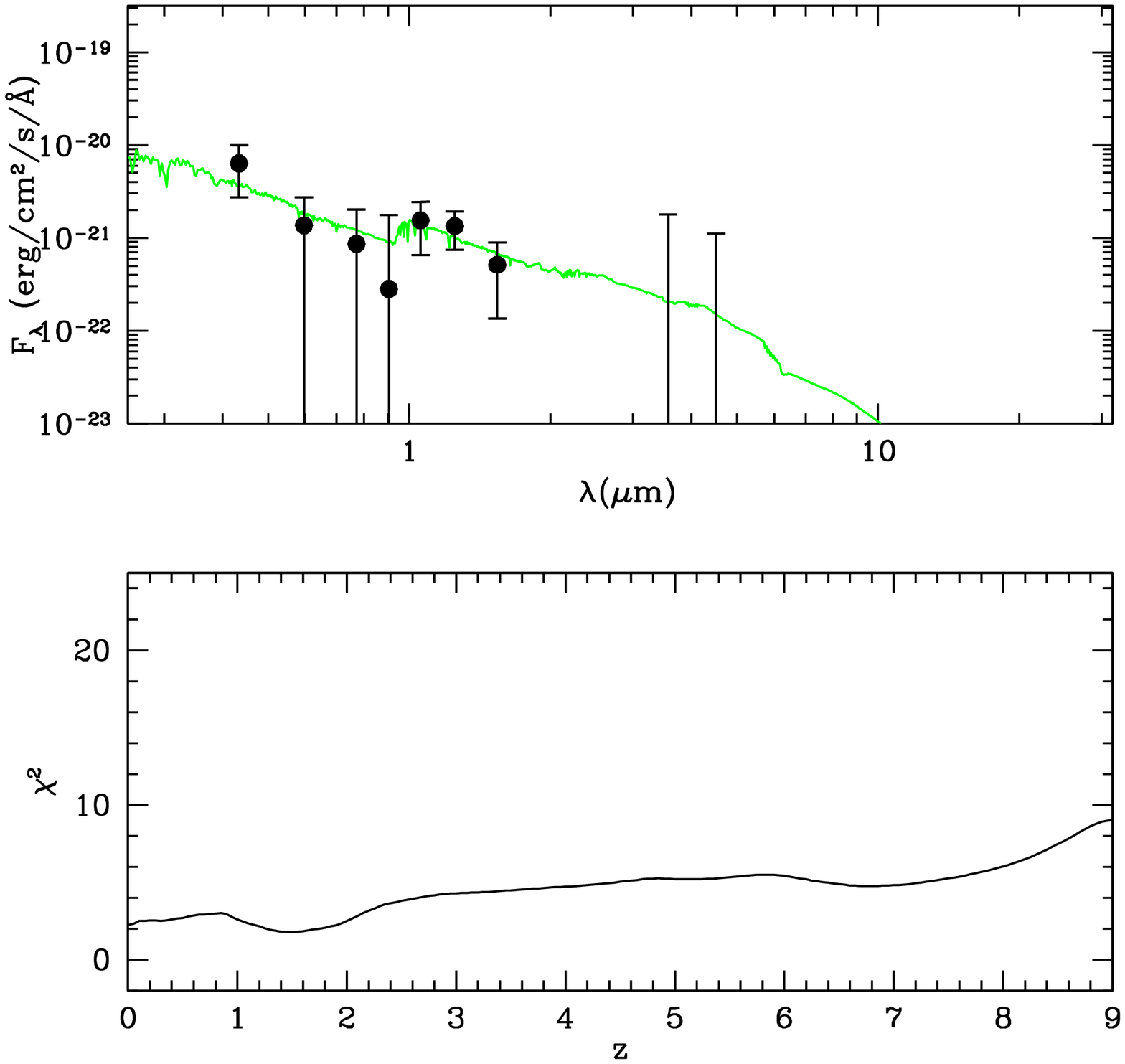}\\
\end{tabular}
\caption{Two different examples of contaminants uncovered within our initial high-redshift galaxy sample. 
The left-hand plot shows what at first sight appears to be the brightest 
$z > 7$ galaxy in the sample. However, this is actually a supernova (or some other transient) at RA 03:32:34.53, Dec -27:47:36.0, 
which has appeared since the original NICMOS F160W imaging of the HUDF. The right-hand plot shows our SED fit to our photometry of UDFz-38537519,
the faintest $z \simeq 7$ galaxy reported by Oesch et al. (2009b). This in fact appears to be a blue, low-mass galaxy at $z \simeq 1.5$, and highlights
the problems that can arise when the WFC3/IR data is pushed to a depth at which little colour discrimination can be provided 
by the ACS optical imaging.}
\end{figure*}

\begin{table*}
\begin{tabular}{lccccrccr}
\hline
{ID} & {RA(J2000)} & {Dec(J2000)} & {$z$} & {$\delta z$} & {$\chi^2$} & {$z_2$} & {$\delta z_2$} & {$\chi_2^2$}\\
\hline
  1735$^i$  & 03 32 39.86 & $-$27 46 19.1 &    5.90 &  (5.70-6.05) &	6.79  & 1.20  & (1.05-1.30)  &  14.64  \\
  1955$^i$  & 03 32 39.46 & $-$27 45 43.4 &    5.90 &  (5.75-6.10) &	2.78  & 1.15  & (1.05-1.30)  &  12.70   \\
  1719$^i$  & 03 32 44.70 & $-$27 46 45.6 &    5.95 &  (5.75-6.15) &	0.77  & 1.15  & (1.05-1.30)  &  10.94  \\
  2217\phantom{$^i$}  & 03 32 40.56 & $-$27 48 02.7 &    5.95 &  (5.40-6.20) &	2.69  & 1.20  & (1.05-1.40)  &  5.67   \\
  \phantom{1}962$^i$   & 03 32 35.05 & $-$27 47 40.1 &    5.95 &  (5.80-6.20) &	3.54  & 1.25  & (1.15-1.35)  &  15.55  \\
  1189$^i$  & 03 32 36.98 & $-$27 45 57.6 &    6.00 &  (5.85-6.15) &	1.70  & 1.20  & (1.10-1.30)  &  16.58  \\
  2830\phantom{$^i$}  & 03 32 34.58 & $-$27 46 58.0 &    6.00 &  (5.45-6.35) &	0.57  & 1.20  & (1.00-1.40)  &  2.28   \\
  2498\phantom{$^i$}  & 03 32 35.04 & $-$27 47 25.8 &    6.00 &  (5.70-6.35) &	2.47  & 1.15  & (0.95-1.35)  &  7.81   \\
  2719\phantom{$^i$}  & 03 32 40.59 & $-$27 45 56.9 &    6.05 &  (5.30-6.90) &	0.65  & 1.25  & (1.00-1.60)  &  2.29   \\
  1625$^i$  & 03 32 43.03 & $-$27 46 23.8 &    6.05 &  (5.90-6.25) &	1.13  & 1.20  & (1.10-1.30)  &  14.78  \\
  1398$^i$  & 03 32 36.63 & $-$27 47 50.1 &    6.10 &  (5.90-6.30) &	0.46  & 1.25  & (1.10-1.35)  &  12.87  \\
  1760\phantom{$^i$}  & 03 32 40.25 & $-$27 46 05.2 &    6.15 &  (5.85-6.45) &	0.24  & 1.25  & (1.05-1.40)  &  6.10    \\
  \phantom{1}934$^i$   & 03 32 37.48 & $-$27 46 32.5 &    6.20 &  (6.00-6.40) &	0.58  & 1.25  & (1.15-1.35)  &  14.82  \\
  2791\phantom{$^i$}  & 03 32 36.64 & $-$27 47 50.2 &    6.25 &  (5.95-6.50) &	1.24  & 1.25  & (1.10-1.40)  &  8.09   \\
  1464\phantom{$^i$}  & 03 32 42.19 & $-$27 46 27.9 &    6.30 &  (5.95-6.75) &	0.55  & 1.30  & (1.15-1.45)  &  7.33   \\
  2003\phantom{$^i$}  & 03 32 36.46 & $-$27 47 32.4 &    6.30 &  (5.65-6.80) &	1.39  & 1.30  & (1.00-1.55)  &  1.67   \\
  2514\phantom{$^i$}  & 03 32 39.79 & $-$27 46 33.8 &    6.30 &  (5.85-6.75) &	0.09  & 1.25  & (0.95-1.50)  &  3.86   \\
  \phantom{1}837$^i$   & 03 32 37.46 & $-$27 46 32.8 &    6.35 &  (6.15-6.55) &	1.18  & 1.30  & (1.20-1.40)  &  25.79  \\
  1855\phantom{$^i$}  & 03 32 43.79 & $-$27 46 33.8 &    6.40 &  (6.15-6.70) &	0.41  & 1.30  & (1.10-1.45)  &  7.39   \\
  1864\phantom{$^i$}  & 03 32 34.52 & $-$27 47 34.8 &    6.40 &  (5.90-6.85) &	0.17  & 1.30  & (1.10-1.50)  &  4.33   \\
  1911$^z$            & 03 32 36.77 & $-$27 47 53.6 &    6.40 &  (6.20-6.60) &	0.97  & 1.30  & (1.15-1.40)  &  13.99  \\
  1915$^z$            & 03 32 39.58 & $-$27 46 56.5 &    6.40 &  (6.15-6.65) &	3.27  & 1.30  & (1.15-1.45)  &  10.31  \\
  2195\phantom{$^i$}  & 03 32 43.05 & $-$27 47 08.1 &    6.45 &  (4.75-7.50) &	2.39  & 1.40  & (0.00-7.65)  &  2.63   \\
  1880$^z$            & 03 32 37.44 & $-$27 46 51.3 &    6.50 &  (6.25-6.80) &	0.36  & 1.30  & (1.15-1.50)  &  7.77   \\
  1958$^z$            & 03 32 36.38 & $-$27 47 16.2 &    6.50 &  (6.25-6.80) &	0.17  & 1.30  & (1.15-1.50)  &  6.93   \\
  2206$^z$            & 03 32 40.58 & $-$27 46 43.6 &    6.50 &  (6.20-6.85) &	0.83  & 1.30  & (1.15-1.55)  &  5.03   \\
  1064\phantom{$^i$}  & 03 32 34.93 & $-$27 47 01.3 &    6.65 &  (6.35-6.90) &	2.04  & 1.35  & (1.25-1.50)  &  14.21  \\
  \phantom{1}688$^z$             & 03 32 42.56 & $-$27 46 56.6 &    6.70 &  (6.50-6.90) &	0.88  & 1.40  & (1.30-1.50)  &  40.82  \\
  2794\phantom{$^i$}  & 03 32 36.75 & $-$27 46 48.2 &    6.75 &  (6.20-7.45) &	1.37  & 1.50  & (1.20-1.70)  &  2.45   \\
  1144$^z$            & 03 32 42.56 & $-$27 47 31.5 &    6.80 &  (6.50-7.10) &	0.40  & 1.40  & (1.25-1.55)  &  12.94  \\
  2395\phantom{$^i$}  & 03 32 44.31 & $-$27 46 45.2 &    6.80 &  (6.40-7.30) &	1.55  & 1.45  & (1.20-1.65)  &  4.54   \\
  1092$^z$            & 03 32 39.55 & $-$27 47 17.5 &    6.85 &  (6.45-7.35) &	3.16  & 1.55  & (1.35-1.75)  &  8.83   \\
  2560$^z$            & 03 32 37.80 & $-$27 47 40.4 &    6.90 &  (6.35-7.45) &	0.52  & 1.45  & (1.10-1.75)  &  3.38   \\
  2826\phantom{$^i$}  & 03 32 37.06 & $-$27 48 15.2 &    6.90 &  (6.30-7.30) &	2.56  & 1.50  & (1.05-1.70)  &  4.90    \\
  1678$^z$            & 03 32 43.14 & $-$27 46 28.6 &    7.05 &  (6.60-7.40) &	1.55  & 1.50  & (1.30-1.70)  &  6.86   \\
  2502$^z$            & 03 32 39.73 & $-$27 46 21.4 &    7.10 &  (6.60-7.65) &	1.85  & 1.50  & (1.20-1.85)  &  6.40    \\
  1574$^z$            & 03 32 37.21 & $-$27 48 06.1 &    7.20 &  (6.55-7.60) &	0.27  & 1.50  & (1.25-1.75)  &  4.46   \\
  \phantom{1}835$^z$             & 03 32 38.81 & $-$27 47 07.2 &    7.20 &  (6.90-7.50) &	0.90  & 1.60  & (1.45-1.75)  &  15.99  \\
  2066$^z$            & 03 32 41.05 & $-$27 47 15.6 &    7.20 &  (6.50-7.80) &	0.60  & 1.50  & (1.15-1.80)  &  3.69   \\
  2888\phantom{$^i$}  & 03 32 44.75 & $-$27 46 45.1 &    7.35 &  (6.10-8.35) &	0.02  & 1.60  & (0.80-2.10)  &  0.88   \\
  2940\phantom{$^i$}  & 03 32 36.52 & $-$27 46 41.9 &    7.40 &  (6.40-8.25) &	0.41  & 1.50  & (0.90-2.00)  &  1.85   \\
  2079$^y$            & 03 32 37.63 & $-$27 46 01.5 &    7.50 &  (6.35-8.25) &	0.67  & 1.65  & (1.10-2.05)  &  1.47   \\
  1107$^z$            & 03 32 44.71 & $-$27 46 44.4 &    7.60 &  (7.30-7.90) &	0.17  & 1.65  & (1.45-1.85)  &  11.01  \\
  1422\phantom{$^i$}  & 03 32 39.52 & $-$27 47 17.3 &    7.60 &  (7.00-8.05) &	0.17  & 1.75  & (1.45-2.00)  &  3.52   \\
  2487\phantom{$^i$}  & 03 32 33.13 & $-$27 46 54.4 &    7.80 &  (7.10-8.30) &	1.12  & 1.75  & (1.15-2.00)  &  3.45   \\
  1765$^y$            & 03 32 42.88 & $-$27 46 34.6 &    7.95 &  (7.35-8.50) &	1.04  & 1.90  & (1.50-2.05)  &  2.83   \\
  2841$^y$            & 03 32 43.09 & $-$27 46 27.9 &    8.10 &  (7.40-8.75) &	1.11  & 1.90  & (1.25-2.20)  &  2.09   \\
  1939$^y$            & 03 32 37.80 & $-$27 46 00.1 &    8.35 &  (7.90-8.70) &	3.79  & 1.90  & (1.70-2.15)  &  4.79   \\
  1721$^y$            & 03 32 38.14 & $-$27 45 54.0 &    8.45 &  (7.75-8.85) &	0.02  & 1.95  & (1.60-2.15)  &  2.97   \\
\hline\end{tabular}
\caption{Our 49 high-redshift galaxy candidates, ranked in order of increasing 
primary photometric redshift. Column 1 gives our catalogue number, and
a $i$, $z$ or $y$ superscript is used to indicate objects which have
been independently selected as $i-$drops, $z$-drops or $Y$-drops by
Bunker et al. (2004), Oesch et al. (2009b) and Bouwens et al.
(2009) respectively. Object positions are given in columns 2 and 3. The primary redshift is listed in column 4, with the 1-$\sigma$ confidence interval 
given in column 5, and the minimum $\chi^2$ achieved given in column 6. Columns  7, 8 and 9 give the equivalent information for the secondary redshift solution
(see Appendix A for plots of $\chi^2$ versus redshift for each object). 
We have 
removed all obvious blended sources from the SExtractor catalogue, but 
note here that we have retained 934 \& 837 as separate objects, 
and 1422 \& 1092 as separate objects, even though in both cases the object
pairs are separated by $\simeq 0.5$ arcsec, and the redshift 
solutions are consistent.}
\end{table*}

\section{The high-redshift galaxy sample}

\subsection{Catalogue extraction and photometry}
The original master object catalogue for this study was produced using
version 2.5.2 of the SExtractor software package (Bertin \& Arnouts
1996). For this catalogue we included all objects
which were detected at $\geq 5\sigma$ significance in at least one of
the near-infrared images in a 0.4-arcsec
diameter aperture. For the purposes of SED analysis and photometric
redshift determination we performed aperture photometry on all objects
within circular apertures of 0.6-arcsec diameter. This aperture
encloses $\simeq 80\%$ of the total flux of a point source in
all three WFC3 images and has a consistent point-souce aperture
correction of $0.22\pm0.03$ magnitudes from $Y_{150}$ through to
$H_{160}$. Moreover, the choice of 0.6-arcsec
diameter apertures avoids biases which can result from very small
aperture photometry, the use of which implicitly assumes that 
high-redshift galaxies are unresolved (and inspection of the 
images shows that a substantial fraction of the high-redshift 
galaxy candidates are resolved by WFC3).

Starting from the master catalogue we confined our attention to high
redshifts by rejecting all sources which were detected at $\geq
2\sigma$ significance in the ACS $i-$band imaging of the UDF. This
resulted in a sample of $\simeq 300$ objects which, while incomplete
for $i$-drops, should retain all potential $i$-dropouts and
higher-redshift galaxy candidates at $z > 6.25$.

To homogenise the photometry, and to avoid confusion over how to aperture-correct 
upper limits on non-detected sources, we corrected the WFC3/IR
0.6-arcsec diameter magnitudes for missing
point-source flux density {\it relative} to the ACS optical imaging.
These corrections were determined empirically from point sources in the
field (common to all images) and amount to corrections of 0.02, 0.04 and 0.08 mag in $Y_{105}$, $J_{125}$ and $H_{160}$ respectively.

\subsection{Redshift estimation}

In contrast to Oesch et al. (2009b) and Bouwens et al. (2009) we have not used 3-band colour information to 
attempt to select high-redshift galaxies in different redshift windows. Rather we performed a full
SED fit to the 7-band photometry for all $\simeq 300$ candidates to derive estimates of, and constraints on their redshifts. 
Inspection of the available IRAC data shows that none of our high-redshift candidates are securely detected at $\geq2 \sigma$
significance in the 3.6$\mu m$ and 4.5$\mu m$ imaging (c.f. Labb\'{e} et al. 2009). However, due to the lack of comparably deep near-imaging 
at wavelengths longer than 1.6$\mu m$ (e.g. $K$-band) it is important
to make use of the information provided by the IRAC non-detections to help rule-out otherwise acceptable redshift solutions at $z \simeq 5 - 6$. 
Consequently, during the SED fitting procedure we adopted $2 \sigma$ upper limits for the non-detections at 3.6$\mu m$ and 4.5$\mu m$ of
26.0 (AB), which are a factor of $\simeq 2$ brighter
than the recent determination by Labb\'{e} et al. (2009). We note that
this is a conservative approach, and that a robust de-confusion of the
available IRAC data (Cirasuolo et al. 2010, in prep) is expected to produce $2 \sigma$ limits which
are fainter, and therefore more able to rule-out potential $z \simeq
5-6$ solutions. Moreover, as in Dunlop et al. (2007), 
rather than simply adopting global upper limits for non-detections in the optical/near-infrared wavebands, we have retained even 
low-significance measurements of flux-density and derived the locally measured r.m.s. within a series of 0.6-arcsec diameter apertures 
for use in the SED fitting process.

The SED fitting procedure we applied to 
derive the photometric redshifts of the high-redshift galaxy candidates
was based largely on the public package Hyperz (Bolzonella et al. 2000).
The observed photometry was fitted with synthetic galaxy templates generated with the 2007 stellar population models of Charlot \& Bruzual. We used a variety
of star-formation histories: instantaneous burst and exponentially declining star-formation with e-folding times $0.1 \le \tau(\rm Gyr) \le 10$, assuming solar
metallicity and a Salpeter initial mass function (IMF). At each redshift step we allowed galaxy templates with ages ranging from 1 Myr to the age of
the Universe at that redshift. Moreover, in addition to the synthetic
stellar populations we also used the composite SED of Lyman-break
galaxies at $z \simeq 3$ derived by Shapley et al. (2003).
For dust reddening we adopted the prescription from Calzetti et
al. (2000), within the range $\rm 0 \le A_V \le 2$, and included
absorption due to $\rm H_I$ clouds along the line of sight according to Madau (1995).

Examination of the resulting SED fits and plots of $\chi^2$ versus $z$ then
allowed us to reject sources with no acceptable redshift solutions at $z > 6$, and identify ``objects'' with peculiar SEDs 
which we then confirmed visually to be artefacts
(e.g. the result of diffraction spikes) or objects for which the photometry was compromised by confusion with nearby galaxies.
As shown in the left-hand panel of Fig. 3, our $\chi^2$ versus $z$ plots also helped us to notice a peculiarly bright and unusually secure 
high-redshift object, which we then noticed was highly stellar, and revealed itself to be a supernova 
or other transient interloper due to its complete absence in the
pre-existing NICMOS infrared images of the UDF. Moreover, in the
right-hand panel of Fig 3 we show our SED fit to UDFz-38537519, the
faintest $z \simeq 7$ galaxy reported by Oesch et
al. (2009b). Adopting $2 \sigma$ upper limits for its non-detection in
$z_{850}$, this object is very close to the edge of the two-colour selection box
employed by Oesch et al. (2009b), and our SED fitting analysis suggests
that it is most likely a low-mass interloper at $z\simeq 1.5$. This
object illustrates the advantage of full SED fitting when dealing with
objects close to the boundaries of colour-colour selection boxes in
the presence of substantial photometric scatter.

The final outcome of this screening process was a reduction in the number of
credible high-redshift ($z \gtsim 5.9$) galaxy candidates from $\simeq 300$ to our final catalogue
of 49 objects. Before moving on to discuss the properties of our final
high-redshift galaxy sample in detail, it is worth
commenting on the potential for contamination by ultra-cool galactic T
dwarf stars. Due to deep absorption bands in their spectra, galactic
T dwarf stars are capable of mimicking the optical--near-infrared colours
of $z\simeq 7$ galaxies, and consequently, have provided a serious
source of contamination for previous ground-based studies in particular.
However, due to the combination of image depth, small field of
view and high angular
resolution (many of our high-redshift candidates are clearly resolved - 
see Appendix A), contamination of our sample by T dwarfs is expected to
be negligible. This is confirmed by considering that the typical
absolute $J-$band (AB) magnitude of T dwarf stars is $J\simeq 19$
(Leggett et al. 2009). At the depth probed by our current sample, a T-dwarf contaminant would
thus have to be located at a distance of $0.5-1.0$\,kpc. Given this
distance is $2 \rightarrow 3$ times the estimated galaxy thin disk
scale-height of $\simeq 300$\,pc (e.g. Reid \& Majewski 1993; Pirzkal
et al. 2009), it is clear that significant contamination is unlikely.
This is not to suggest that dwarf stars cannot be found at such
distances as, for example, Stanway et al.  (2008) report the discovery
of M dwarfs out to distances of $\simeq 10$\,kpc. However, in the context
of the present study, the surface density is low, with the integrated
surface density over all M dwarf types contained within $\simeq
1$\,kpc amounting to $\simeq 0.07$ per arcmin$^2$.  Extrapolating
these results to T dwarfs is somewhat uncertain, but a comparable
surface density for L and T dwarf stars is supported by the search for
such stars in deep fields undertaken by Ryan et al. (2005). The
results of this study suggest that the 4-arcmin$^2$ field-of-view
covered by the new WFC3/IR data should contain $\leq 0.5$ T dwarf
stars down to a magnitude limit of $z_{850} = 29$.

\begin{table*}
\begin{tabular}{lcccccc}
\hline
{ID} & {RA(J2000)} & {Dec(J2000)} & {$z_{850}$} & {$Y_{105}$} & {$J_{125}$} & {$H_{160}$}\\
\hline
  1735$^i$ &  03    32     39.86  &  $-$27 46     19.1   &   $27.81	 \pm  0.21 $  &  $28.17  \pm  0.22$   &  $  28.00   \pm  0.18 $ &  $   28.14   \pm   0.20  $ \\
  1955$^i$ &  03    32     39.46  &  $-$27 45     43.4   &   $27.90	 \pm  0.22 $  &  $28.28  \pm  0.24$   &  $  28.10   \pm  0.19 $ &  $   28.22   \pm   0.21 $ \\
  1719$^i$ &  03    32     44.70  &  $-$27 46     45.6   &   $28.14	 \pm  0.27 $  &  $28.08  \pm  0.21$   &  $  28.09  \pm  0.19 $ &  $   28.48   \pm   0.26 $ \\
  2217\phantom{$^z$} &  03    32     40.56  &  $-$27 48     02.7   &   $28.54	 \pm  0.36 $  &  $28.67  \pm  0.32$   &  $  28.18  \pm  0.21 $ &  $   28.40    \pm   0.24 $ \\
  \phantom{1}962$^i$  &  03    32     35.05  &  $-$27 47     40.1   &   $27.90	 \pm  0.22 $  &  $27.69  \pm  0.16$   &  $  27.25  \pm  0.12 $ &  $   27.33   \pm   0.12 $ \\
  1189$^i$ &  03    32     36.98  &  $-$27 45     57.6   &   $27.72	 \pm  0.20  $  &  $27.70   \pm  0.16$   &  $  27.59  \pm  0.14 $ &  $   27.58   \pm   0.14 $ \\
  2830\phantom{$^z$} &  03    32     34.58  &  $-$27 46     58.0   &   $28.77	 \pm  0.44 $  &  $28.69  \pm  0.32$   &  $  28.80   \pm  0.32 $ &  $   28.73   \pm   0.31 $ \\
  2498\phantom{$^z$} &  03    32     35.04  &  $-$27 47     25.8   &   $28.81	 \pm  0.45 $  &  $28.27  \pm  0.24$   &  $  29.00   \pm  0.38 $ &  $   29.12   \pm   0.42 $ \\
  2719\phantom{$^z$} &  03    32     40.59  &  $-$27 45     56.9   &   $29.24	 \pm  0.65 $  &  $29.07  \pm  0.44$   &  $  28.75  \pm  0.31 $ &  $   29.00    \pm   0.39 $ \\
  1625$^i$ &  03    32     43.03  &  $-$27 46     23.8   &   $27.98	 \pm  0.24 $  &  $27.85  \pm  0.18$   &  $  28.08  \pm  0.19 $ &  $   28.19   \pm   0.21 $ \\
  1398$^i$ &  03    32     36.63  &  $-$27 47     50.1   &   $27.96	 \pm  0.23 $  &  $27.73  \pm  0.16$   &  $  27.89  \pm  0.17 $ &  $   27.91   \pm   0.17 $ \\
  1760\phantom{$^z$} &  03    32     40.25  &  $-$27 46     05.2   &   $28.57	 \pm  0.37 $  &  $28.25  \pm  0.23$   &  $  28.47  \pm  0.25 $ &  $   28.45   \pm   0.25 $ \\
  \phantom{1}934$^i$  &  03    32     37.48  &  $-$27 46     32.5   &   $27.88	 \pm  0.22 $  &  $27.30   \pm  0.13$   &  $  27.36  \pm  0.12 $ &  $   27.26   \pm   0.12 $ \\
  2791\phantom{$^z$} &  03    32     36.64  &  $-$27 47     50.2   &   $28.56	 \pm  0.37 $  &  $28.09  \pm  0.21$   &  $  28.36  \pm  0.23 $ &  $   28.44   \pm   0.25 $ \\
  1464\phantom{$^z$} &  03    32     42.19  &  $-$27 46     27.9   &   $28.72	 \pm  0.42 $  &  $28.03  \pm  0.20 $   &  $  27.91  \pm  0.17 $ &  $   27.90    \pm   0.17 $ \\
  2003\phantom{$^z$} &  03    32     36.46  &  $-$27 47     32.4   &   $29.37	 \pm  0.73 $  &  $28.50   \pm  0.28$   &  $  28.71  \pm  0.30  $ &  $   28.36   \pm   0.23 $ \\
  2514\phantom{$^z$} &  03    32     39.79  &  $-$27 46     33.8   &   $29.32	 \pm  0.69 $  &  $28.70   \pm  0.33$   &  $  28.91  \pm  0.35 $ &  $   29.10    \pm   0.42 $ \\
  \phantom{1}837$^i$  &  03    32     37.46  &  $-$27 46     32.8   &   $28.01	 \pm  0.24 $  &  $27.23  \pm  0.12$   &  $  27.24  \pm  0.12 $ &  $   27.07   \pm   0.11 $ \\
  1855\phantom{$^z$} &  03    32     43.79  &  $-$27 46     33.8   &   $28.84	 \pm  0.46 $  &  $28.02  \pm  0.20 $   &  $  28.09  \pm  0.19 $ &  $   28.30    \pm   0.22 $ \\
  1864\phantom{$^z$} &  03    32     34.52  &  $-$27 47     34.8   &   $29.14	 \pm  0.59 $  &  $28.37  \pm  0.25$   &  $  28.48  \pm  0.26 $ &  $   28.43   \pm   0.25 $ \\
  1911$^z$           &  03    32     36.77  &  $-$27 47     53.6   &   $28.61	 \pm  0.39 $  &  $27.76  \pm  0.17$   &  $  28.08  \pm  0.19 $ &  $   28.15   \pm   0.20  $ \\
  1915$^z$           &  03    32     39.58  &  $-$27 46     56.5   &   $29.02	 \pm  0.54 $  &  $27.97  \pm  0.19$   &  $  28.20   \pm  0.21 $ &  $   28.54   \pm   0.27 $ \\
  2195\phantom{$^z$} &  03    32     43.05  &  $-$27 47     08.1   &   $>29.3 \phantom{\pm 0.55} $  &  $29.27  \pm  0.52$   &  $  30.72  \pm  1.66 $ &  $   29.23   \pm   0.46 $ \\
  1880$^z$           &  03    32     37.44  &  $-$27 46     51.3   &   $29.20	 \pm  0.62 $  &  $28.02  \pm  0.20 $   &  $  28.17  \pm  0.20  $ &  $   28.36   \pm   0.23 $ \\
  1958$^z$           &  03    32     36.38  &  $-$27 47     16.2   &   $29.25	 \pm  0.65 $  &  $28.07  \pm  0.20 $   &  $  28.26  \pm  0.22 $ &  $   28.35   \pm   0.23 $ \\
  2206$^z$           &  03    32     40.58  &  $-$27 46     43.6   &   $29.59	 \pm  0.87 $  &  $28.35  \pm  0.25$   &  $  28.66  \pm  0.29 $ &  $   28.76   \pm   0.32 $ \\
  1064\phantom{$^z$} &  03    32     34.93  &  $-$27 47     01.3   &   $28.58	 \pm  0.37 $  &  $27.43  \pm  0.14$   &  $  27.35  \pm  0.12 $ &  $   27.41   \pm   0.13 $ \\
  \phantom{1}688$^z$            &  03    32     42.56  &  $-$27 46     56.6   &   $28.14	 \pm  0.27 $  &  $26.84  \pm  0.10 $   &  $  26.69  \pm  0.09 $ &  $   26.68   \pm   0.09 $ \\
  2794\phantom{$^z$} &  03    32     36.75  &  $-$27 46     48.2   &   $29.84	 \pm  1.08 $  &  $28.59  \pm  0.30 $   &  $  28.30   \pm  0.22 $ &  $   28.41   \pm   0.24 $ \\
  1144$^z$           &  03    32     42.56  &  $-$27 47     31.5   &   $29.29	 \pm  0.67 $  &  $27.64  \pm  0.15$   &  $  27.59  \pm  0.14 $ &  $   27.53   \pm   0.14 $ \\

  2395\phantom{$^z$} &  03    32     44.31  &  $-$27 46     45.2   &$> 29.3\phantom{\pm0.55} $  &  $28.63  \pm  0.31$   &  $  28.36  \pm  0.23 $ &  $   28.84   \pm   0.34 $ \\

  1092$^z$           &  03    32     39.55  &  $-$27 47     17.5   &   $29.69	 \pm  0.96 $  &  $28.05  \pm  0.20 $   &  $  27.70   \pm  0.15 $ &  $   27.48   \pm   0.13 $ \\
  2560$^z$           &  03    32     37.80  &  $-$27 47     40.4   &   $ > 29.3 \phantom{\pm 0.55}           $  &	 $28.86  \pm  0.37$   &  $  28.87  \pm  0.34 $ &  $   29.13   \pm   0.43 $ \\
  2826\phantom{$^z$} &  03    32     37.06  &  $-$27 48     15.2   &   $ > 29.3 \phantom{\pm 0.55}           $  &	 $28.34  \pm  0.25$   &  $  28.43  \pm  0.25 $ &  $   27.89   \pm   0.17 $ \\
  1678$^z$           &  03    32     43.14  &  $-$27 46     28.6   &   $ > 29.3 \phantom{\pm 0.55}           $  &	 $28.30   \pm  0.24$   &  $  28.22  \pm  0.21 $ &  $   28.08   \pm   0.19 $ \\
  2502$^z$           &  03    32     39.73  &  $-$27 46     21.4   &   $ > 29.3\phantom{\pm 0.55}            $  &	 $28.85  \pm  0.37$   &  $  28.54  \pm  0.27 $ &  $   28.95   \pm   0.37 $ \\
  1574$^z$           &  03    32     37.21  &  $-$27 48     06.1   &   $ > 29.3\phantom{\pm 0.55}            $  &	 $28.52  \pm  0.28$   &  $  28.26  \pm  0.22 $ &  $   28.13   \pm   0.20  $ \\
  \phantom{1}835$^z$            &  03    32     38.81  &  $-$27 47 07.2   &   $> 29.3 \phantom{\pm 0.55} $  &  $27.74  \pm  0.16$   &  $  27.28  \pm  0.12 $ &  $   27.03   \pm   0.10  $ \\
  2066$^z$           &  03    32     41.05  &  $-$27 47     15.6   &   $> 29.3 \phantom{\pm 0.55} $  &  $28.92  \pm  0.39$   &  $  28.40   \pm  0.24 $ &  $   28.73   \pm   0.31 $ \\
  2888\phantom{$^z$} &  03    32     44.75  &  $-$27 46     45.1   &   $ > 29.3\phantom{\pm 0.55}            $  &	 $29.57  \pm  0.66$   &  $  28.98  \pm  0.37 $ &  $   28.83   \pm   0.34 $ \\
  2940\phantom{$^z$} &  03    32     36.52  &  $-$27 46     41.9   &   $ > 29.3\phantom{\pm 0.55}	           $  &	 $29.68  \pm  0.73$   &  $  29.01  \pm  0.38 $ &  $   29.51   \pm   0.58 $ \\
  2079$^y$           &  03    32     37.63  &  $-$27 46     01.5   &   $ > 29.3\phantom{\pm 0.55}            $  &	 $29.62  \pm  0.69$   &  $  29.01  \pm  0.38 $ &  $   29.33   \pm   0.50  $ \\
  1107$^z$           &  03    32     44.71  &  $-$27 46     44.4   &   $ > 29.3\phantom{\pm 0.55}            $  &	 $28.30   \pm  0.24$   &  $  27.58  \pm  0.14 $ &  $   27.70    \pm   0.15 $ \\
  1422\phantom{$^z$} &  03    32     39.52  &  $-$27 47     17.3   &   $ > 29.3\phantom{\pm 0.55}            $  &	 $28.90   \pm  0.38$   &  $  28.07  \pm  0.19 $ &  $   27.83   \pm   0.16 $ \\
  2487\phantom{$^z$} &  03    32     33.13  &  $-$27 46     54.4   & $ > 29.3 \phantom{\pm 0.55} $  &  $29.73  \pm  0.76$   &  $  28.59  \pm  0.28 $ &  $   29.18   \pm   0.44 $ \\
  1765$^y$           &  03    32     42.88  &  $-$27 46     34.6   &   $ > 29.3\phantom{\pm 0.55}            $  &  $29.44  \pm  0.60 $   &  $  28.23  \pm  0.21 $ &  $   28.11   \pm   0.20  $ \\

  2841$^y$           &  03    32     43.09  &  $-$27 46     27.9   &   $ > 29.3\phantom{\pm 0.55}            $  &  $> 30.0 \phantom{\pm 0.55} $   &  $  28.98  \pm  0.37 $ &  $   29.41   \pm   0.54 $ \\
  1939$^y$           &  03    32     37.80  &  $-$27 46     00.1   &   $> 29.3 \phantom{\pm 0.55} $  & $> 30.0 \phantom{\pm 0.55} $  &  $  28.33  \pm  0.23 $ &  $   28.49   \pm   0.26 $ \\
  1721$^y$           &  03    32     38.14  &  $-$27 45     54.0   &   $ > 29.3\phantom{\pm 0.55}            $  &  $> 30.0 \phantom{\pm 0.55}$   &  $  28.41  \pm  0.24 $ &  $   28.16   \pm   0.20  $\\ 
\hline\end{tabular}
\caption{Near-infrared photometric information for all 49 galaxies in our high-redshift sample. As a result of our selection criteria, all objects
are by definition undetected (at $> 2\sigma$) in the $i$-band, and so here we list the AB magnitudes measured at longer wavelengths, i.e. $z_{850}$, $Y_{105}$, $J_{125}$ and $H_{160}$. 
All magnitudes were measured 
through a 0.6-arcsec diameter aperture, with the
WFC3/IR values decreased (i.e. brightened) by the small systematic relative 
correction factors given in Section 4.1. The quoted limits are all $2\sigma$.
}
\end{table*}

\begin{figure*}
\includegraphics[width=0.75\textwidth]{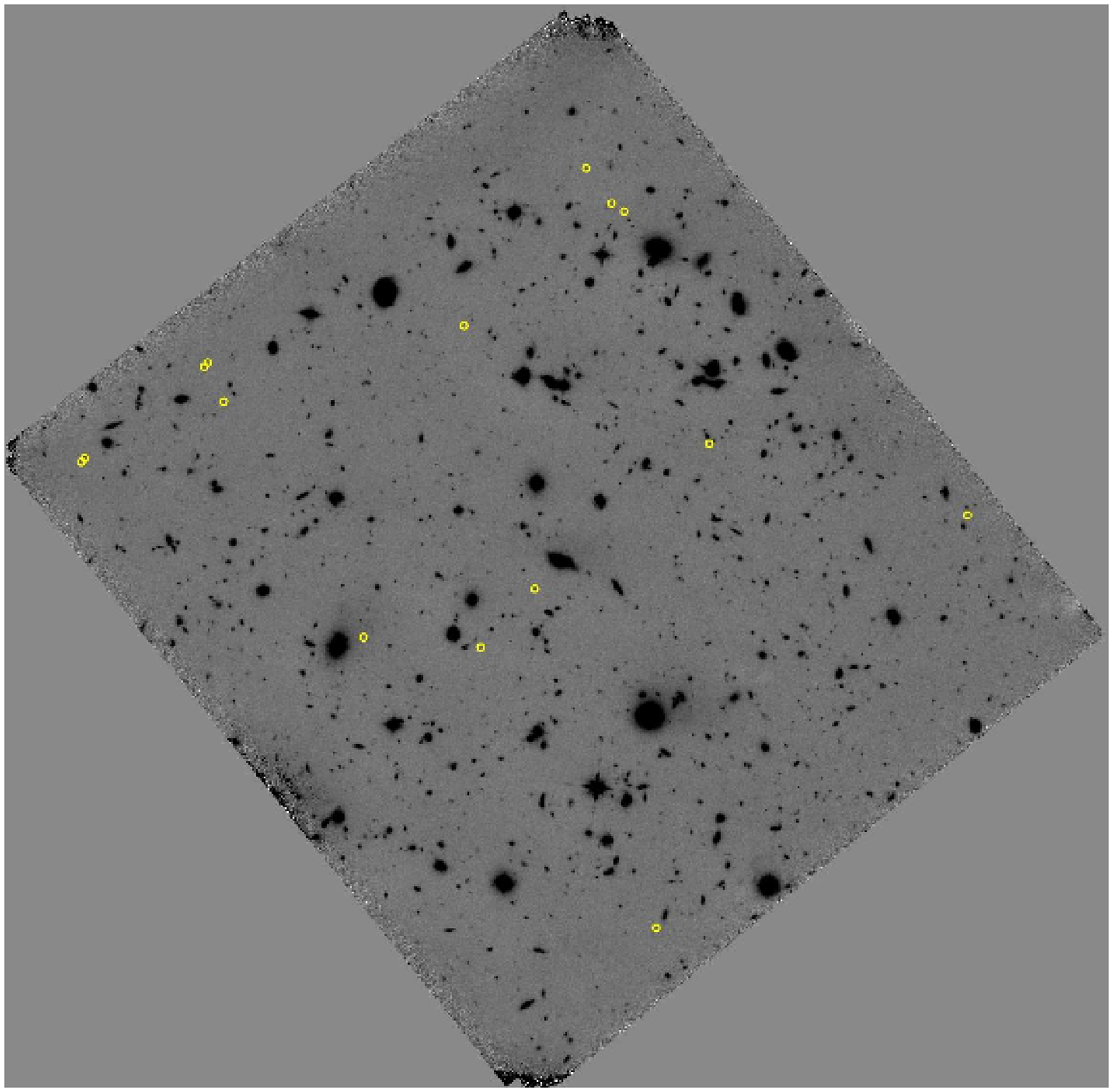}
\caption{The positions of the 15 galaxies in our sample with primary
  photometric redshift solutions at $z > 7$ overplotted on the new WFC3/IR 
$H_{160}$-band image of the HUDF (North is up, and East is to the left). The 4 most distant galaxies in the sample (at $z > 7.8$) are confined to the groups 
at the north and north-east of the image.}
\end{figure*}

\subsection{The sample}

Details on the final high-redshift sample of 49 galaxies are 
provided in Tables 2 and 3. The galaxies are ranked in order 
of increasing redshift, and Table 2 summarises the results of the redshift 
estimation process, giving primary and secondary redshift solutions, along with
1$\sigma$ confidence intervals, and the minimum value of $\chi^2$ at each 
solution. With the number of degrees of freedom involved in the fitting 
process, a solution with $\chi^2 < 9$ must be regarded as statistically
acceptable. Postage stamp images, SED fits, and plots of $\chi^2$ versus $z$
for all 49 galaxies are given in Appendix A, with the galaxy candidates 
again ranked in the same order of increasing primary redshift. Galaxies which 
have been independently discovered as $z_{850}$-drops by Oesch et al. (2009b)
are marked with a superscript $z$, while the 5 candidates 
 independently discovered as $Y_{105}$-drops by Bouwens et al. (2009)
are marked with the superscript $y$.

Table 3 gives the $z_{850}$, $Y_{105}$, $J_{125}$ and $H_{160}$ 
magnitudes for each galaxy, as measured 
through a 0.6-arcsec diameter aperture, with the
WFC3/IR values decreased (i.e. brightened) by the small systematic relative 
correction factors given above in Section 4.1.

Although we are primarily interested in galaxies at $z > 6.5$ we have retained 
all galaxy candidates from our original sample which have a primary 
redshift solution $z > 5.9$, because the redshift probability distributions
for these objects frequently extend well beyond $z \simeq 6$, and this information 
requires to be factored into our determination of the galaxy luminosity function at $z \simeq 7$. Our insistence
on 2$\sigma$ non-detections in the $i$-band means that the sample cannot be 
regarded as reasonably complete until $z > 6.25$. At higher redshifts
this sample can be regarded as a master list of all convincing candidates 
for galaxies at $z > 6.25$, brighter than a magnitude $\simeq 29$ (AB). in at least one near-infrared passband in the 4.5\,arcmin$^2$ area covered by the WFC3/IR image 
of the HUDF. It does {\it not} follow that all candidates are equally robust,
as can be judged from the plots presented in Appendix A, and the alternative 
redshift solutions quantified in Table 2. 

Our independent list, selected 
without recourse to hard optical-infrared colour criteria, contains all 
but the faintest of the 16 $z \sim 7$ galaxies listed by 
Oesch et al. (2009b) (which in turn includes
the 4 credible $z > 6$ galaxies from Bouwens et al. (2004) listed in Table 1). 
Our sample also includes all 5 of the $z \sim 8$ galaxies listed by Bouwens et al. (2009). 

Our list contains an additional 15 objects at $z > 6.25$ which have 
evaded selection by Oesch et al. (2009b) and Bouwens et al. (2009). 
In general, inspection of the additional objects reveals that they lie just 
outside or on the edge of the colour selection boxes adopted in these alternative studies. However, use of our own 0.6-arcsec photometry would also lead us 
to locate several of the objects selected by Oesch et al. and 
Bouwens et al. outside their adopted selection
boxes, and in general the $\chi^2$ plots shown in Appendix A do not suggest
that our additional candidates are any less robust than those selected 
via strict colour criteria. 

In addition to providing a new, more complete list of candidates, 
our analysis of course 
yields quantitative constraints on the redshifts of all sources, 
and information to allow the reader to properly assess the robustness 
of each individual high-redshift galaxy candidate.

\begin{figure*}
\includegraphics[width=0.85\textwidth]{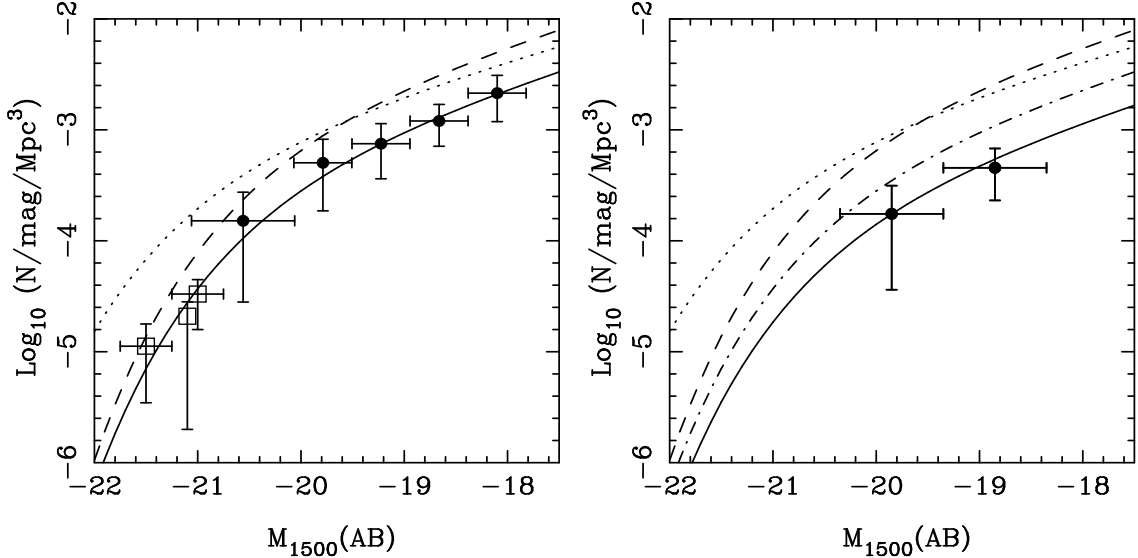}
\caption{Our new determinations of the ultraviolet (UV) galaxy luminosity function at $z \simeq 7$ and 
$z \simeq 8$. The left-hand panel shows our $V/V_{\rm max}$ estimate of the redshift
$z=7$ luminosity function. The absolute UV
magnitudes have been calculated at a rest-frame wavelength
of 1500\AA. The filled circles are the data-points from this study,
where the horizontal error bars indicate the width of the magnitude
bins adopted (either $\Delta m=0.5$ or $\Delta m=1.0$
depending on signal-to-noise) and the vertical error bars indicate
the uncertainty due to simple Poisson statistics. The open squares
are taken from the ground-based study of the Subaru Deep Field and
GOODS-North field by Ouchi et al. (2009). The dotted and dashed curves
are the Schechter function fits to the $z=5$ and $z=6$ luminosity
function from McLure et al. (2009). The solid line is our best-fitting
Schechter function to the $z=7$ luminosity function. The right-hand panel shows our corresponding
estimate of the luminosity function at redshift $z=8$. In this panel
the dashed and dot-dashed curves are the $z=6$ and $z=7$ Schechter
function fits from McLure et al. (2009) and this work
respectively. The solid line is the best-fitting $z=7$ Schechter
function with $\phi^{\star}$ reduced by a factor of two
(see text for a discussion).}
\end{figure*}

\section{Results}

\subsection{Galaxies at z = 7 -- 8.5}

Given the natural interest in the possible discovery of galaxies at $z > 7$, we focus first on the existence and robustness 
of potential galaxies at $z \simeq 8$. As discussed above, our sample contains 
$\simeq 15$ galaxies at $z > 7$, of which nominally 3 
lie at $z > 8$. Of the sub-sample of 8 galaxies with primary 
redshift solutions at $z > 7.5$, five are the 
$z \simeq 8$ galaxy candidates reported by 
Bouwens et al. (2009), and one was selected as a $z$-drop by 
Oesch et al. (2009b). Thus the majority of these apparently extremely distant 
galaxy candidates have been identified more than once, in independent reductions and
analyses of the combined 
WFC3/IR+ACS dataset. However, we caution that, given 
the number of degrees of freedom in our SED fitting, 
any redshift solution with $\chi^2 < 9$ is acceptable at the $2\sigma$ level. Thus, given the available data, all 8-10 objects apparently at
$z > 7.5$ could also lie at $z \simeq 1.6$, at which point the 4000\AA\ or Balmer break in a galaxy spectrum has begun to move into the F105W bandpass.

The problem here is one of dynamic range. The WFC3/IR imaging is so deep that a non-detection in the ACS $z_{850}$ and other 
HUDF optical images is not sufficient to discriminate between the extreme Lyman-break at 1215\AA\ anticipated at $z \simeq 8$, from the more modest
4000\AA /Balmer break in a low-luminosity galaxy at much lower redshift. For the faintest proposed $z \sim 8$ galaxy candidates, the 
available photometry produces redshift constraints which are inevitably rather weak.

Nonetheless, we note here something else of at least qualitative interest about these $z \simeq 8$ galaxy candidates, which is that they appear to be strongly
clustered on the sky (see Fig. 4). For example, the four apparently most distant 
objects in the sample occur in two pairs with projected galaxy
separations $< 10$\,arcsec ($\simeq 50$ kpc at $z\simeq 8$). However,
it should be cautioned that the suggestion of clustering does not
necessarily strengthen the case that these objects are genuine $z
\simeq 8$ galaxies, given that the most probable interloper population
(i.e. galaxies at $1<z<2$) could also be clustered.
 
Finally, we stress that, even if the high-redshift solutions are adopted for these objects, we have yet to uncover a completely compelling example 
of a galaxy at $z > 8$ in the WFC3/IR image of the HUDF.

\subsection{The ultraviolet galaxy luminosity function at z = 7 - 8}
Based on our new dataset it is possible to make a preliminary
estimate of the ultraviolet (UV) galaxy luminosity function at $z=7$
and $z=8$. In order to exploit the redshift information available for our
sample, when deriving the luminosity function we adopt the method
outlined in McLure et al. (2009) whereby, rather than simply adopting the
primary photometric redshift solution, each object is represented by
its normalized redshift probability density function. In addition to
making better use of the available information, this method makes it
possible to construct the luminosity function using the classic
$V/V_{\rm max}$ estimator of Schmidt (1968), and deals with the problem
of multiple redshift solutions in a transparent fashion.

In Fig. 5 we show our estimates of the $z=7$ and $z=8$ 
galaxy luminosity functions, which were calculated within the redshift intervals
$6.5<z<7.5$ and $7.5<z<8.5$ respectively. In both panels the
data-points from this study are shown as the filled circles, and in the
left-hand panel we also plot the $z=7$ data-points at bright magnitudes from
the ground-based study of the Subaru Deep Field and GOODS-North field
by Ouchi et al. (2009). Although a full analysis of the evolving
UV-selected luminosity function is deferred until the remainder of the
deep HUDF09 WFC3/IR data are obtained over the coming year, we have
performed a simple fit to the $z=7$ luminosity function,
including the Ouchi et al. (2009) data-points at the bright end. The
best-fitting Schechter function is shown as the solid line in the
left-hand panel of Fig. 5 and yields the following parameter values:
$M_{1500}^{\star}=-20.11, \phi^{\star}=0.0007 {\rm Mpc}^{-3}$ and
$\alpha=-1.72$. Due to the large uncertainties on the data-points it
is not possible to place meaningful constraints on either
$M_{1500}^{\star}$ or $\alpha$ at $z=7$. However, a comparison with
the best-fitting Schechter luminosity function parameters at $z=6$
($M_{1500}^{\star}=-20.04\pm0.12, \phi^{\star}=0.0018\pm0.0005\,{\rm
  Mpc}^{-3}$ and $\alpha=-1.71\pm0.11$) from McLure et al. (2009),
suggests that the $z=6$ and $z=7$ luminosity functions have very
similar shapes, and can be reconciled via an evolution in
number density by a factor of $\simeq 2.5$.

The limited dynamical range in absolute luminosity makes it
impossible, at present, to place any meaningful constraints on the shape of the
UV-selected galaxy luminosity function at $z=8$. All that can 
be said with the current dataset is that, at a characteristic
luminosity of $M_{1500}\simeq -19.5$, the number density of
high-redshift galaxies is a further factor of $\simeq 2$ smaller than
at $z=7$. To illustrate this point we show in the right-hand panel of
Fig. 5 that the best-fitting Schechter function at $z=7$ can be brought
into good agreement with the $z=8$ data-points simply by reducing
$\phi^{\star}$ by a factor of two.  Although the uncertainties
associated with the current dataset, along with degeneracies between
Schechter function parameters, prevent any
definitive conclusions, it appears that the $z=7$ and $z=8$ luminosity
functions are consistent with having the same overall shape as at
$z=6$, but with $\phi^{\star}$ a factor of $\simeq 2.5$ and $\simeq 5$ lower respectively.

\subsection{Implications for re-ionisation}
Given the apparent decline in the number density of star-forming
galaxies  over the redshift range $z\simeq 6-8$, it is prudent to
examine the ability  of the observed population of galaxies to achieve
reionisation.  We adopt the method of Bolton \& Haehnelt (2007),
converting the observed luminosity functions at $z>6$ to ionising emissivities
assuming a Salpeter  IMF, a metallicity of 0.2 Z$_\odot$, and  an
escape fraction for ionising photons of 20\%.  For simplicity, we
ignore any  possible mass dependence in the escape fraction.  We consider
relatively low values for the hydrogen clumping factors in our analysis
($C\simeq 2-5$), following the results of recent simulations
(e.g. Bolton  \& Haehnelt 2007).  At $z<6$, we adopt the comoving ionising 
photon density implied by the Ly$\alpha$ opacity data presented in 
Bolton \& Haehnelt (2007), where $\dot{N}_{ion}(z)$=10$^{50.5-0.06(z-6)}$ $s^{-1}$
Mpc$^{-3}$.

For the luminosity function at $z\simeq 7$ and 8, we adopt the
$z\simeq 6$ Schechter parameters from McLure et al. (2009), with the
normalisation  ($\phi^\star$) lowered by a factor of 2.5 and 5,
respectively, consistent with  the measurements presented above.  To
provide an upper bound on the possible  contribution of the
star-forming galaxy population to reionisation, we first assume  the
luminosity function remains constant over $8<z<15$ and compute the
evolution of the filling factor of ionised hydrogen, $Q_{HII}$, as a
function  of redshift. To do this, we integrate the ionising
emissivities forward from $z=15$, arbitrarily assuming that the
ionising flux decreases to zero at $z>15$.  Defining the end of
reionisation as the redshift when $Q_{HII}=1$, we find that if we
include only sources brighter than $M=-18$  (roughly the detection
limit of the WFC3 data at $z=7$), the IGM would not  achieve a unity
HII filling factor until $z=4.2$, at odds with the  high ionisation
fraction implied by measurements of the Gunn Peterson optical depth
from quasars at $z\ltsim 6.2$ (e.g. Fan et al. 2006, Becker et
al. 2007). Even if sources as faint as $M=-10$ are included,
reionisation would only be completed at  $z=4.5$ (with $Q_{HII}=0.53$
at $z=6.0$), at odds with the transmission seen in quasar spectra at
these redshifts.  Furthermore, each of these computations assumes that
the ionising emissivity does not decrease beyond $z=8$.  If the
normalisation of the luminosity function instead continues to decline
at $z>8$, consistent with the trends at $6<z<8$, then the galaxy
population would undoubtedly be hard-pressed to reionise the IGM  by
$z\simeq 6$ without significant variations in the escape fraction.
While caution  must surely be exercised in interpreting these results
given the uncertainties in the form of the luminosity functions, our
results indicate that it is difficult for the observed population of
star-forming galaxies to achieve reionisation by $z \simeq 6$ without a significant
contribution from galaxies well below the detection limits, in addition
to  alterations in the escape fraction of ionising photons and/or
continued vigorous star formation at $z>15$.

\section{SUMMARY}

Following the first public data release of the raw WFC3/IR  
data taken within the HUDF (as part of HST Treasury programme GO-11563), we have reduced the
new deep $Y_{105}$, $J_{125}$, and $H_{160}$ images, and analysed them in tandem with the existing deep optical $B,V,i,z$ ACS imaging,
to conduct a new systematic search for high-redshift galaxies at $z = 6 - 9$.
After registering the new infrared data to the ACS optical image reference frame, and establishing our own 
photometric zero-points, we used SExtractor to select a complete parent catalogue consisting of all objects
which were detected at $\geq 5\sigma$ significance in at least one of the near-infrared wavebands in a 0.4-arcsec
diameter aperture. For the purposes of SED analysis and photometric
redshift determination we performed aperture photometry on all objects
within circular apertures of 0.6-arcsec diameter, applying relative (to ACS resolution) point-source aperture corrections 
to the derived $Y_{105}$, $J_{125}$, and $H_{160}$ magnitudes.

Starting from the master catalogue we confined our attention to high
redshifts by rejecting all sources which were detected at $\geq
2\sigma$ significance in the ACS $i-$band imaging. 
We then performed full SED fitting
to the optical+infrared photometry of the remaining $\simeq 300$ objects in the catalogue, and utilised this information 
to inform the rejection of a variety of contaminants (stars, diffraction spikes, transients) and lower-redshift
galaxy interlopers. The final result is 
a sample of 49 galaxies with primary photometric redshift 
solutions $z > 5.9$, within the 4.5 arcmin$^2$ field covered by the new near-infrared imaging.
Our sample, selected without recourse to specific 
colour cuts, re-selects all but the faintest one of the 16 $z_{850}$-drops selected by Oesch et al. (2009),
recovers all 5 of the $Y_{105}$-drops reported by Bouwens et al. (2009), and adds a further 29 
equally plausible galaxy candidates, of which 12 lie beyond $z \simeq 6.3$, and 4 lie beyond $z \simeq 7.0$. However, we also present 
confidence intervals on our photometric redshift estimates, marginalising over all other fitted parameters, 
and highlight the prevalence of alternative secondary redshift
solutions (especially for the highest redshift galaxy candidates). 
As a result of this analysis we caution that acceptable low-redshift ($z < 2$) solutions
exist for 28 out of the 37 galaxies at $z > 6.3$, and in particular for all 8 
of the galaxy candidates reported here at $z > 7.5$. Nevertheless, we note that the
very highest redshift candidates appear to be strongly clustered in
the field. 

Based on our photometric redshift analysis we derive new estimates of the ultraviolet galaxy luminosity function at $z \simeq 7$ and $z \simeq 8$.
Where our results are most robust, at a characteristic luminosity $M_{1500} \simeq -19.5 (AB)$, we find that the comoving 
number density of galaxies declines by a factor of $\simeq 2.5$ between $z \simeq 6$ and $z \simeq 7$, and by a further 
factor of 2 by $z \simeq 8$. These results suggest 
that it is difficult for the observed population of
high-redshift 
star-forming galaxies to achieve reionisation without a significant
contribution from galaxies well below the detection limits plus
alterations in the escape fraction of ionising photons and/or
continued vigorous star formation at $z>15$.

\section*{ACKNOWLEDGEMENTS}

The authors would like to thank the anonymous referee whose comments
and suggestions significantly improved the final version of this
manuscript. This work is based primarily on observations made with the NASA/ESA {\it Hubble Space Telescope}, which is operated by the Association 
of Universities for Research in Astronomy, Inc, under NASA contract NAS5-26555.
This work is based in part on observations made with the {\it Spitzer Space Telescope}, which is operated by the Jet Propulsion Laboratory, 
California Institute of Technology under NASA contract 1407. 
RJM and JSD acknowledge the support of the Royal Society through a University Research Fellowship and a Wolfson Research Merit award respectively. 
MC and DS acknowledge the support of STFC through the award of an Advanced Fellowship and Post-Doctoral Fellowship respectively. TAT acknowledges
the support of NSERC.

{}

\appendix

\section{SED fits and Redshift Estimates}

In this appendix we present detailed information on each of our 49 high-redshift galaxy candidates, to facilitate 
comparison with other studies and to allow the reader to judge for themselves the 
robustness of the high-redshift solutions. The galaxies are ranked in order of increasing primary redshift.
For each galaxy we provide optical and near-infrared postage 
stamp grey-scale images, and also a plot showing the best fitting galaxy SED, and the dependence of $\chi^2$ on redshift
(marginalised over all other fitted parameters).

\begin{figure*}
\begin{tabular}{llll}
\hspace*{0.8cm}
\includegraphics[width=0.097\textwidth, angle=270]{1735_z_stamp.ps}
\includegraphics[width=0.097\textwidth, angle=270]{1735_y_stamp.ps}
\includegraphics[width=0.097\textwidth, angle=270]{1735_j_stamp.ps}
\includegraphics[width=0.097\textwidth, angle=270]{1735_h_stamp.ps}&
\hspace*{0.8cm}
\includegraphics[width=0.097\textwidth, angle=270]{1955_z_stamp.ps}
\includegraphics[width=0.097\textwidth, angle=270]{1955_y_stamp.ps}
\includegraphics[width=0.097\textwidth, angle=270]{1955_j_stamp.ps}
\includegraphics[width=0.097\textwidth, angle=270]{1955_h_stamp.ps}\\
\includegraphics[width=0.47\textwidth]{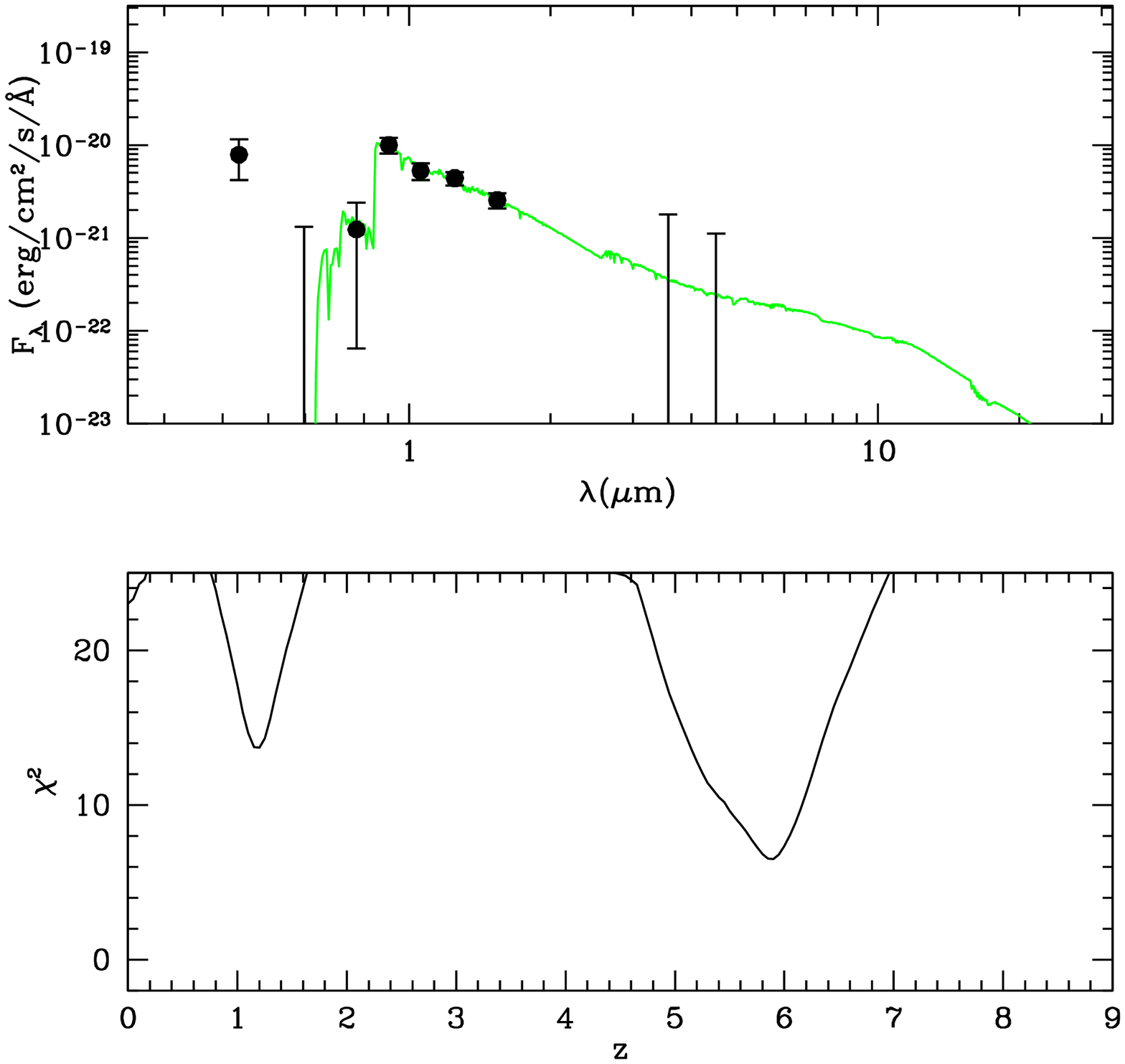}&
\includegraphics[width=0.47\textwidth]{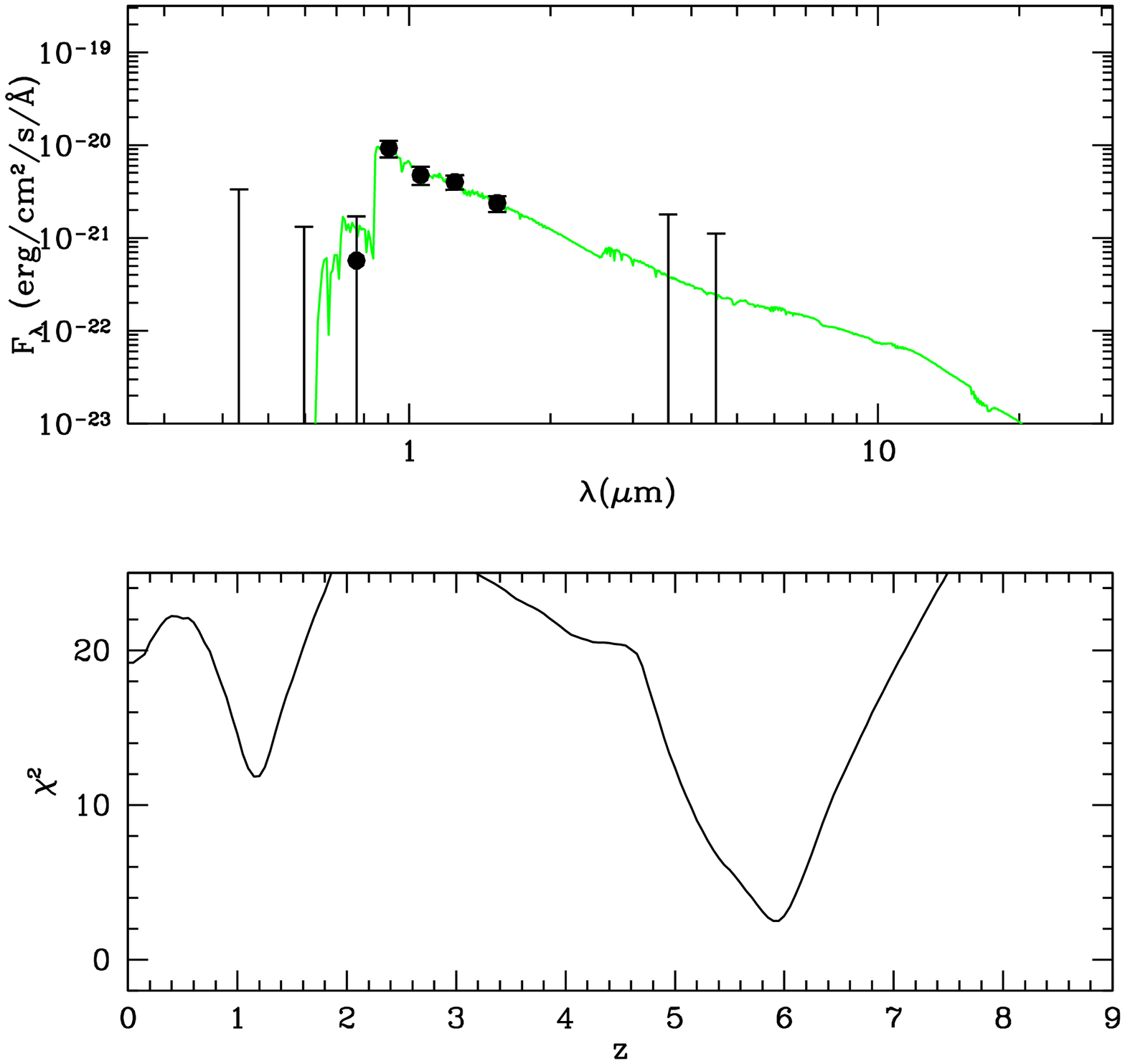}\\
\hspace*{1cm}{\bf 1735:} $z_{\rm est} = 5.90\, (5.70-6.05)$&
\hspace*{1cm}{\bf 1955:} $z_{\rm est} = 5.90\, (5.75-6.10)$\\
\\
\\
\\
\\
\hspace*{0.8cm}
\includegraphics[width=0.097\textwidth, angle=270]{1719_z_stamp.ps}
\includegraphics[width=0.097\textwidth, angle=270]{1719_y_stamp.ps}
\includegraphics[width=0.097\textwidth, angle=270]{1719_j_stamp.ps}
\includegraphics[width=0.097\textwidth, angle=270]{1719_h_stamp.ps}&
\hspace*{0.8cm}
\includegraphics[width=0.097\textwidth, angle=270]{2217_z_stamp.ps}
\includegraphics[width=0.097\textwidth, angle=270]{2217_y_stamp.ps}
\includegraphics[width=0.097\textwidth, angle=270]{2217_j_stamp.ps}
\includegraphics[width=0.097\textwidth, angle=270]{2217_h_stamp.ps}\\
\includegraphics[width=0.47\textwidth]{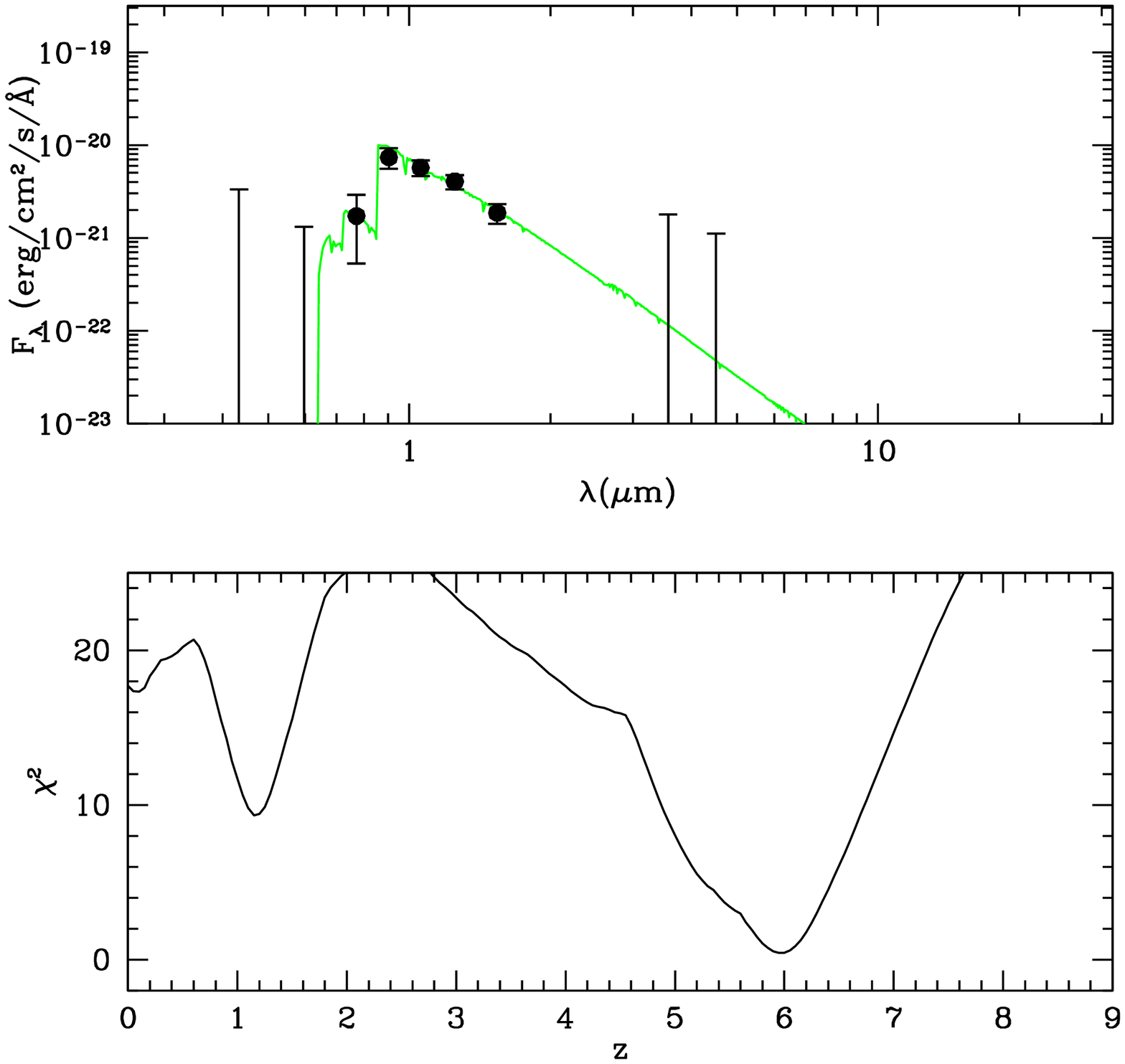}&
\includegraphics[width=0.47\textwidth]{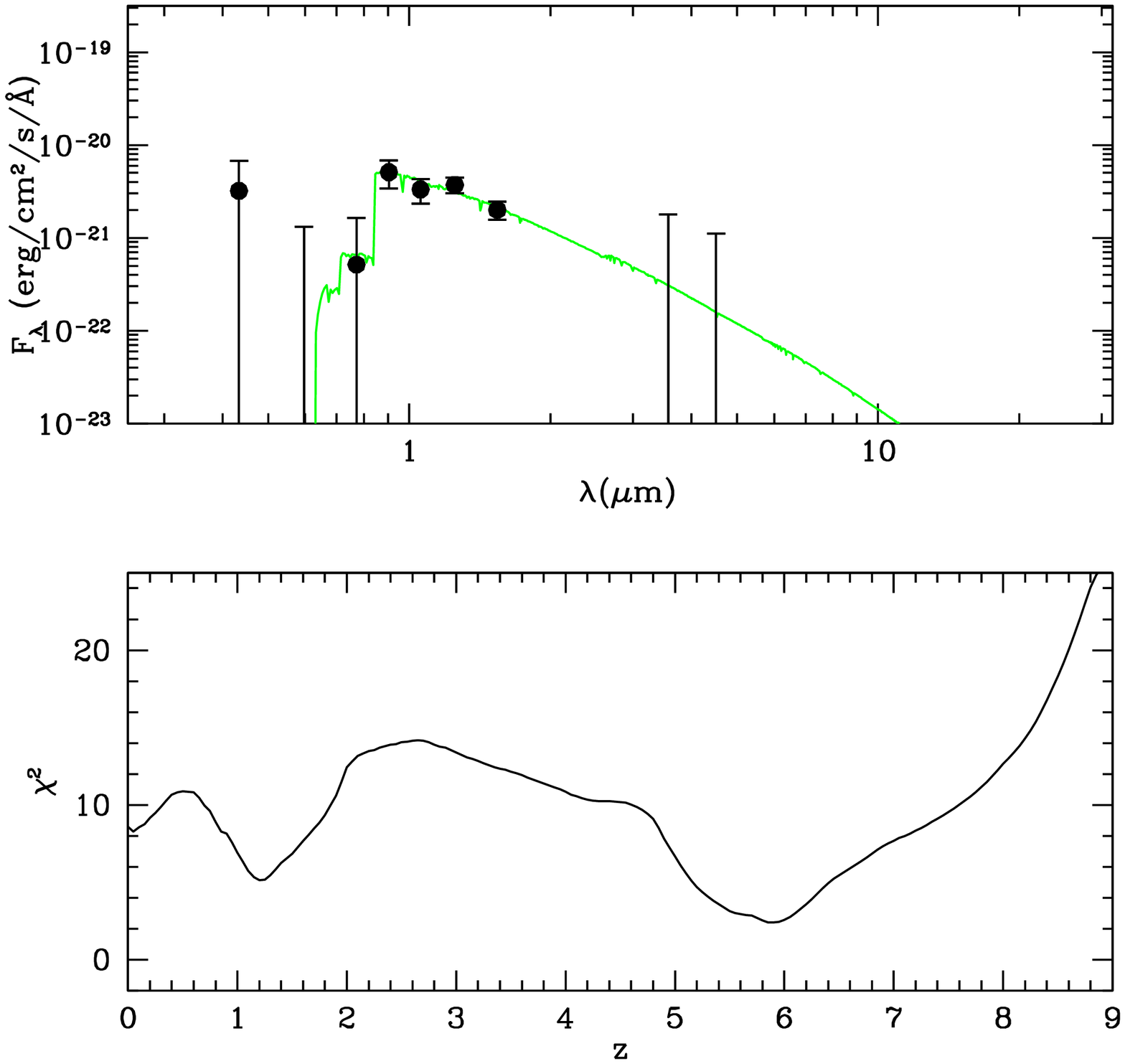}\\
\hspace*{1cm}{\bf 1719:} $z_{\rm est} = 5.95\, (5.75-6.15)$&
\hspace*{1cm} {\bf 2217:} $z_{\rm est} = 5.95\, (5.40-6.20)$\\
\end{tabular}
\caption{3 $\times$ 3 arcsec $z_{850}$, $Y_{105}$, $J_{125}$, $H_{160}$ (left to right) greyscale (-2$\sigma$ -- +6$\sigma$) postage-stamp
images are shown for each galaxy candidate above a plot which shows the galaxy SED
which best fits the HST ACS optical, WFC3/IR near-infrared
and Spitzer 3.6${\rm \mu m}$ and 4.5${\rm \mu m}$
photometric limits. The lower panel for each object shows how $\chi^2$ varies with redshift, marginalised over galaxy age,
star-formation history, and dust reddening. The ID number 
of each source, and its estimated redshift (with $1\sigma$ error range) are given under each plot.}
\end{figure*}

\begin{figure*}
\begin{tabular}{llll}
\hspace*{0.8cm}
\includegraphics[width=0.097\textwidth, angle=270]{962_z_stamp.ps}
\includegraphics[width=0.097\textwidth, angle=270]{962_y_stamp.ps}
\includegraphics[width=0.097\textwidth, angle=270]{962_j_stamp.ps}
\includegraphics[width=0.097\textwidth, angle=270]{962_h_stamp.ps}&
\hspace*{0.8cm}
\includegraphics[width=0.097\textwidth, angle=270]{1189_z_stamp.ps}
\includegraphics[width=0.097\textwidth, angle=270]{1189_y_stamp.ps}
\includegraphics[width=0.097\textwidth, angle=270]{1189_j_stamp.ps}
\includegraphics[width=0.097\textwidth, angle=270]{1189_h_stamp.ps}\\
\includegraphics[width=0.47\textwidth]{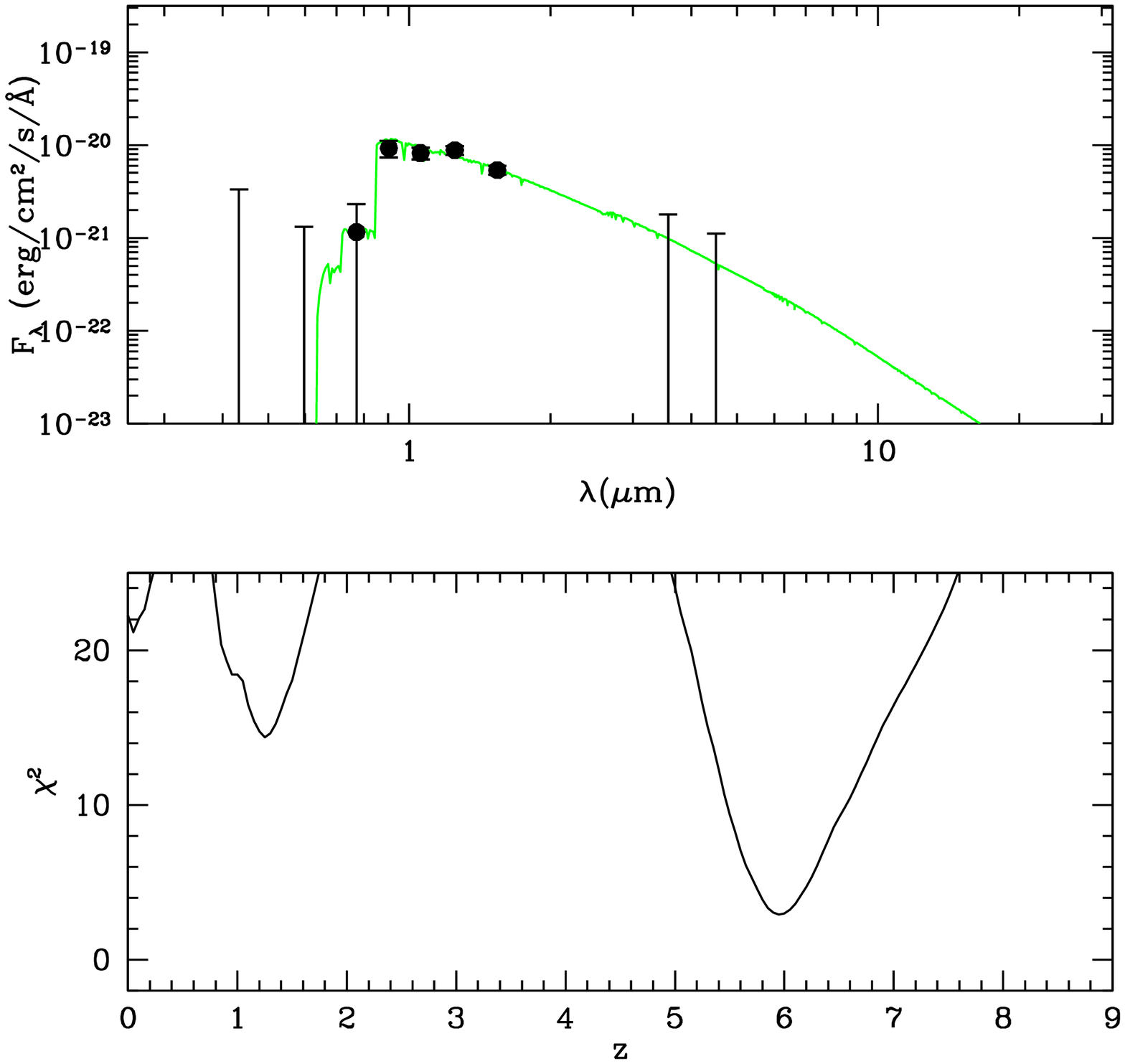}&
\includegraphics[width=0.47\textwidth]{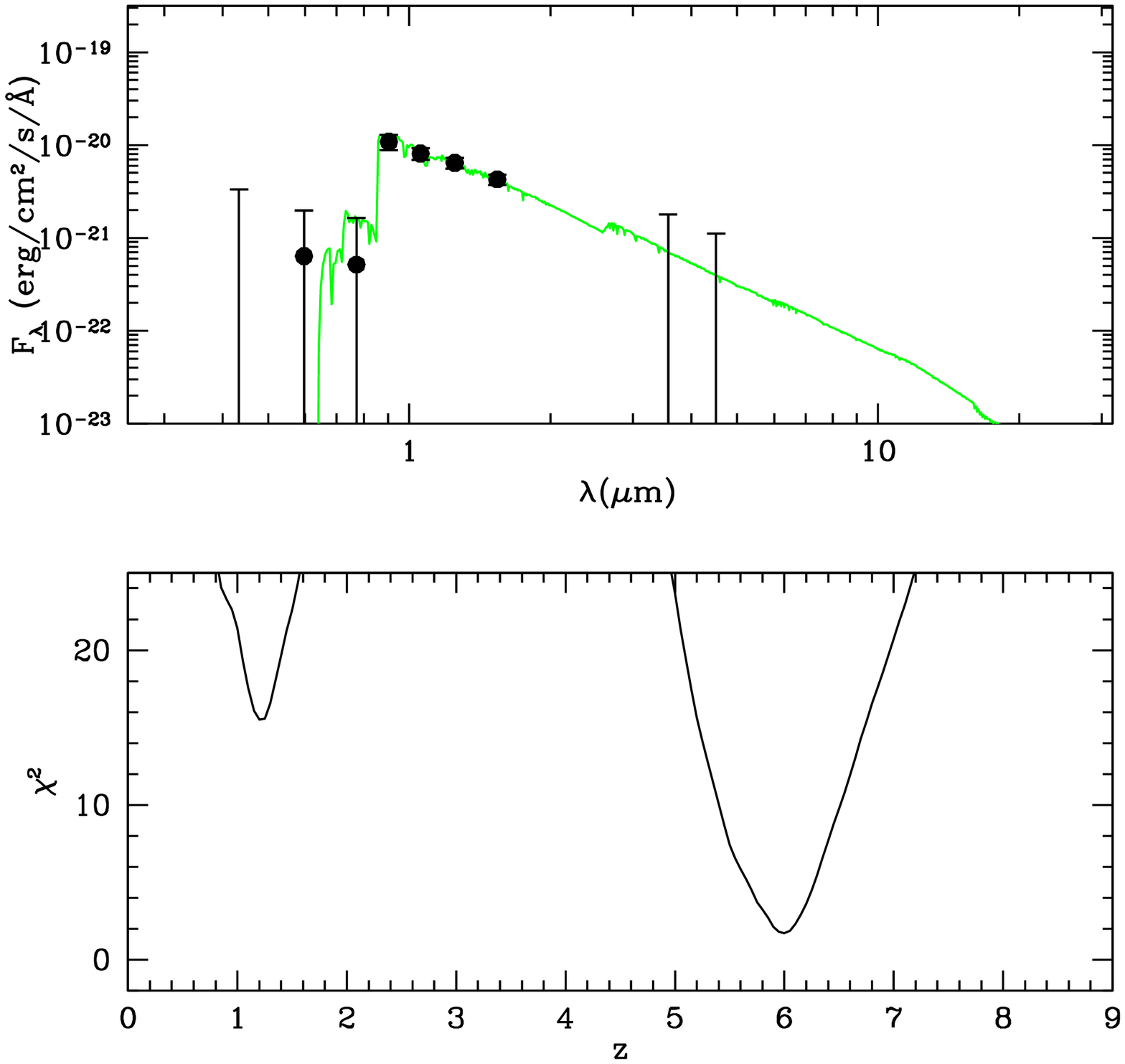}\\
\hspace*{1cm} {\bf 962:} $z_{\rm est} = 5.95\, (5.80-6.20)$&
\hspace*{1cm}{\bf 1189:} $z_{\rm est} = 6.00\, (5.85-6.15)$\\
\\
\\
\\
\\
\hspace*{0.8cm}
\includegraphics[width=0.097\textwidth, angle=270]{2830_z_stamp.ps}
\includegraphics[width=0.097\textwidth, angle=270]{2830_y_stamp.ps}
\includegraphics[width=0.097\textwidth, angle=270]{2830_j_stamp.ps}
\includegraphics[width=0.097\textwidth, angle=270]{2830_h_stamp.ps}&
\hspace*{0.8cm}
\includegraphics[width=0.097\textwidth, angle=270]{2498_z_stamp.ps}
\includegraphics[width=0.097\textwidth, angle=270]{2498_y_stamp.ps}
\includegraphics[width=0.097\textwidth, angle=270]{2498_j_stamp.ps}
\includegraphics[width=0.097\textwidth, angle=270]{2498_h_stamp.ps}\\
\includegraphics[width=0.47\textwidth]{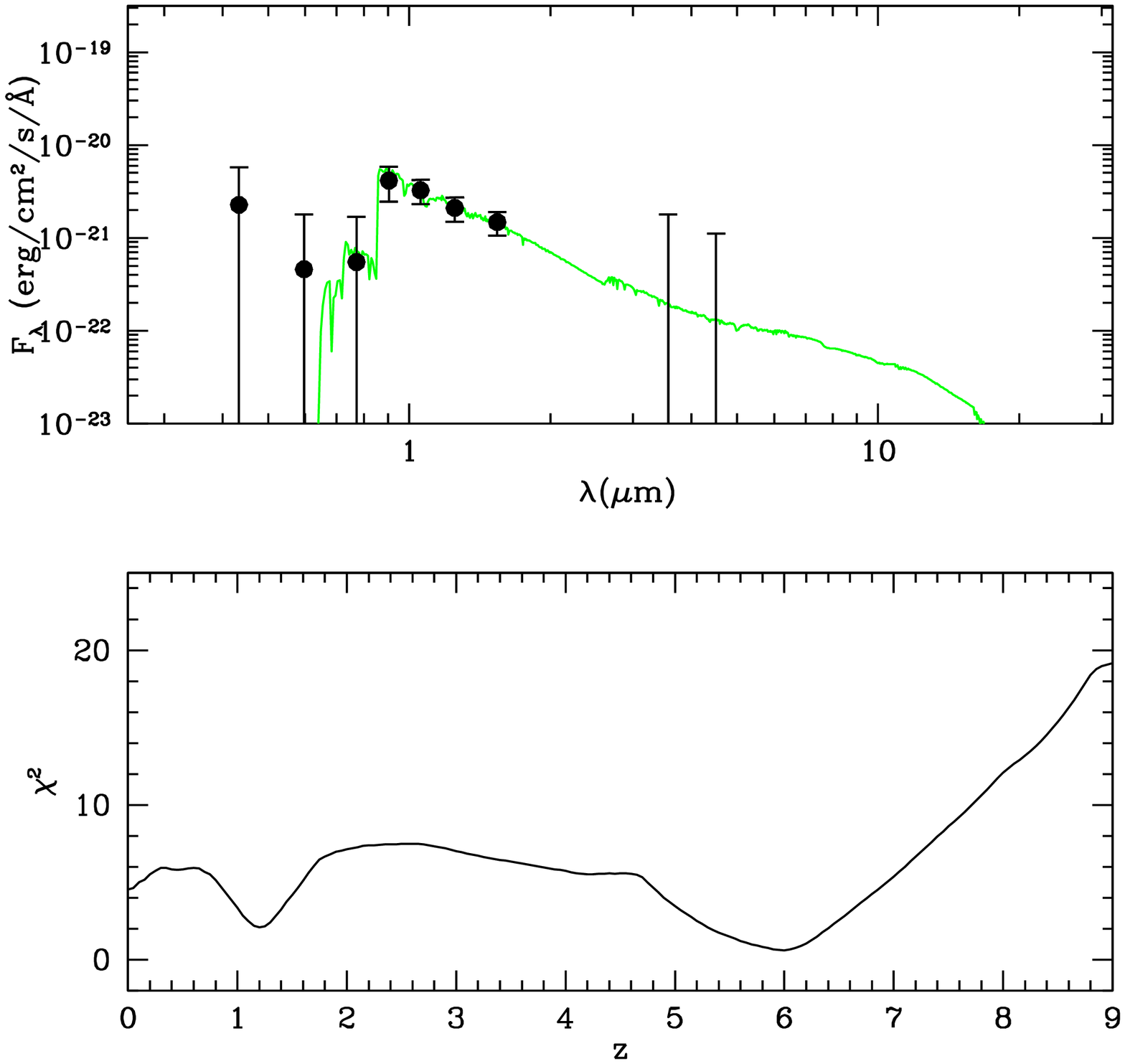}&
\includegraphics[width=0.47\textwidth]{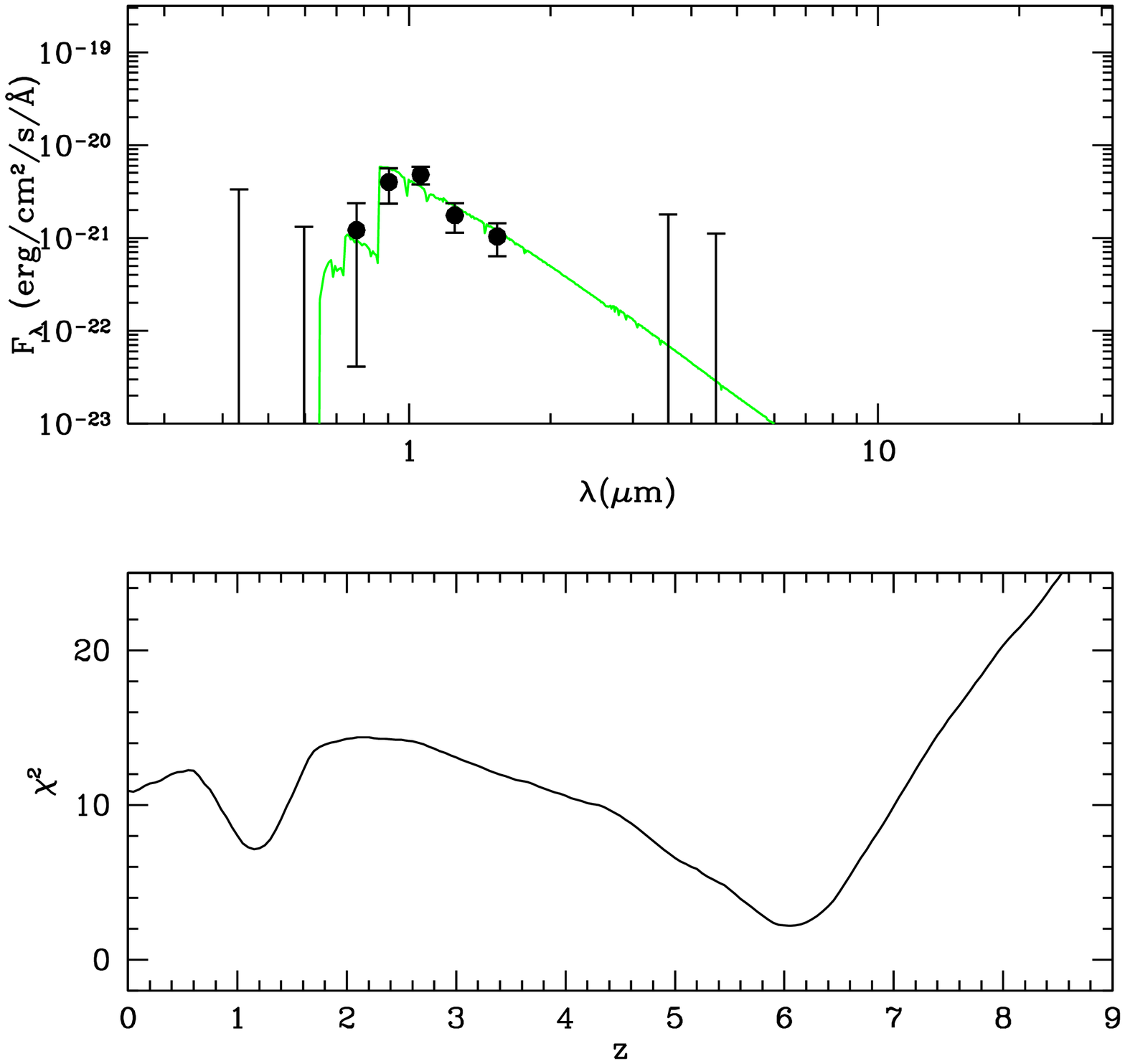}\\
\hspace*{1cm}{\bf 2830:} $z_{\rm est} = 6.00\, (5.45-6.35)$&
\hspace*{1cm} {\bf 2498:} $z_{\rm est} = 6.00\, (5.70-6.35)$\\
\\
\\
\end{tabular}
\addtocounter{figure}{-1}
\caption{continued.}
\end{figure*}

\begin{figure*}
\begin{tabular}{llll}
\hspace*{0.8cm}
\includegraphics[width=0.097\textwidth, angle=270]{2719_z_stamp.ps}
\includegraphics[width=0.097\textwidth, angle=270]{2719_y_stamp.ps}
\includegraphics[width=0.097\textwidth, angle=270]{2719_j_stamp.ps}
\includegraphics[width=0.097\textwidth, angle=270]{2719_h_stamp.ps}&
\hspace*{0.8cm}
\includegraphics[width=0.097\textwidth, angle=270]{1625_z_stamp.ps}
\includegraphics[width=0.097\textwidth, angle=270]{1625_y_stamp.ps}
\includegraphics[width=0.097\textwidth, angle=270]{1625_j_stamp.ps}
\includegraphics[width=0.097\textwidth, angle=270]{1625_h_stamp.ps}\\
\includegraphics[width=0.47\textwidth]{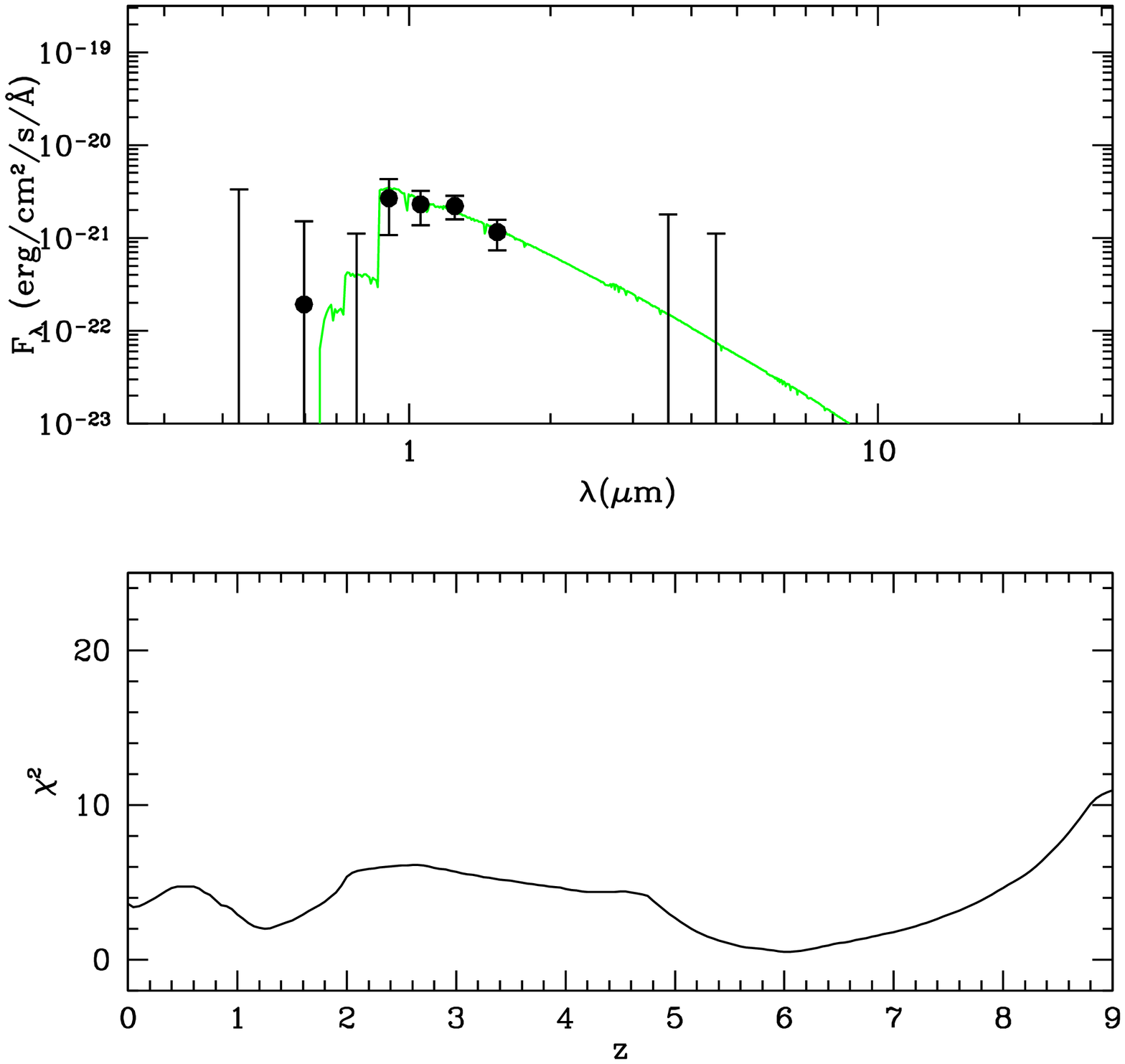}&
\includegraphics[width=0.47\textwidth]{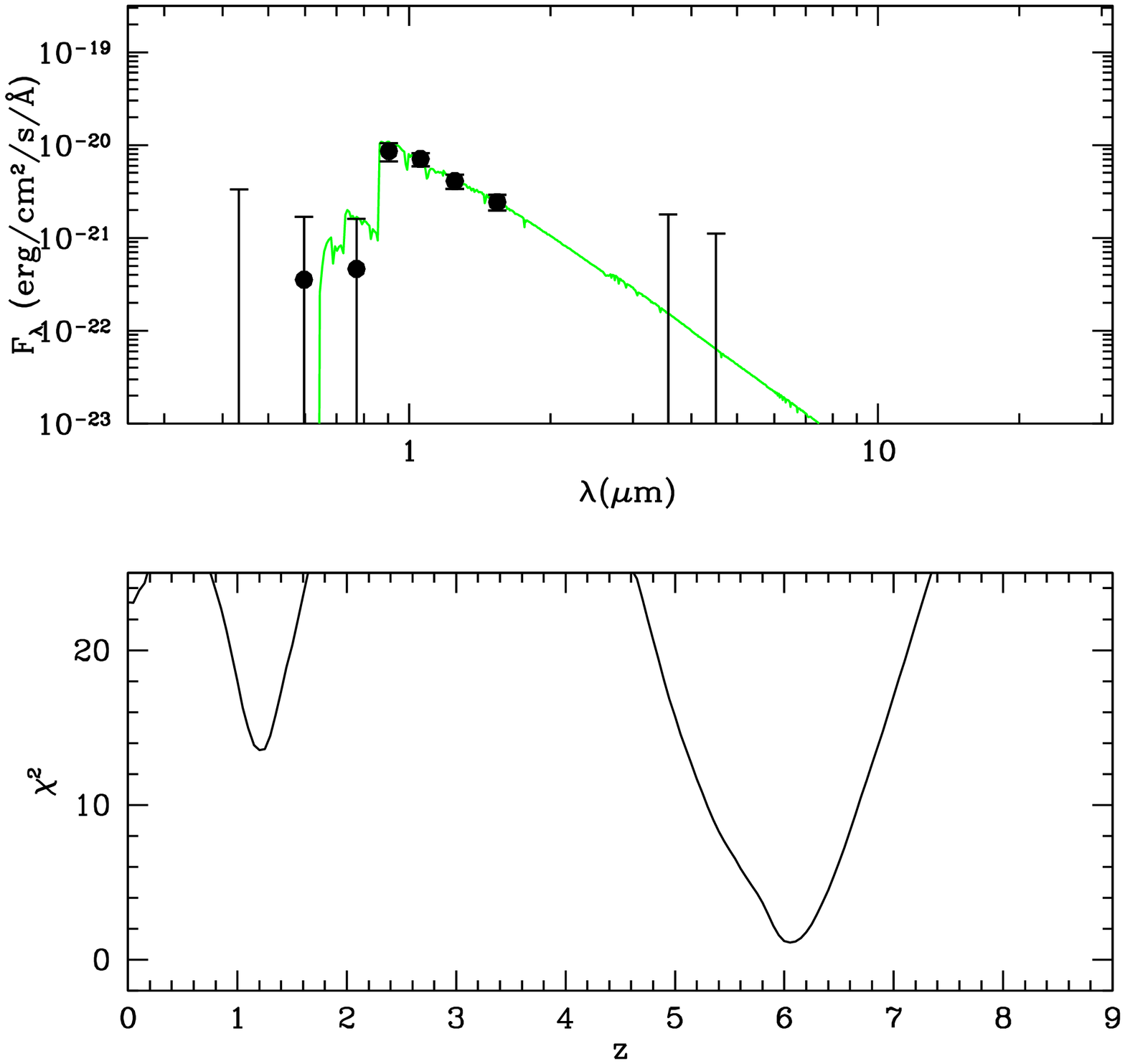}\\
\hspace*{1cm}{\bf 2719:} $z_{\rm est} = 6.05\, (5.30-6.90)$&
\hspace*{1cm} {\bf 1625:} $z_{\rm est} = 6.05\, (5.90-6.25)$\\
\\
\\
\\
\\
\hspace*{0.8cm}
\includegraphics[width=0.097\textwidth, angle=270]{1398_z_stamp.ps}
\includegraphics[width=0.097\textwidth, angle=270]{1398_y_stamp.ps}
\includegraphics[width=0.097\textwidth, angle=270]{1398_j_stamp.ps}
\includegraphics[width=0.097\textwidth, angle=270]{1398_h_stamp.ps}&
\hspace*{0.8cm}
\includegraphics[width=0.097\textwidth, angle=270]{1760_z_stamp.ps}
\includegraphics[width=0.097\textwidth, angle=270]{1760_y_stamp.ps}
\includegraphics[width=0.097\textwidth, angle=270]{1760_j_stamp.ps}
\includegraphics[width=0.097\textwidth, angle=270]{1760_h_stamp.ps}\\
\includegraphics[width=0.47\textwidth]{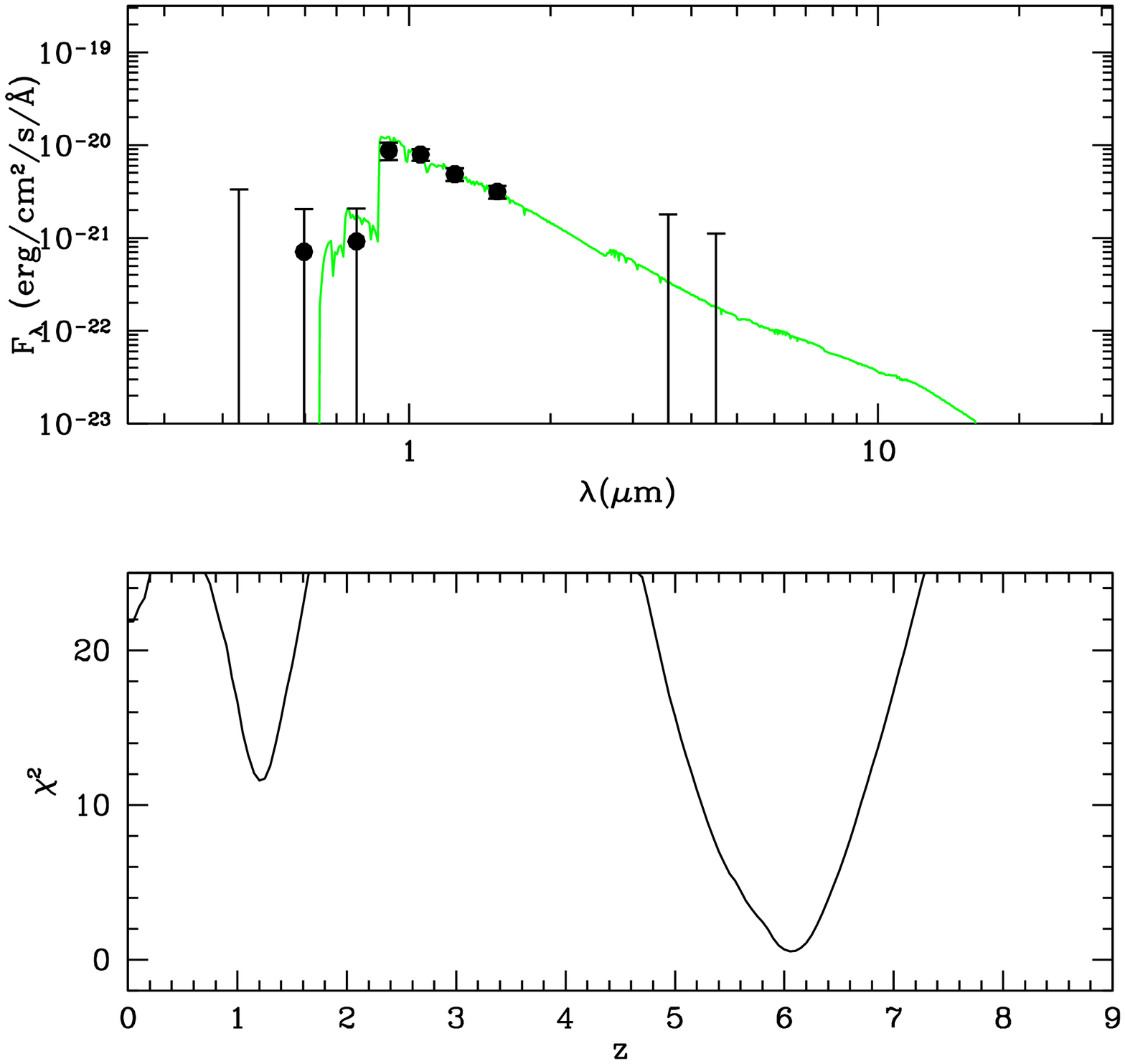}&
\includegraphics[width=0.47\textwidth]{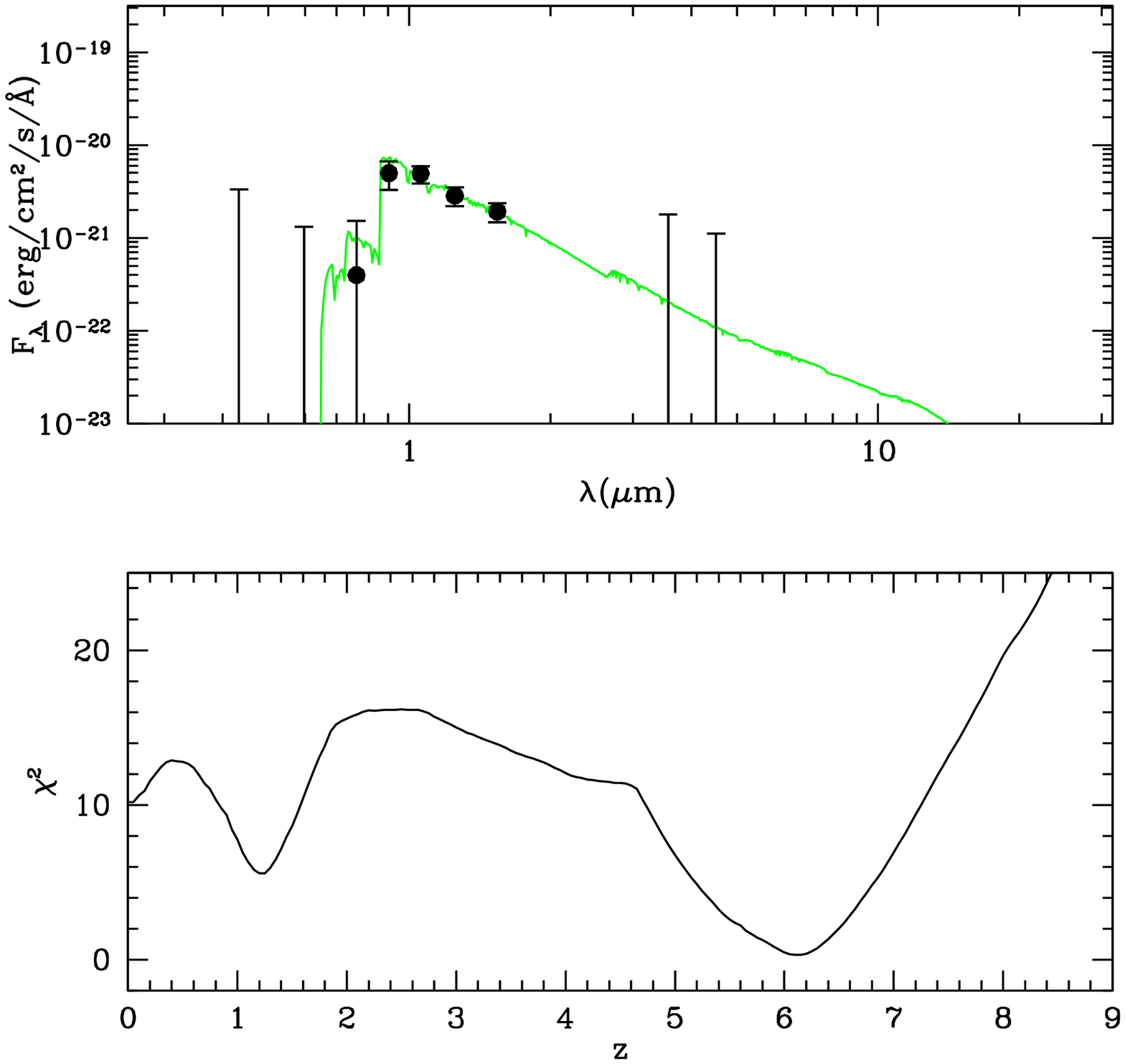}\\
\hspace*{1cm} {\bf 1398:} $z_{\rm est} = 6.10\, (5.90-6.30)$&
\hspace*{1cm}{\bf 1760:} $z_{\rm est} = 6.15\, (5.85-6.45)$\\
\\
\\
\end{tabular}
\addtocounter{figure}{-1}
\caption{continued.}
\end{figure*}

\begin{figure*}
\begin{tabular}{llll}
\hspace*{0.8cm}
\includegraphics[width=0.097\textwidth, angle=270]{934_z_stamp.ps}
\includegraphics[width=0.097\textwidth, angle=270]{934_y_stamp.ps}
\includegraphics[width=0.097\textwidth, angle=270]{934_j_stamp.ps}
\includegraphics[width=0.097\textwidth, angle=270]{934_h_stamp.ps}&
\hspace*{0.8cm}
\includegraphics[width=0.097\textwidth, angle=270]{2791_z_stamp.ps}
\includegraphics[width=0.097\textwidth, angle=270]{2791_y_stamp.ps}
\includegraphics[width=0.097\textwidth, angle=270]{2791_j_stamp.ps}
\includegraphics[width=0.097\textwidth, angle=270]{2791_h_stamp.ps}\\
\includegraphics[width=0.47\textwidth]{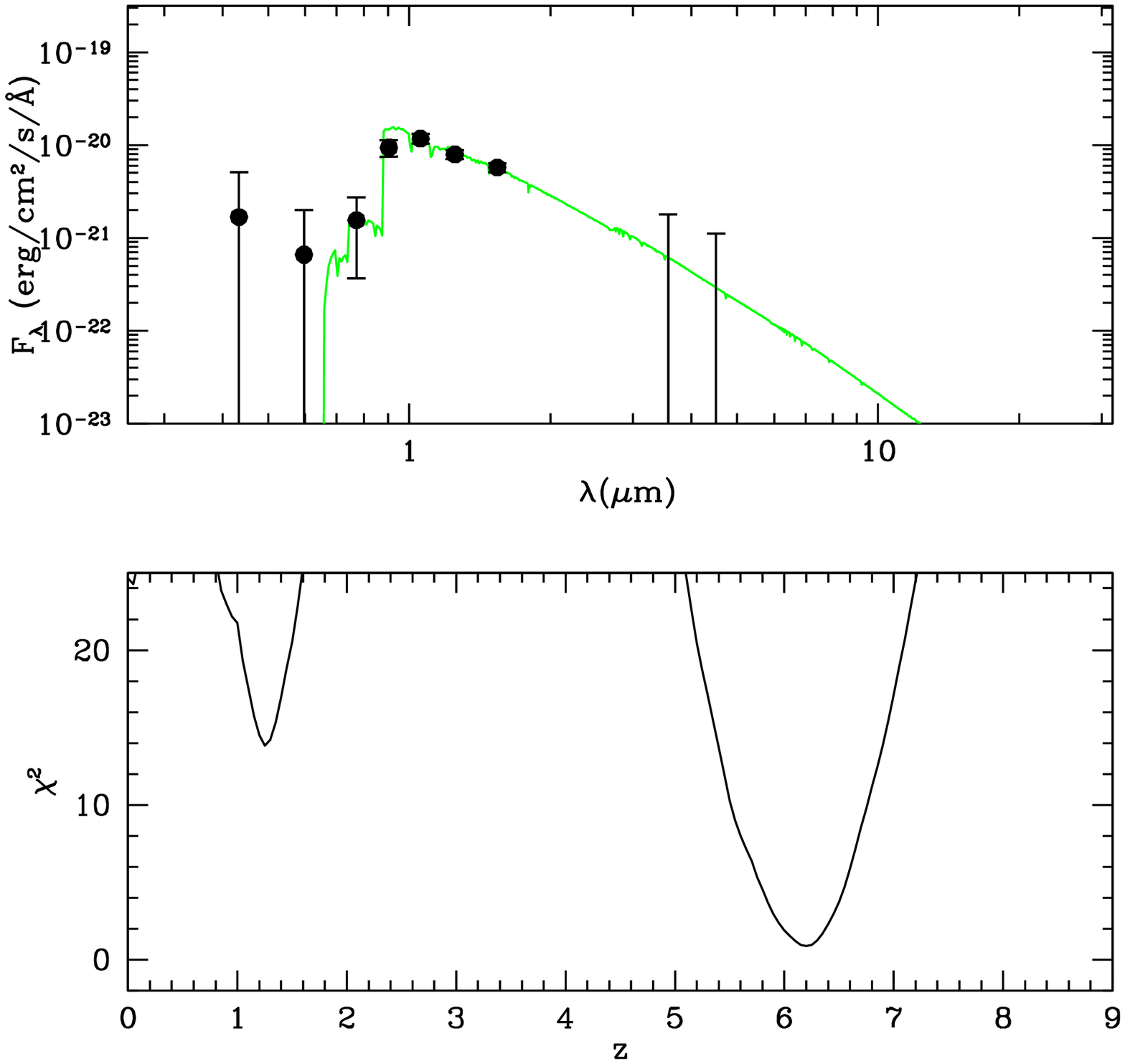}&
\includegraphics[width=0.47\textwidth]{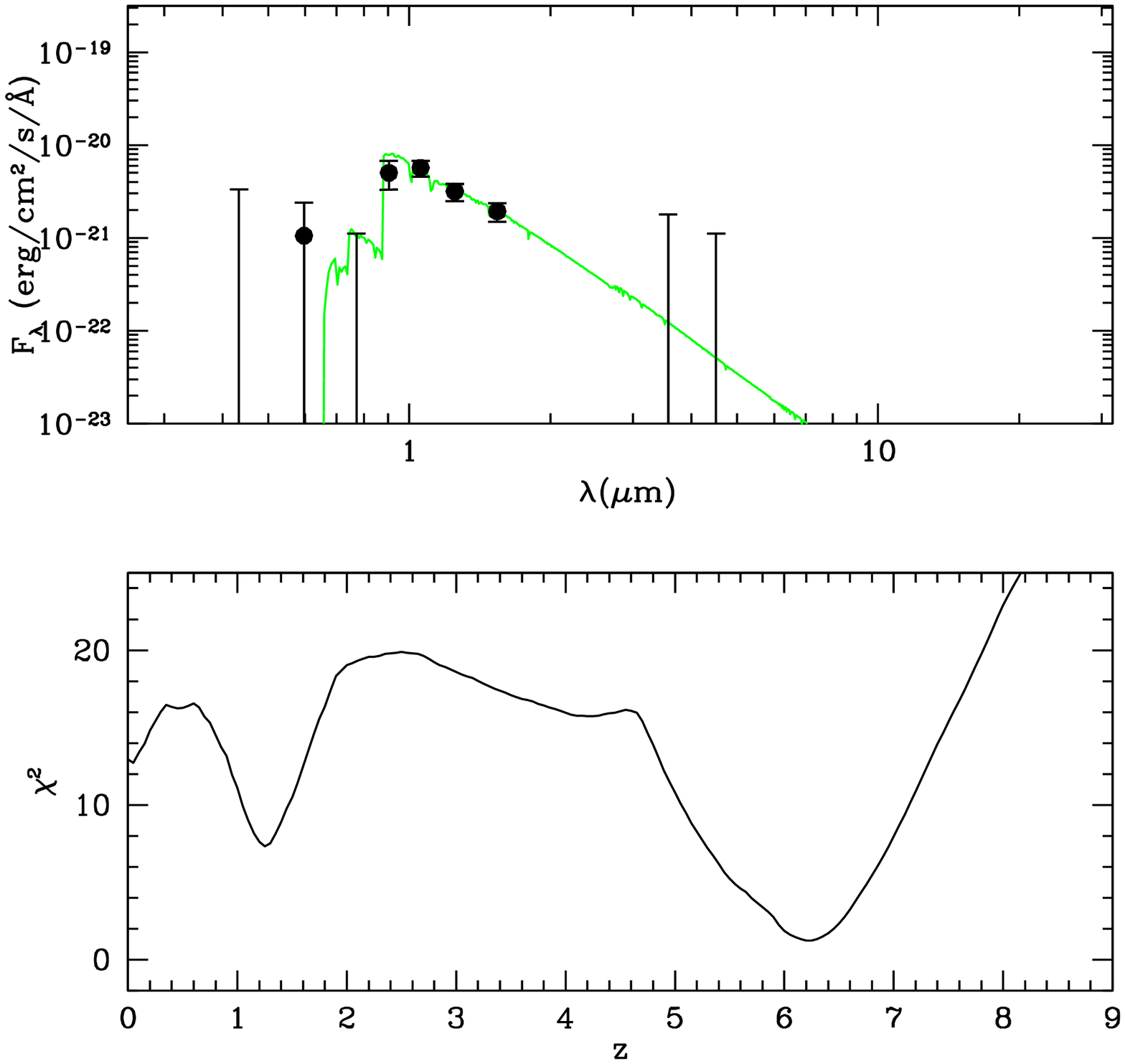}\\
\hspace*{1cm} {\bf 934:} $z_{\rm est} = 6.20\, (6.00-6.40)$&
\hspace*{1cm}{\bf 2791:} $z_{\rm est} = 6.25\, (5.95-6.50)$\\
\\
\\
\\
\\
\hspace*{0.8cm}
\includegraphics[width=0.097\textwidth, angle=270]{1464_z_stamp.ps}
\includegraphics[width=0.097\textwidth, angle=270]{1464_y_stamp.ps}
\includegraphics[width=0.097\textwidth, angle=270]{1464_j_stamp.ps}
\includegraphics[width=0.097\textwidth, angle=270]{1464_h_stamp.ps}&
\hspace*{0.8cm}
\includegraphics[width=0.097\textwidth, angle=270]{2003_z_stamp.ps}
\includegraphics[width=0.097\textwidth, angle=270]{2003_y_stamp.ps}
\includegraphics[width=0.097\textwidth, angle=270]{2003_j_stamp.ps}
\includegraphics[width=0.097\textwidth, angle=270]{2003_h_stamp.ps}\\
\includegraphics[width=0.47\textwidth]{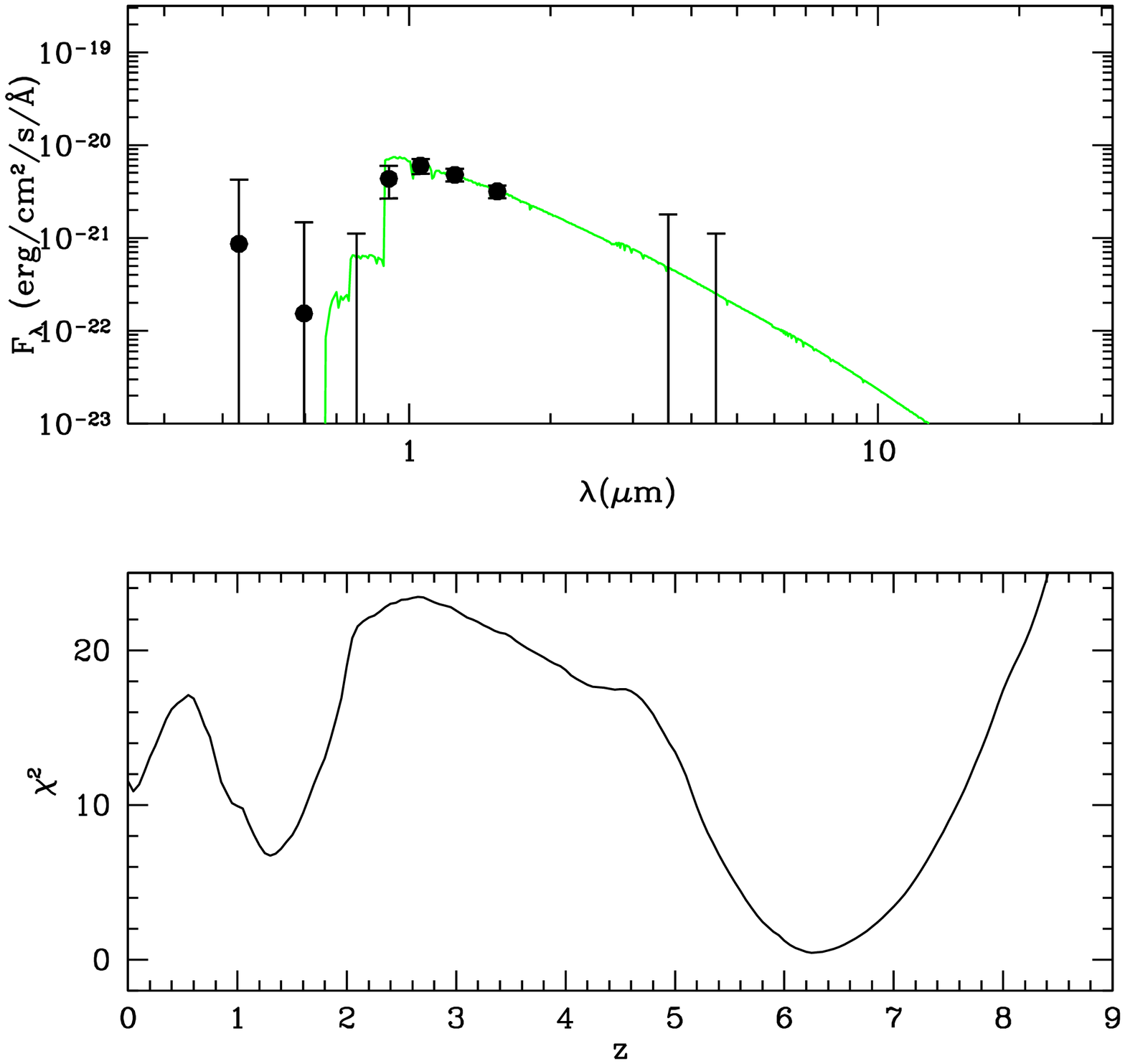}&
\includegraphics[width=0.47\textwidth]{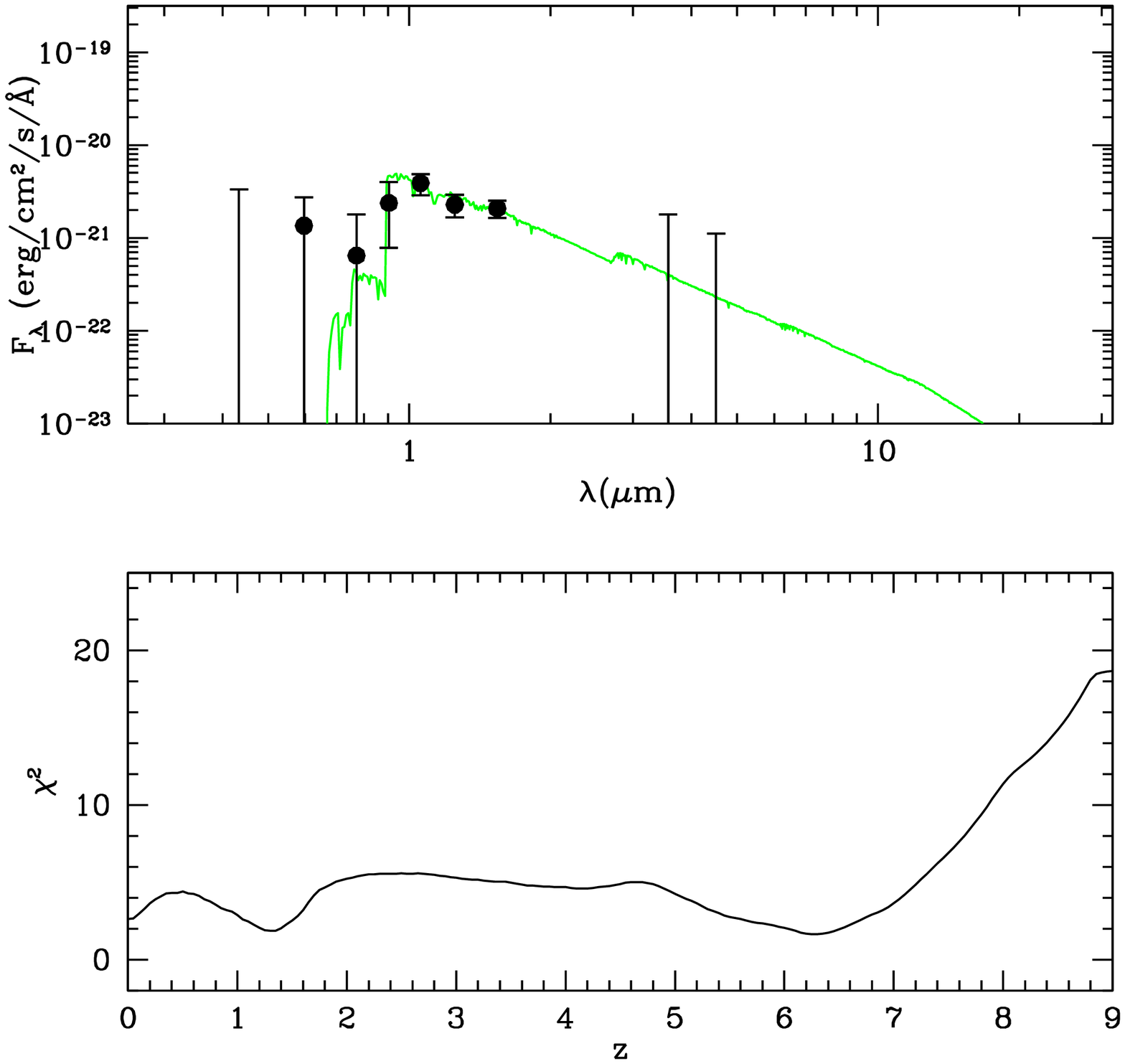}\\
\hspace*{1cm} {\bf 1464:} $z_{\rm est} = 6.30\, (5.95-6.75)$&
\hspace*{1cm}{\bf 2003:} $z_{\rm est} = 6.30\, (5.65-6.80)$\\ 
\\
\\
\end{tabular}
\addtocounter{figure}{-1}
\caption{continued.}
\end{figure*}

\begin{figure*}
\begin{tabular}{llll}
\hspace*{0.8cm}
\includegraphics[width=0.097\textwidth, angle=270]{2514_z_stamp.ps}
\includegraphics[width=0.097\textwidth, angle=270]{2514_y_stamp.ps}
\includegraphics[width=0.097\textwidth, angle=270]{2514_j_stamp.ps}
\includegraphics[width=0.097\textwidth, angle=270]{2514_h_stamp.ps}&
\hspace*{0.8cm}
\includegraphics[width=0.097\textwidth, angle=270]{837_z_stamp.ps}
\includegraphics[width=0.097\textwidth, angle=270]{837_y_stamp.ps}
\includegraphics[width=0.097\textwidth, angle=270]{837_j_stamp.ps}
\includegraphics[width=0.097\textwidth, angle=270]{837_h_stamp.ps}\\
\includegraphics[width=0.47\textwidth]{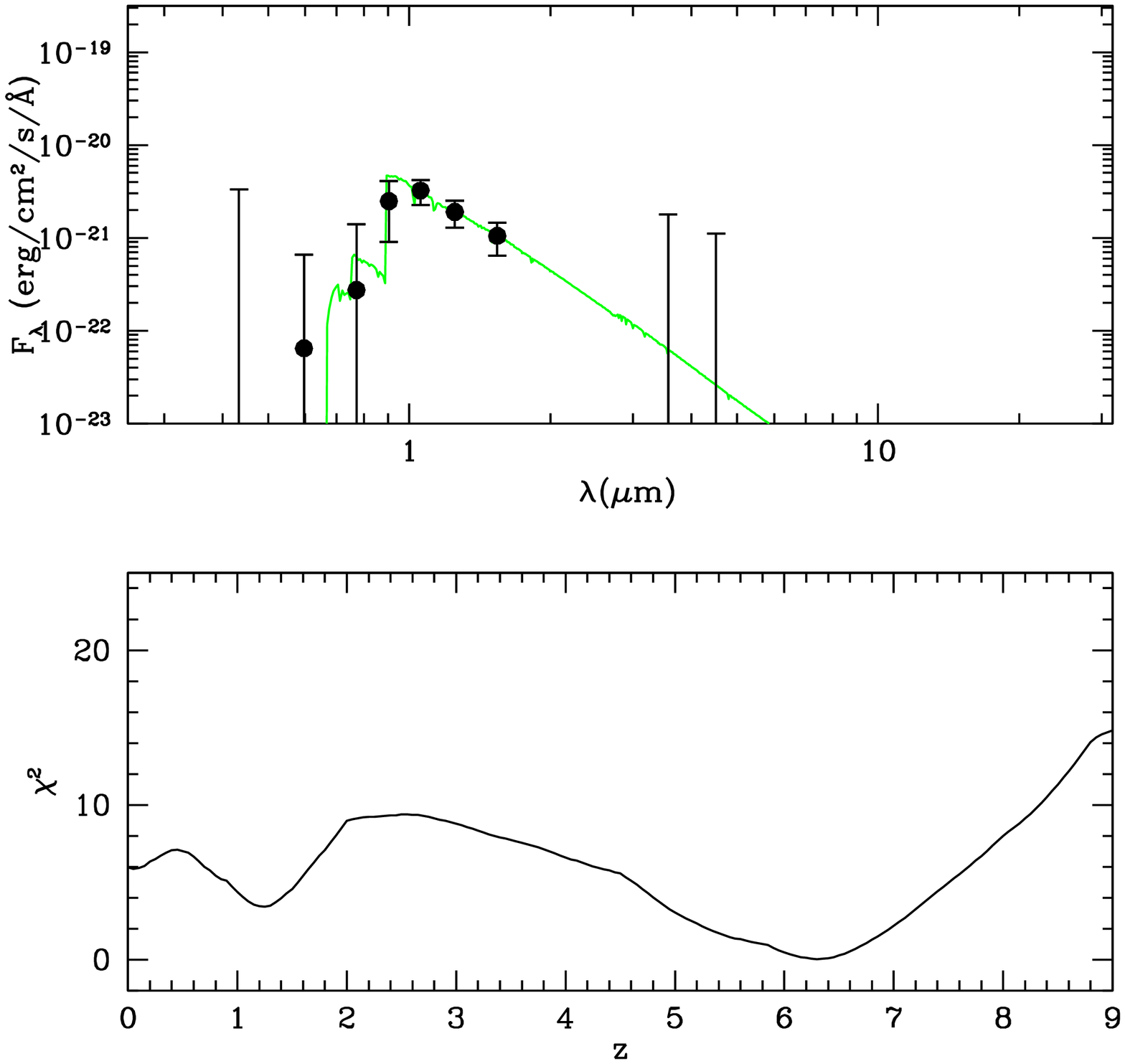}&
\includegraphics[width=0.47\textwidth]{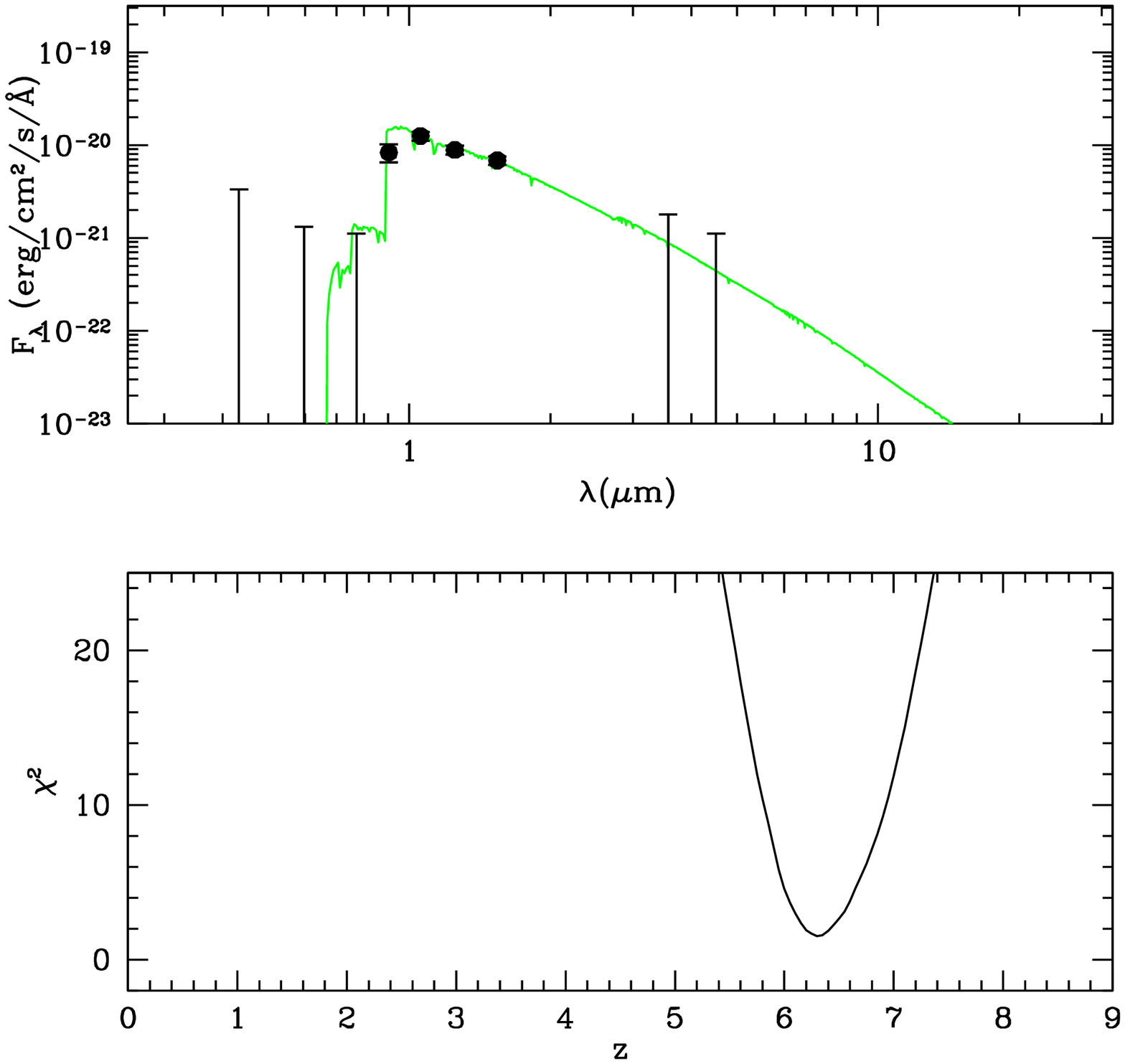}\\
\hspace*{1cm} {\bf 2514:} $z_{\rm est} = 6.30\, (5.85-6.75)$&
\hspace*{1cm}{\bf 837:} $z_{\rm est} = 6.35\, (6.15-6.55)$\\ 
\\
\\
\\
\\
\hspace*{0.8cm}
\includegraphics[width=0.097\textwidth, angle=270]{1855_z_stamp.ps}
\includegraphics[width=0.097\textwidth, angle=270]{1855_y_stamp.ps}
\includegraphics[width=0.097\textwidth, angle=270]{1855_j_stamp.ps}
\includegraphics[width=0.097\textwidth, angle=270]{1855_h_stamp.ps}&
\hspace*{0.8cm}
\includegraphics[width=0.097\textwidth, angle=270]{1864_z_stamp.ps}
\includegraphics[width=0.097\textwidth, angle=270]{1864_y_stamp.ps}
\includegraphics[width=0.097\textwidth, angle=270]{1864_j_stamp.ps}
\includegraphics[width=0.097\textwidth, angle=270]{1864_h_stamp.ps}\\
\includegraphics[width=0.47\textwidth]{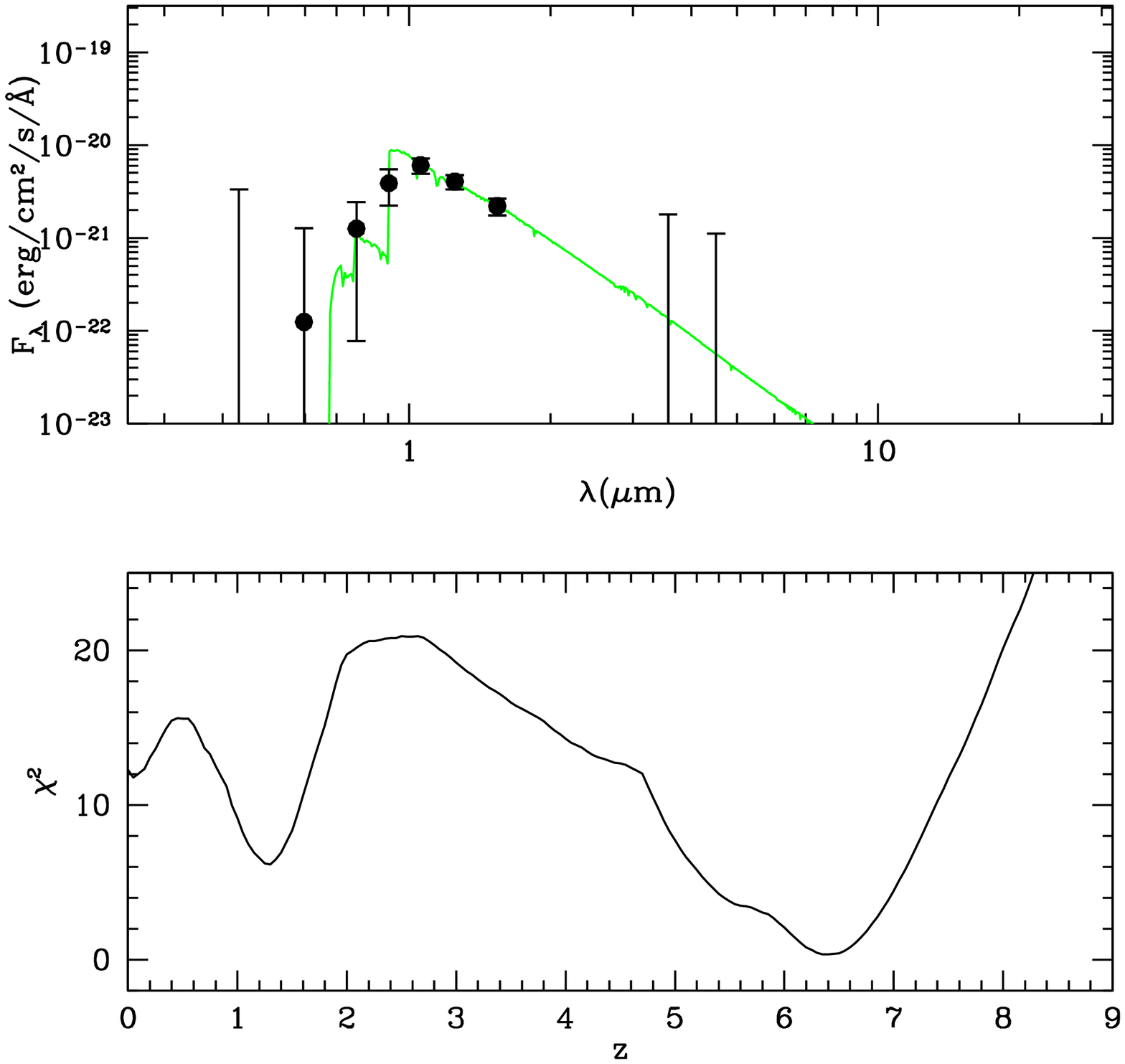}&
\includegraphics[width=0.47\textwidth]{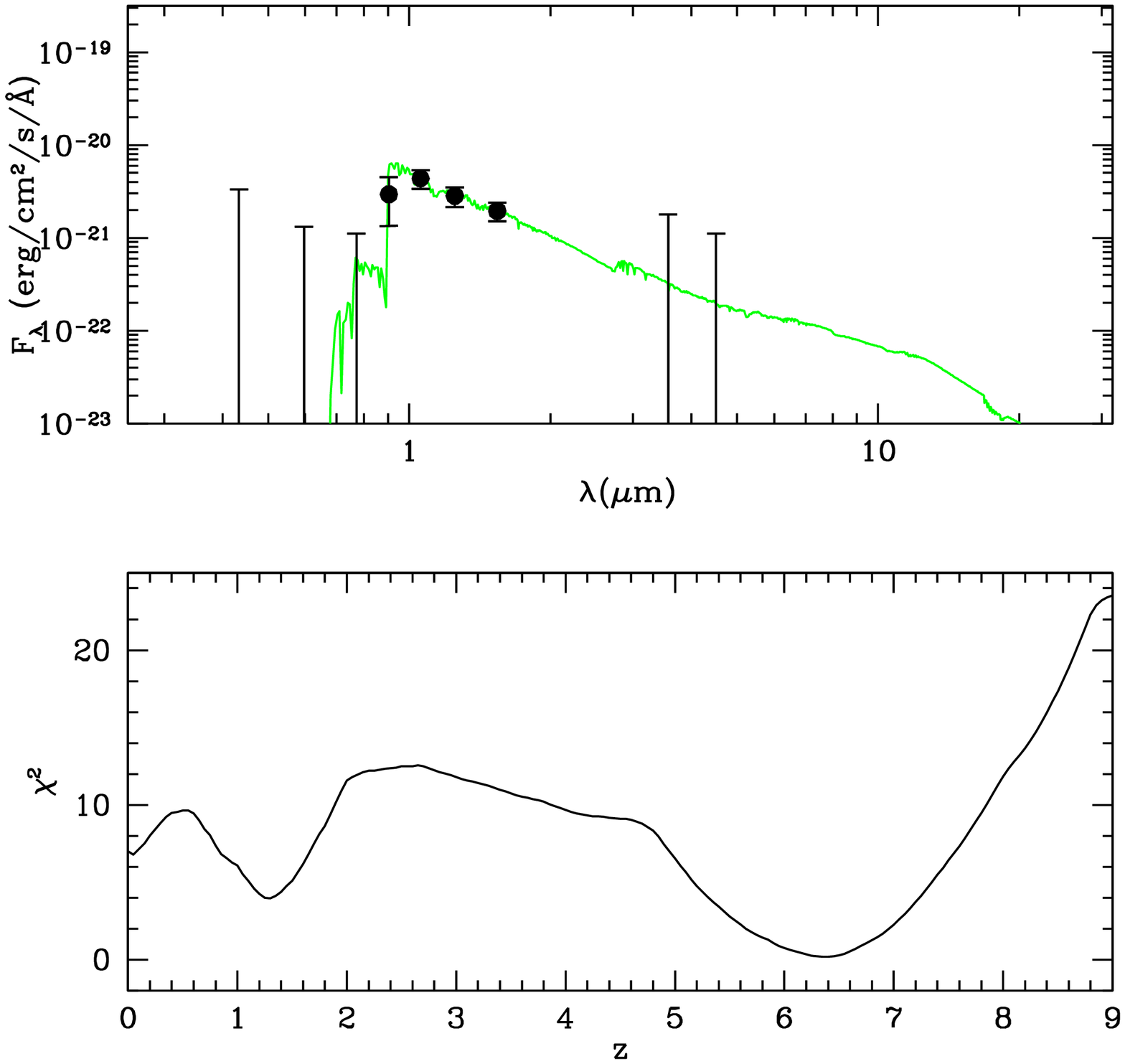}\\
\hspace*{1cm} {\bf 1855:} $z_{\rm est} = 6.40\, (6.05-6.70)$&
\hspace*{1cm}{\bf 1864:} $z_{\rm est} = 6.40\, (5.90-6.85)$\\
\\
\\
\end{tabular}
\addtocounter{figure}{-1}
\caption{continued.}
\end{figure*}

\begin{figure*}
\begin{tabular}{llll}
\hspace*{0.8cm}
\includegraphics[width=0.097\textwidth, angle=270]{1911_z_stamp.ps}
\includegraphics[width=0.097\textwidth, angle=270]{1911_y_stamp.ps}
\includegraphics[width=0.097\textwidth, angle=270]{1911_j_stamp.ps}
\includegraphics[width=0.097\textwidth, angle=270]{1911_h_stamp.ps}&
\hspace*{0.8cm}
\includegraphics[width=0.097\textwidth, angle=270]{1915_z_stamp.ps}
\includegraphics[width=0.097\textwidth, angle=270]{1915_y_stamp.ps}
\includegraphics[width=0.097\textwidth, angle=270]{1915_j_stamp.ps}
\includegraphics[width=0.097\textwidth, angle=270]{1915_h_stamp.ps}\\
\includegraphics[width=0.47\textwidth]{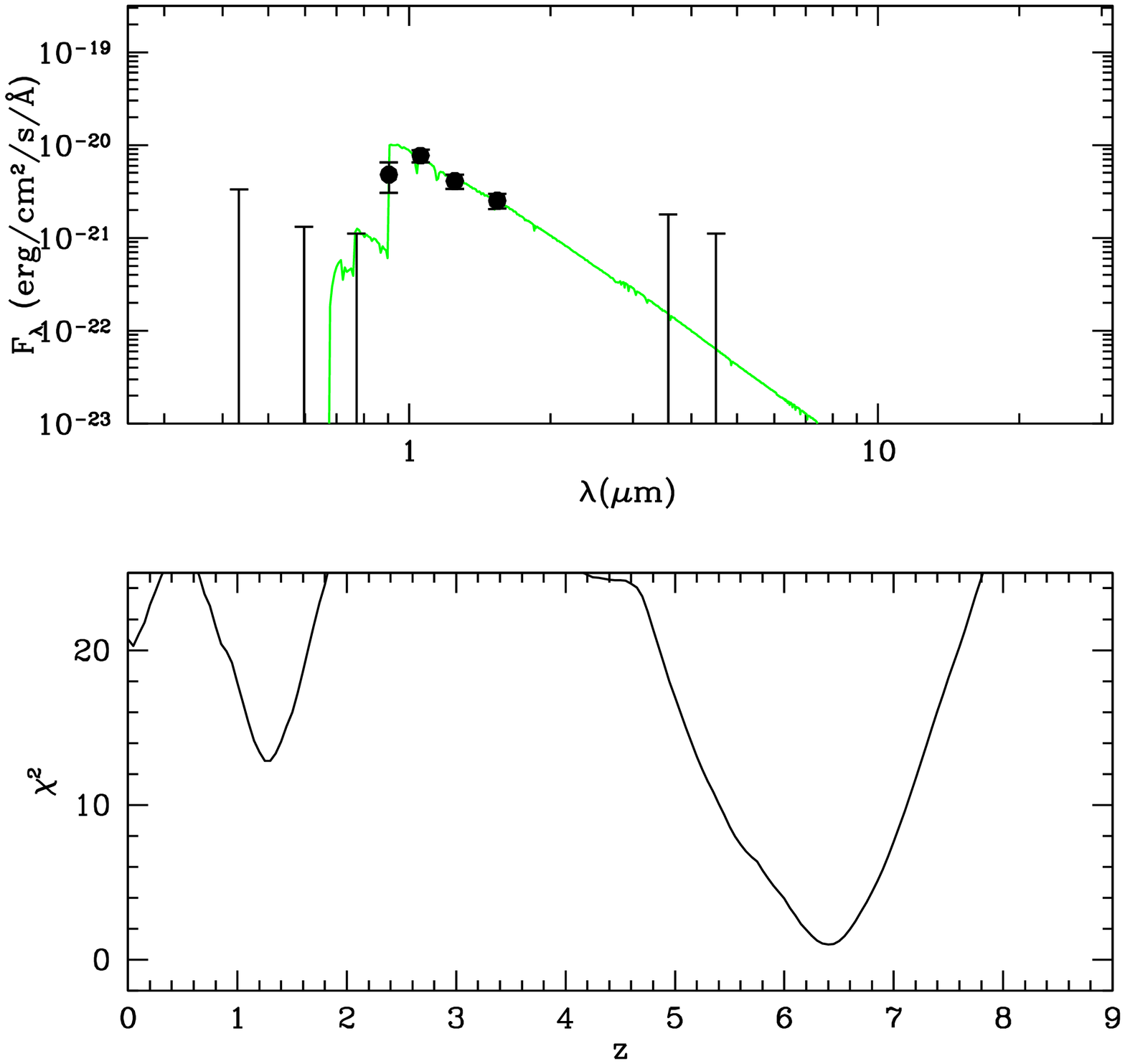}&
\includegraphics[width=0.47\textwidth]{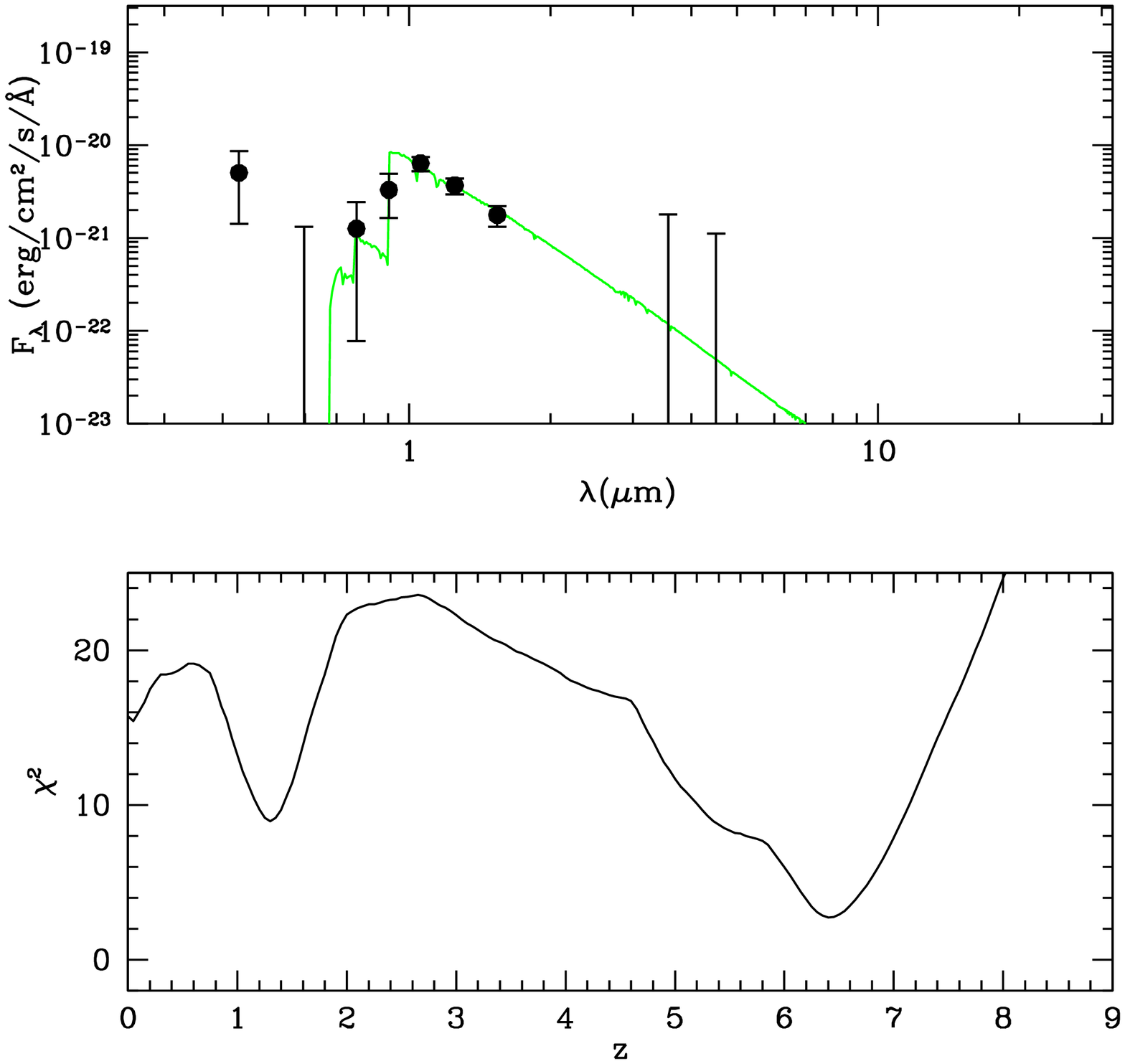}\\
\hspace*{1cm} {\bf 1911:} $z_{\rm est} = 6.40\, (6.20-6.60)$&
\hspace*{1cm}{\bf 1915:} $z_{\rm est} = 6.40\, (6.15-6.65)$\\
\\
\\
\\
\\
\hspace*{0.8cm}
\includegraphics[width=0.097\textwidth, angle=270]{2195_z_stamp.ps}
\includegraphics[width=0.097\textwidth, angle=270]{2195_y_stamp.ps}
\includegraphics[width=0.097\textwidth, angle=270]{2195_j_stamp.ps}
\includegraphics[width=0.097\textwidth, angle=270]{2195_h_stamp.ps}&
\hspace*{0.8cm}
\includegraphics[width=0.097\textwidth, angle=270]{1880_z_stamp.ps}
\includegraphics[width=0.097\textwidth, angle=270]{1880_y_stamp.ps}
\includegraphics[width=0.097\textwidth, angle=270]{1880_j_stamp.ps}
\includegraphics[width=0.097\textwidth, angle=270]{1880_h_stamp.ps}\\
\includegraphics[width=0.47\textwidth]{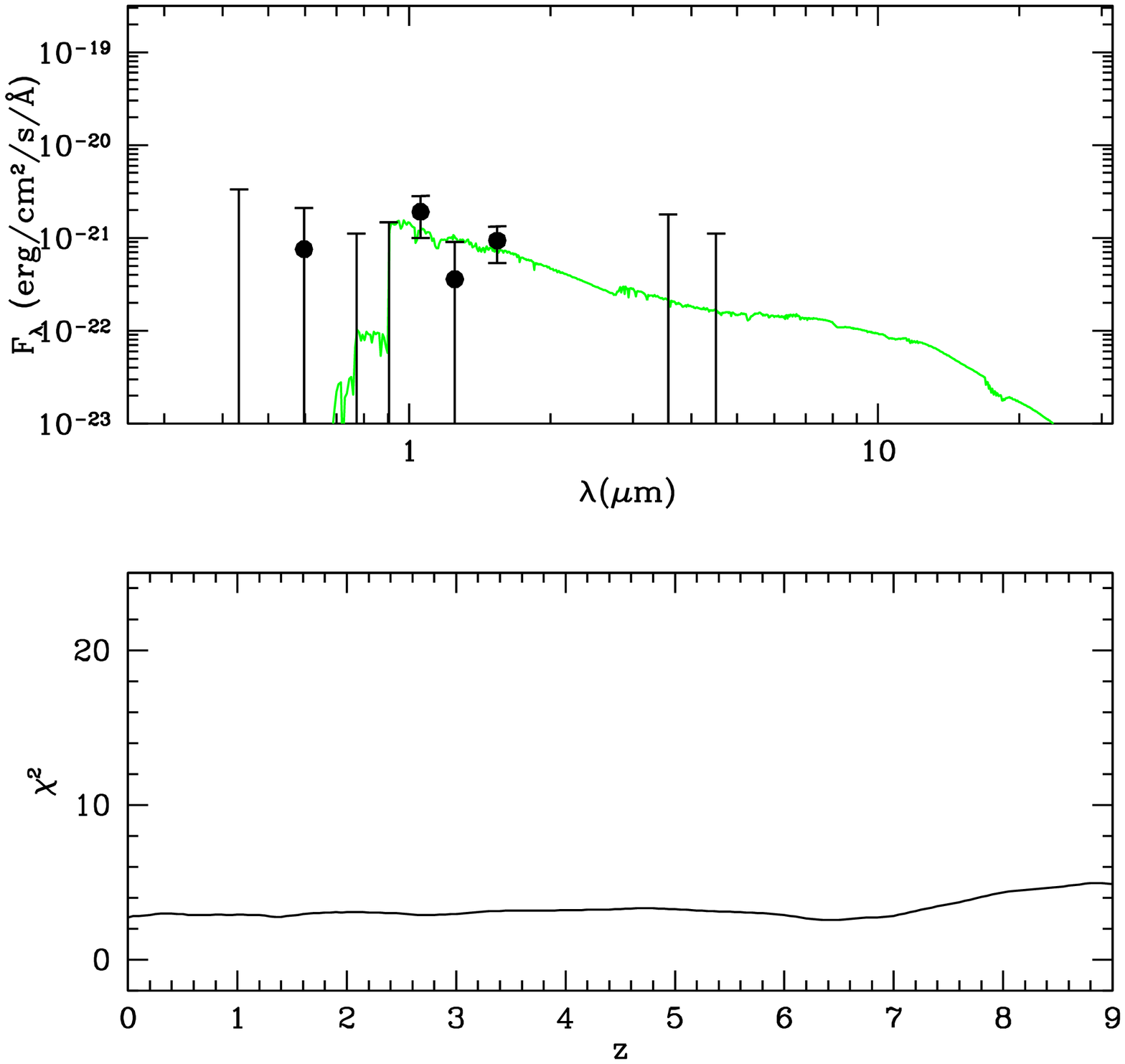}&
\includegraphics[width=0.47\textwidth]{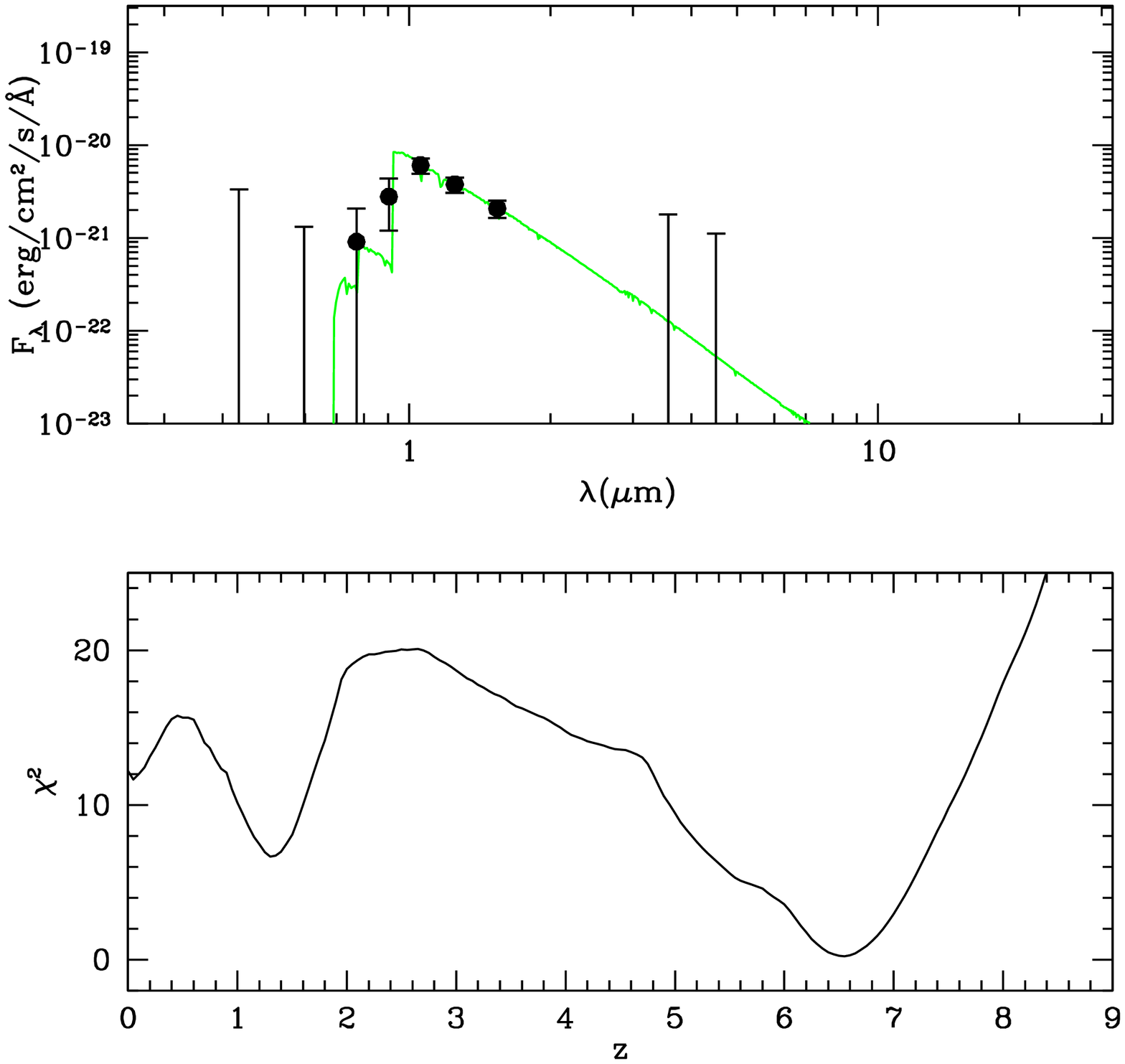}\\
\hspace*{1cm} {\bf 2195:} $z_{\rm est} = 6.45\, (4.75-7.50)$&
\hspace*{1cm}{\bf 1880:} $z_{\rm est} = 6.50\, (6.25-6.80)$\\
\\
\\
\\

\end{tabular}
\addtocounter{figure}{-1}
\caption{continued.}
\end{figure*}

\begin{figure*}
\begin{tabular}{llll}
\hspace*{0.8cm}
\includegraphics[width=0.097\textwidth, angle=270]{1958_z_stamp.ps}
\includegraphics[width=0.097\textwidth, angle=270]{1958_y_stamp.ps}
\includegraphics[width=0.097\textwidth, angle=270]{1958_j_stamp.ps}
\includegraphics[width=0.097\textwidth, angle=270]{1958_h_stamp.ps}&
\hspace*{0.8cm}
\includegraphics[width=0.097\textwidth, angle=270]{2206_z_stamp.ps}
\includegraphics[width=0.097\textwidth, angle=270]{2206_y_stamp.ps}
\includegraphics[width=0.097\textwidth, angle=270]{2206_j_stamp.ps}
\includegraphics[width=0.097\textwidth, angle=270]{2206_h_stamp.ps}\\
\includegraphics[width=0.47\textwidth]{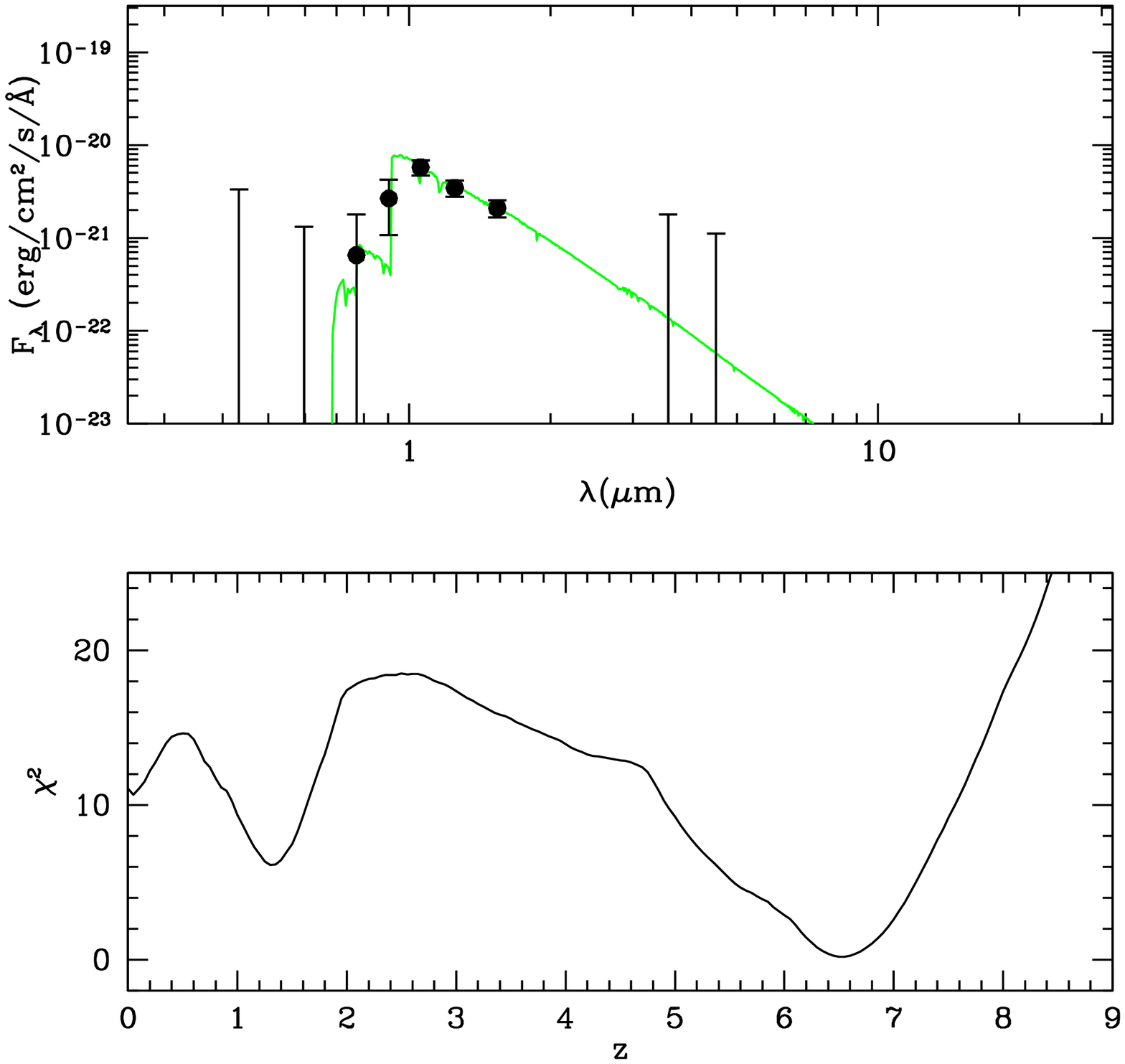}&
\includegraphics[width=0.47\textwidth]{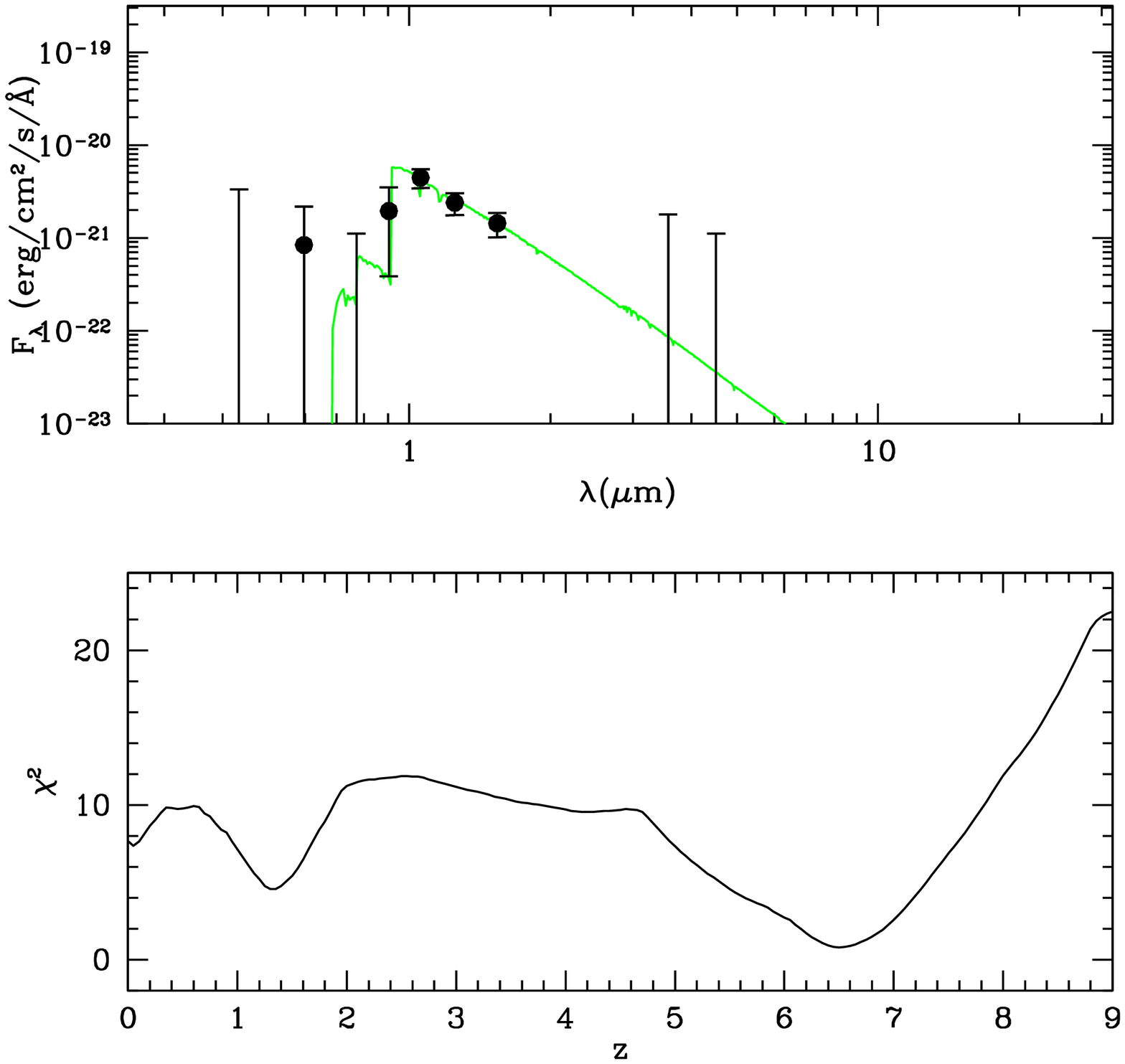}\\
\hspace*{1cm} {\bf 1958:} $z_{\rm est} = 6.50\, (6.25-6.80)$&
\hspace*{1cm}{\bf 2206:} $z_{\rm est} = 6.50\, (6.20-6.85)$\\
\\
\\
\\
\\
\hspace*{0.8cm}
\includegraphics[width=0.097\textwidth, angle=270]{1064_z_stamp.ps}
\includegraphics[width=0.097\textwidth, angle=270]{1064_y_stamp.ps}
\includegraphics[width=0.097\textwidth, angle=270]{1064_j_stamp.ps}
\includegraphics[width=0.097\textwidth, angle=270]{1064_h_stamp.ps}&
\hspace*{0.8cm}
\includegraphics[width=0.097\textwidth, angle=270]{688_z_stamp.ps}
\includegraphics[width=0.097\textwidth, angle=270]{688_y_stamp.ps}
\includegraphics[width=0.097\textwidth, angle=270]{688_j_stamp.ps}
\includegraphics[width=0.097\textwidth, angle=270]{688_h_stamp.ps}\\
\includegraphics[width=0.47\textwidth]{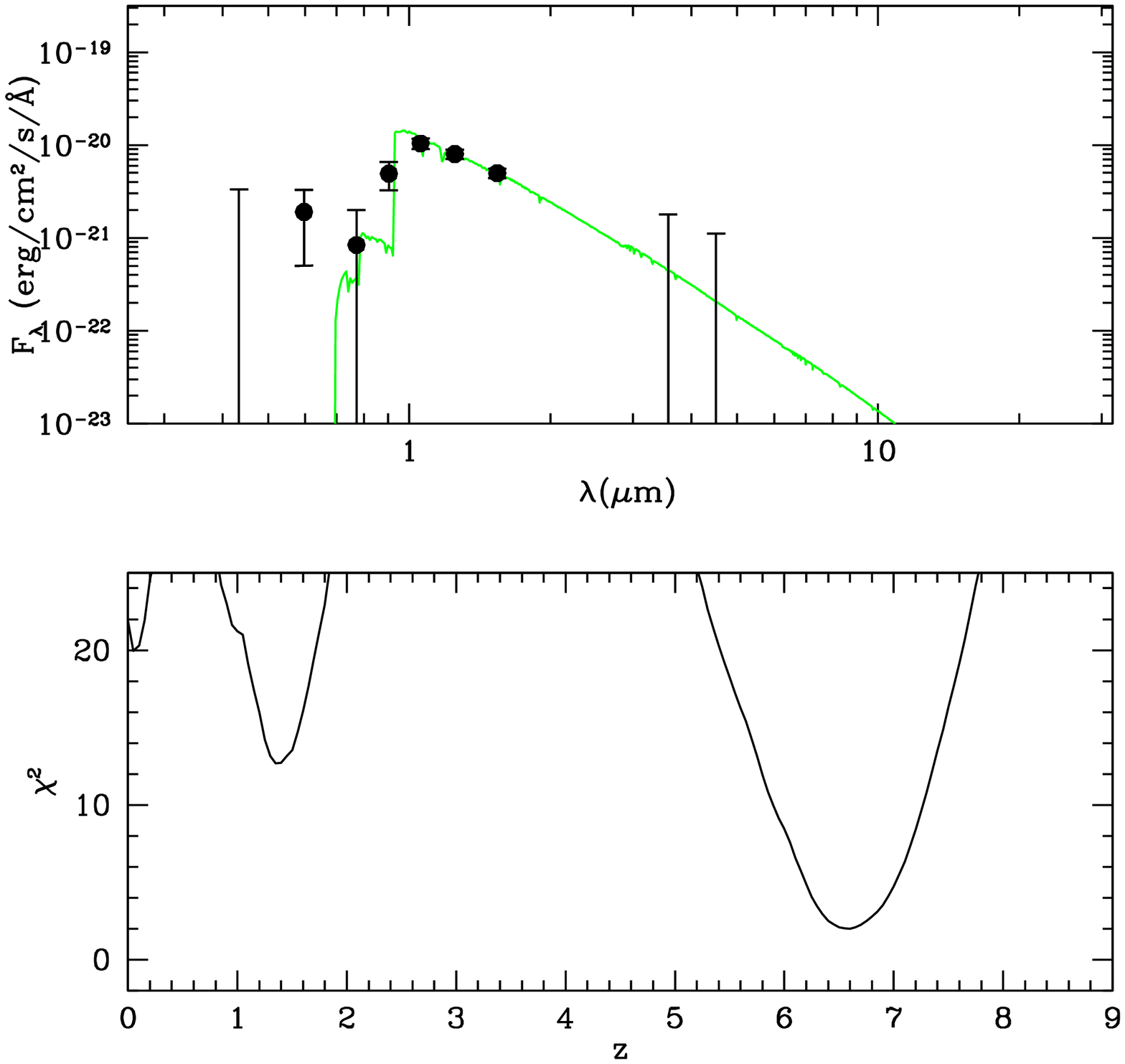}&
\includegraphics[width=0.47\textwidth]{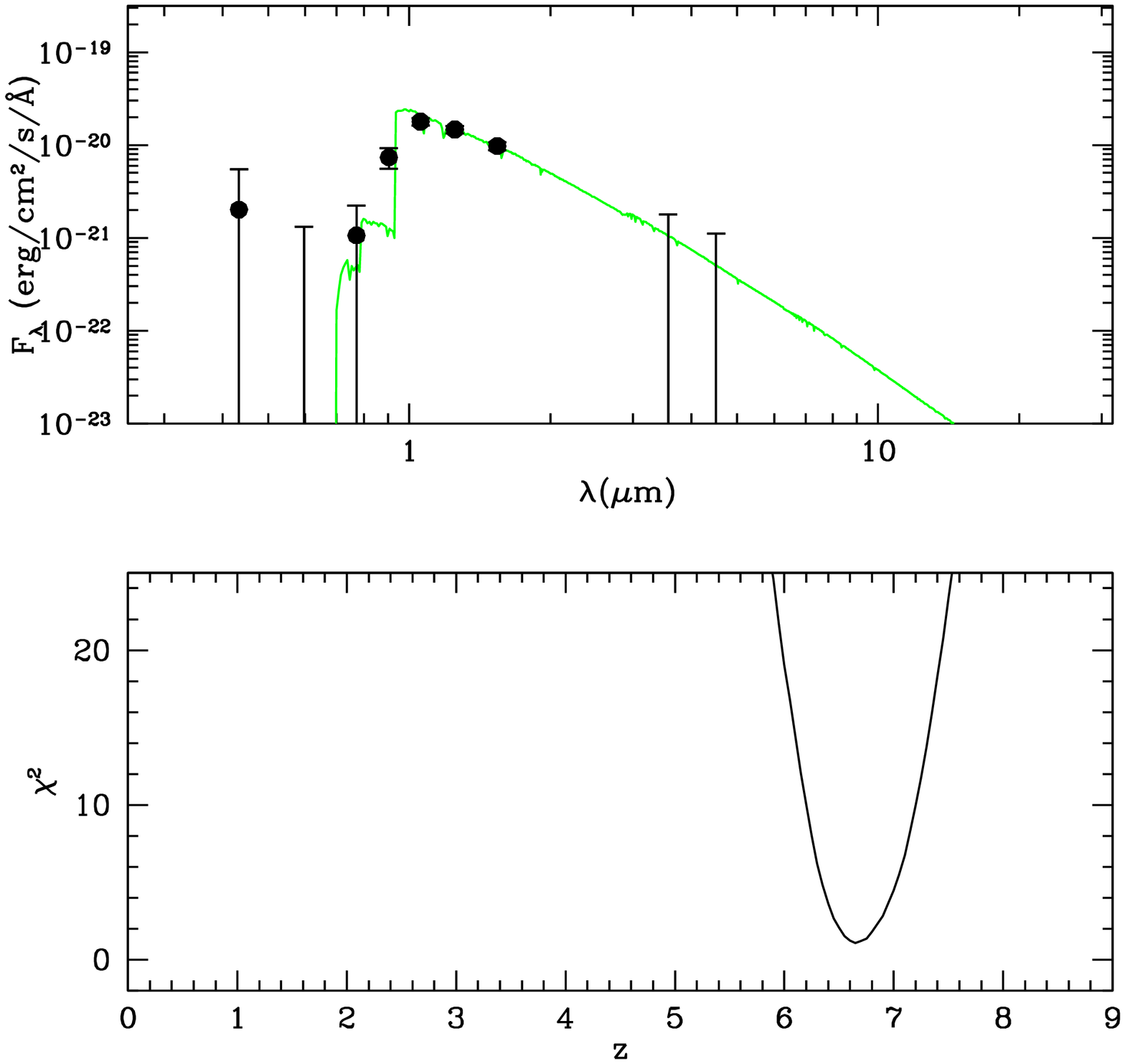}\\
\hspace*{1cm} {\bf 1064:} $z_{\rm est} = 6.65\, (6.35-6.90)$&
\hspace*{1cm}{\bf 688:} $z_{\rm est} = 6.70\, (6.50-6.90)$\\
\\
\\
\\
\end{tabular}
\addtocounter{figure}{-1}
\caption{continued.}
\end{figure*}

\begin{figure*}
\begin{tabular}{llll}
\hspace*{0.8cm}
\includegraphics[width=0.097\textwidth, angle=270]{2794_z_stamp.ps}
\includegraphics[width=0.097\textwidth, angle=270]{2794_y_stamp.ps}
\includegraphics[width=0.097\textwidth, angle=270]{2794_j_stamp.ps}
\includegraphics[width=0.097\textwidth, angle=270]{2794_h_stamp.ps}&
\hspace*{0.8cm}
\includegraphics[width=0.097\textwidth, angle=270]{1144_z_stamp.ps}
\includegraphics[width=0.097\textwidth, angle=270]{1144_y_stamp.ps}
\includegraphics[width=0.097\textwidth, angle=270]{1144_j_stamp.ps}
\includegraphics[width=0.097\textwidth, angle=270]{1144_h_stamp.ps}\\
\includegraphics[width=0.47\textwidth]{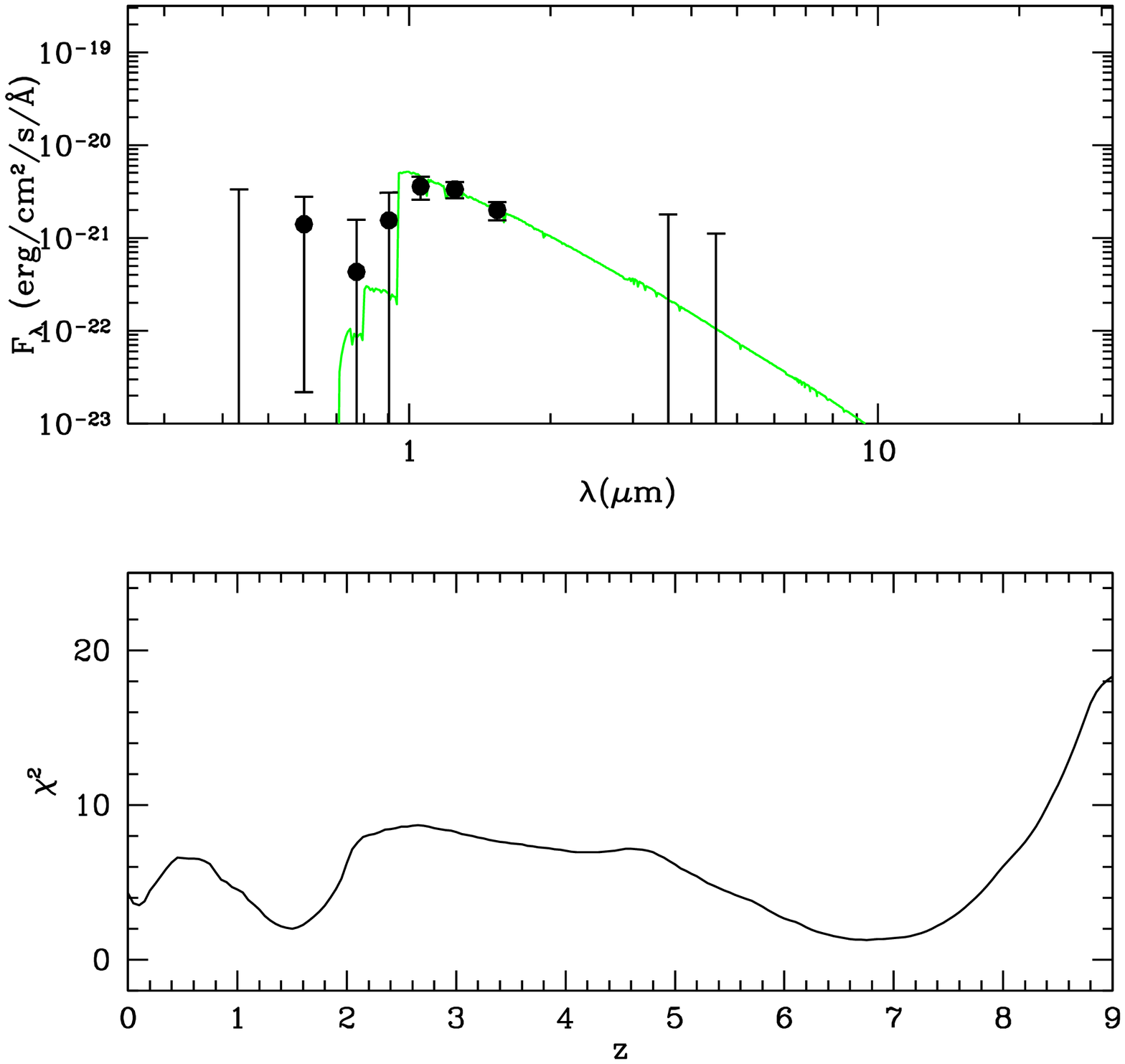}&
\includegraphics[width=0.47\textwidth]{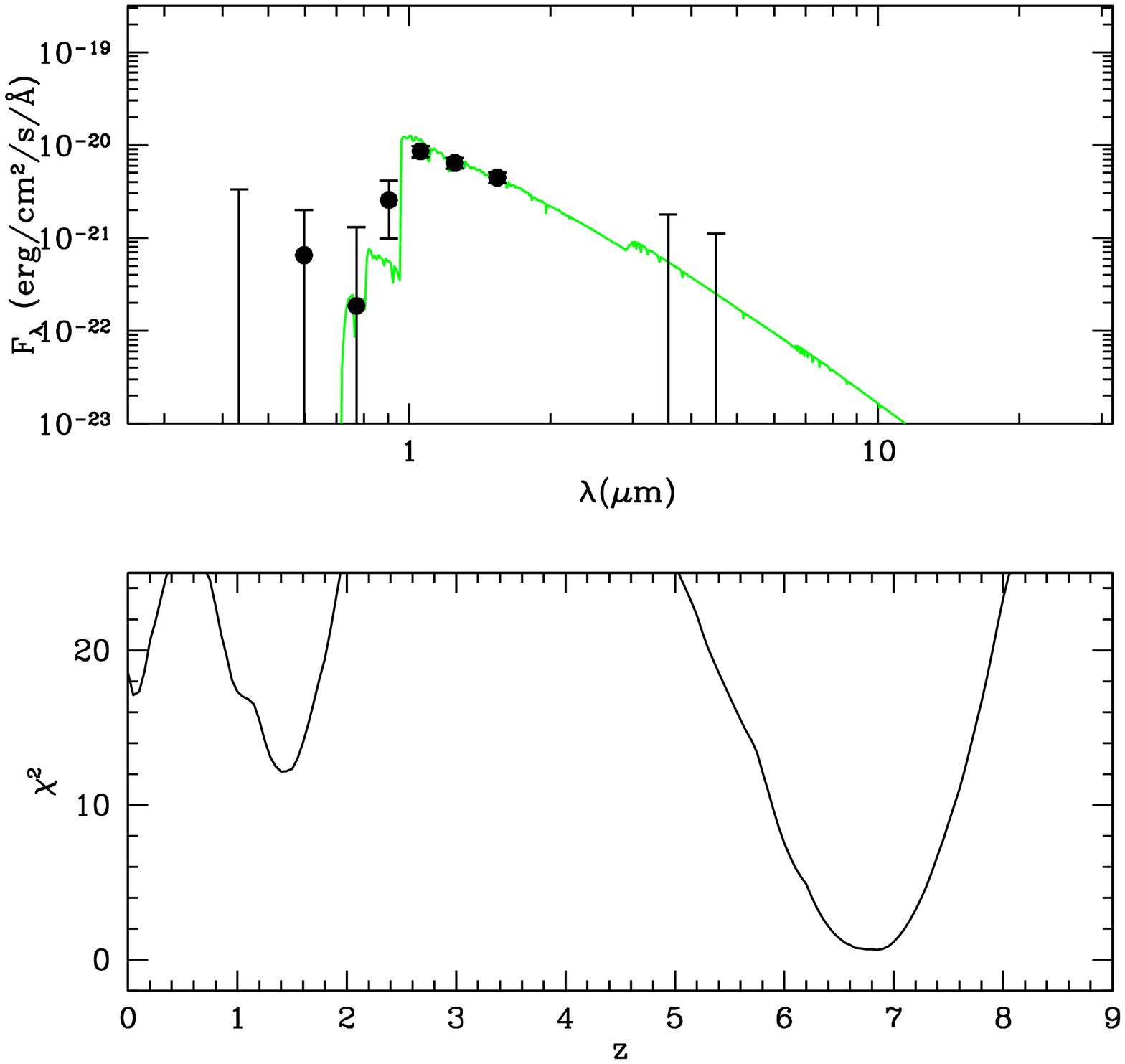}\\
\hspace*{1cm} {\bf 2794:} $z_{\rm est} = 6.75\, (6.20-7.45)$&
\hspace*{1cm}{\bf 1144:} $z_{\rm est} = 6.80\, (6.50-7.10)$\\
\\
\\
\\
\\
\hspace*{0.8cm}
\includegraphics[width=0.097\textwidth, angle=270]{2395_z_stamp.ps}
\includegraphics[width=0.097\textwidth, angle=270]{2395_y_stamp.ps}
\includegraphics[width=0.097\textwidth, angle=270]{2395_j_stamp.ps}
\includegraphics[width=0.097\textwidth, angle=270]{2395_h_stamp.ps}&
\hspace*{0.8cm}
\includegraphics[width=0.097\textwidth, angle=270]{1092_z_stamp.ps}
\includegraphics[width=0.097\textwidth, angle=270]{1092_y_stamp.ps}
\includegraphics[width=0.097\textwidth, angle=270]{1092_j_stamp.ps}
\includegraphics[width=0.097\textwidth, angle=270]{1092_h_stamp.ps}\\
\includegraphics[width=0.47\textwidth]{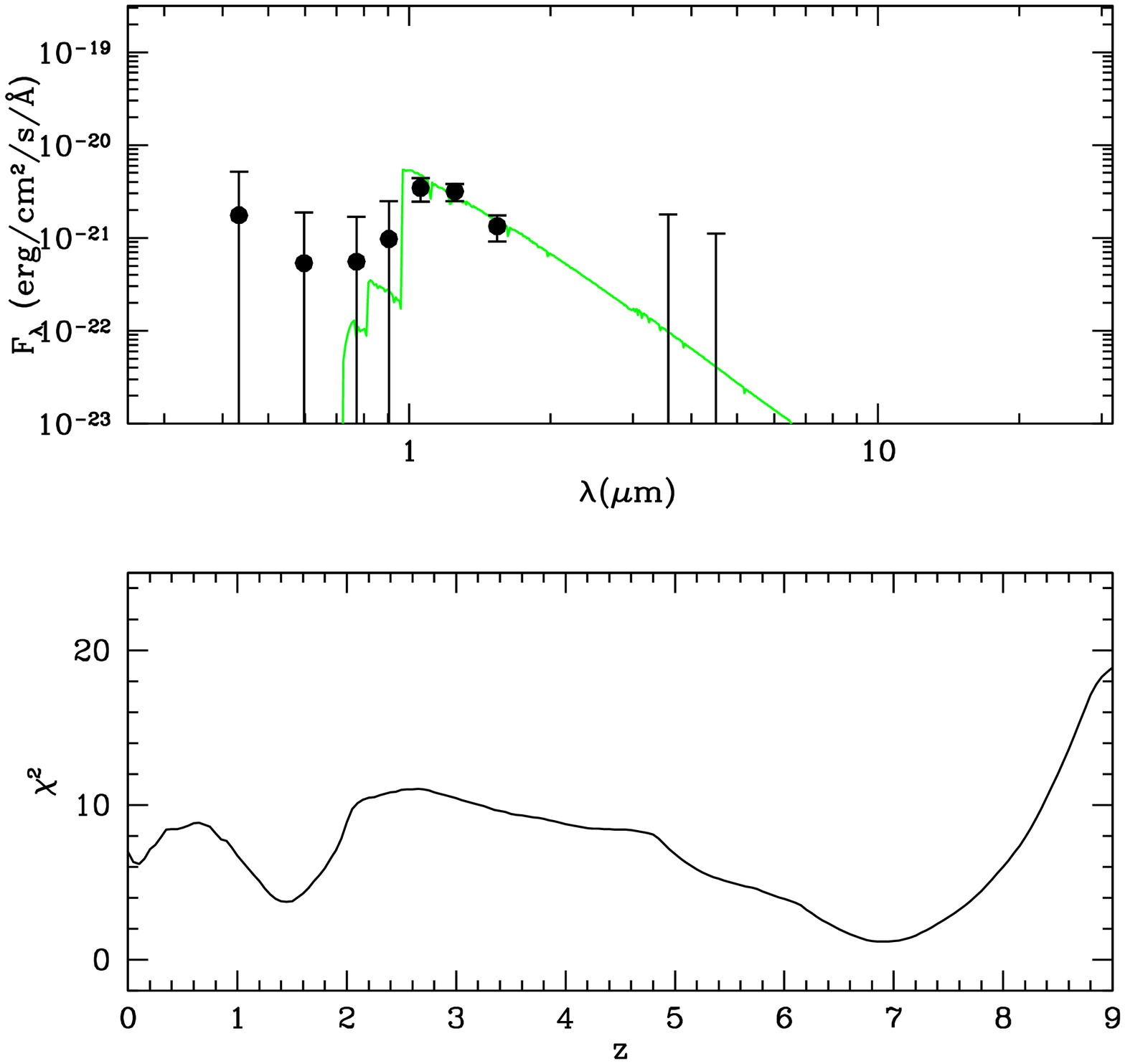}&
\includegraphics[width=0.47\textwidth]{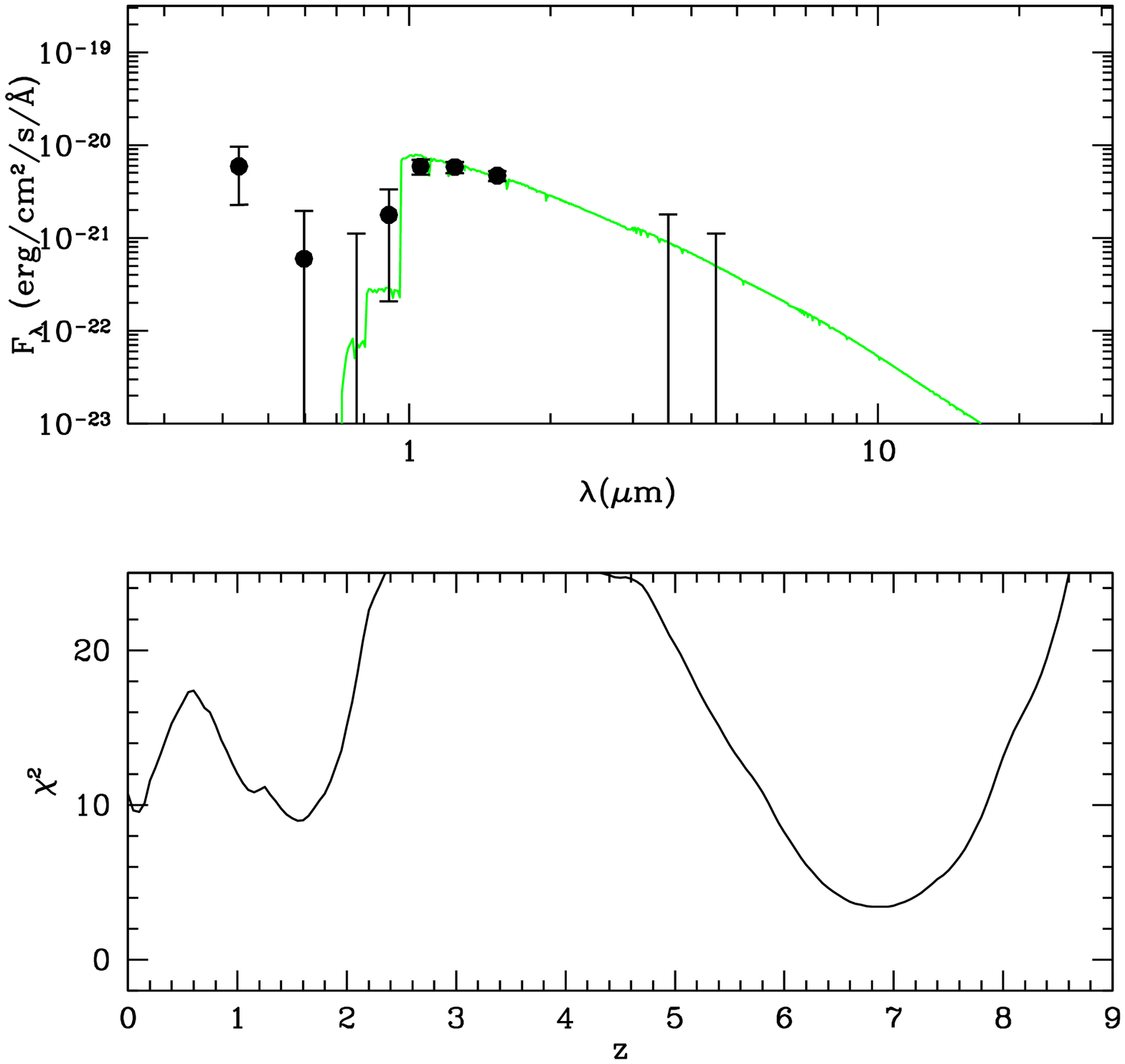}\\
\hspace*{1cm} {\bf 2395:} $z_{\rm est} = 6.80\, (6.40-7.30)$&
\hspace*{1cm}{\bf 1092:} $z_{\rm est} = 6.85\, (6.45-7.35)$\\
\\
\\
\end{tabular}
\addtocounter{figure}{-1}
\caption{continued.}
\end{figure*}

\begin{figure*}
\begin{tabular}{llll}
\hspace*{0.8cm}
\includegraphics[width=0.097\textwidth, angle=270]{2560_z_stamp.ps}
\includegraphics[width=0.097\textwidth, angle=270]{2560_y_stamp.ps}
\includegraphics[width=0.097\textwidth, angle=270]{2560_j_stamp.ps}
\includegraphics[width=0.097\textwidth, angle=270]{2560_h_stamp.ps}&
\hspace*{0.8cm}
\includegraphics[width=0.097\textwidth, angle=270]{2826_z_stamp.ps}
\includegraphics[width=0.097\textwidth, angle=270]{2826_y_stamp.ps}
\includegraphics[width=0.097\textwidth, angle=270]{2826_j_stamp.ps}
\includegraphics[width=0.097\textwidth, angle=270]{2826_h_stamp.ps}\\
\includegraphics[width=0.47\textwidth]{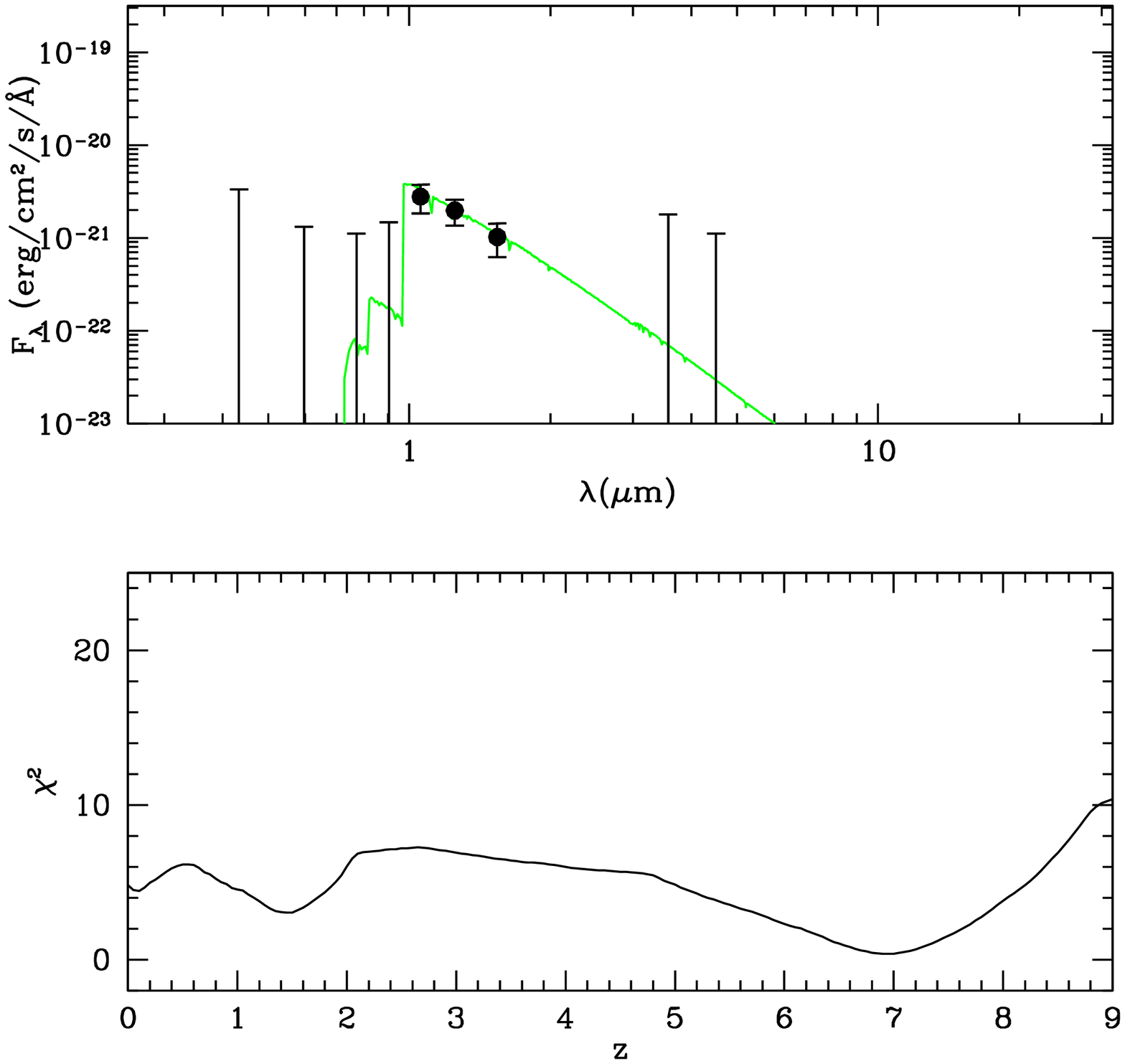}&
\includegraphics[width=0.47\textwidth]{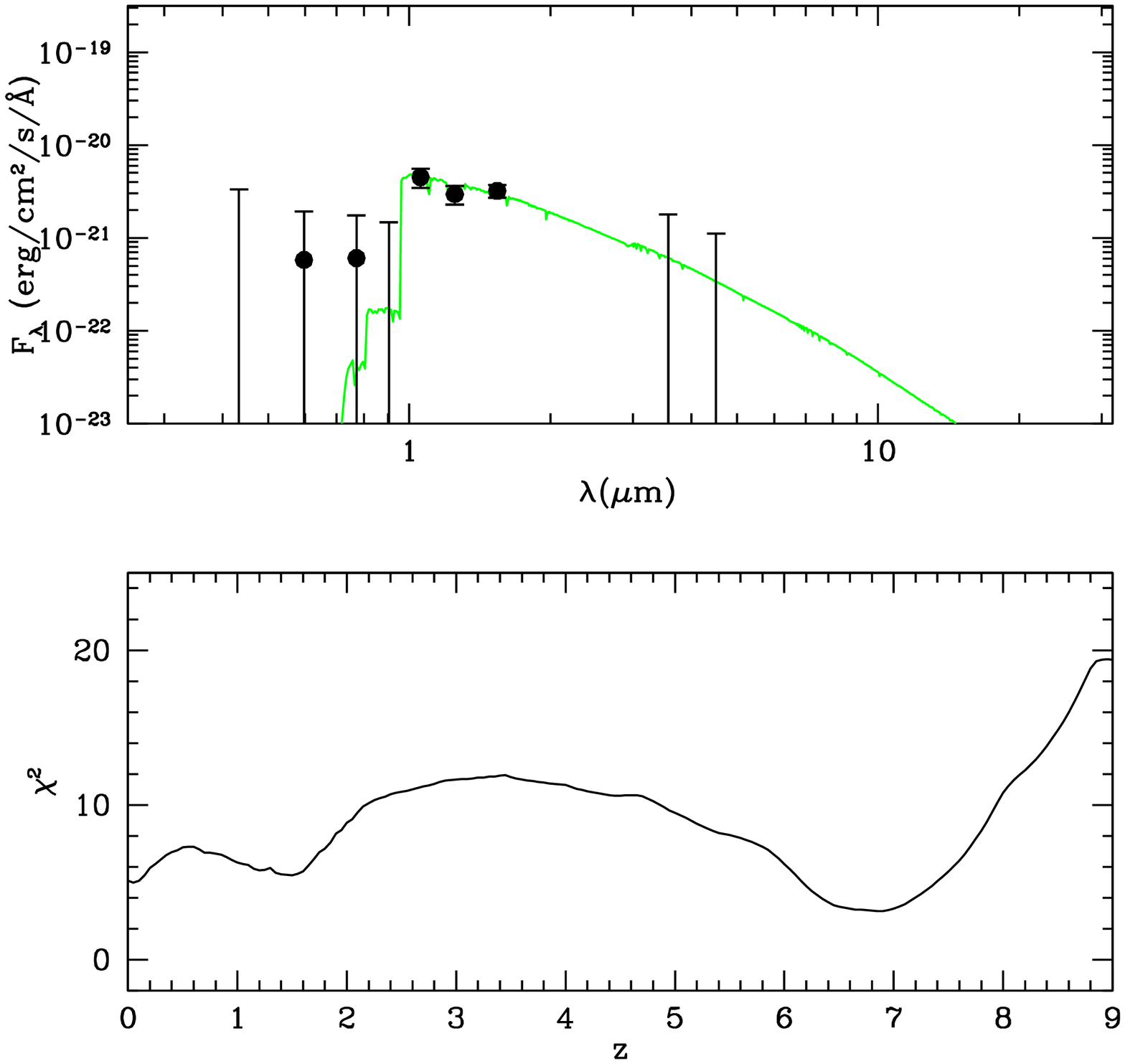}\\
\hspace*{1cm} {\bf 2560:} $z_{\rm est} = 6.90\, (6.35-7.45)$&
\hspace*{1cm}{\bf 2826:} $z_{\rm est} = 6.90\, (6.30-7.30)$\\
\\
\\
\\
\\
\hspace*{0.8cm}
\includegraphics[width=0.097\textwidth, angle=270]{1678_z_stamp.ps}
\includegraphics[width=0.097\textwidth, angle=270]{1678_y_stamp.ps}
\includegraphics[width=0.097\textwidth, angle=270]{1678_j_stamp.ps}
\includegraphics[width=0.097\textwidth, angle=270]{1678_h_stamp.ps}&
\hspace*{0.8cm}
\includegraphics[width=0.097\textwidth, angle=270]{2502_z_stamp.ps}
\includegraphics[width=0.097\textwidth, angle=270]{2502_y_stamp.ps}
\includegraphics[width=0.097\textwidth, angle=270]{2502_j_stamp.ps}
\includegraphics[width=0.097\textwidth, angle=270]{2502_h_stamp.ps}\\
\includegraphics[width=0.47\textwidth]{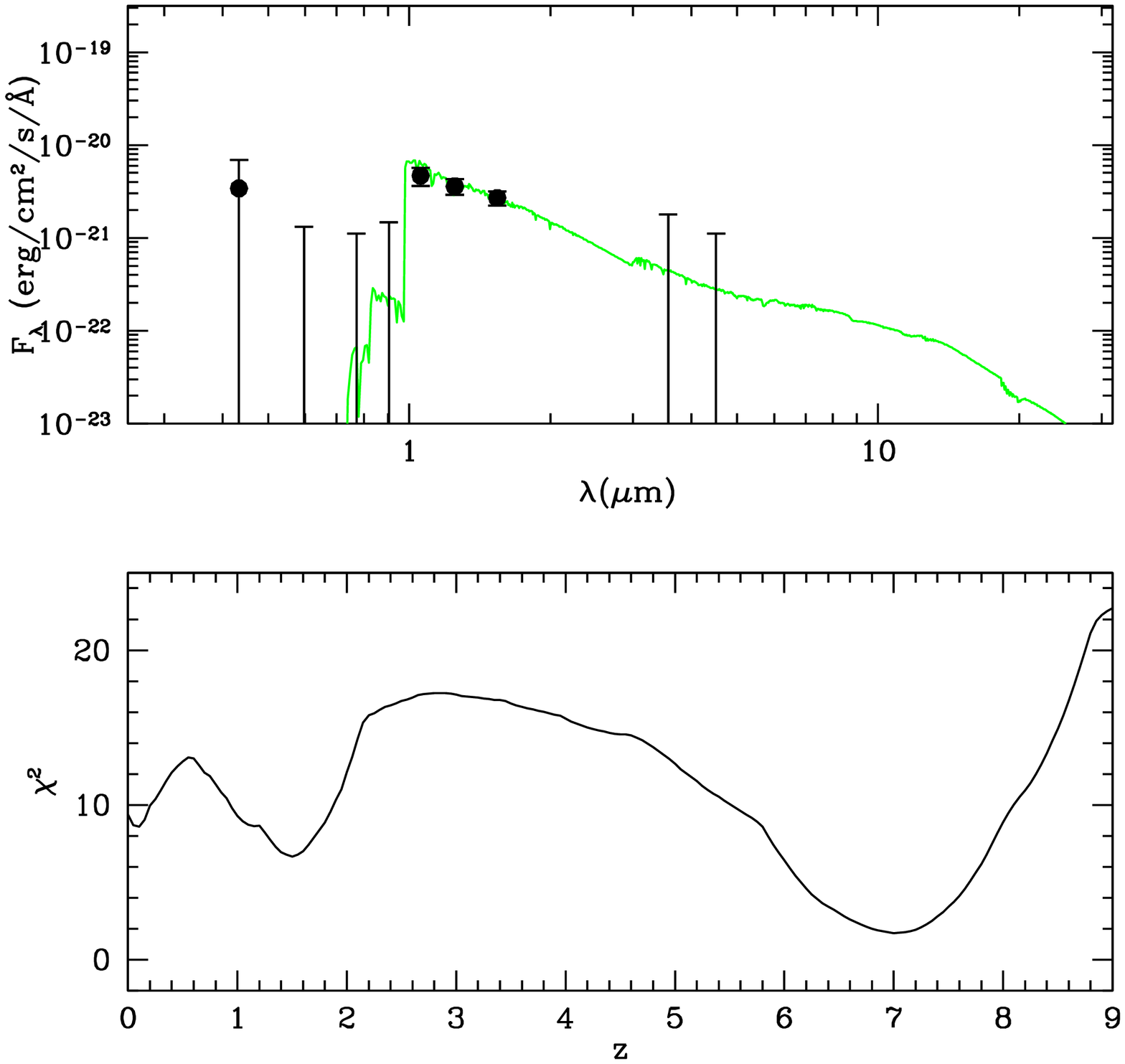}&
\includegraphics[width=0.47\textwidth]{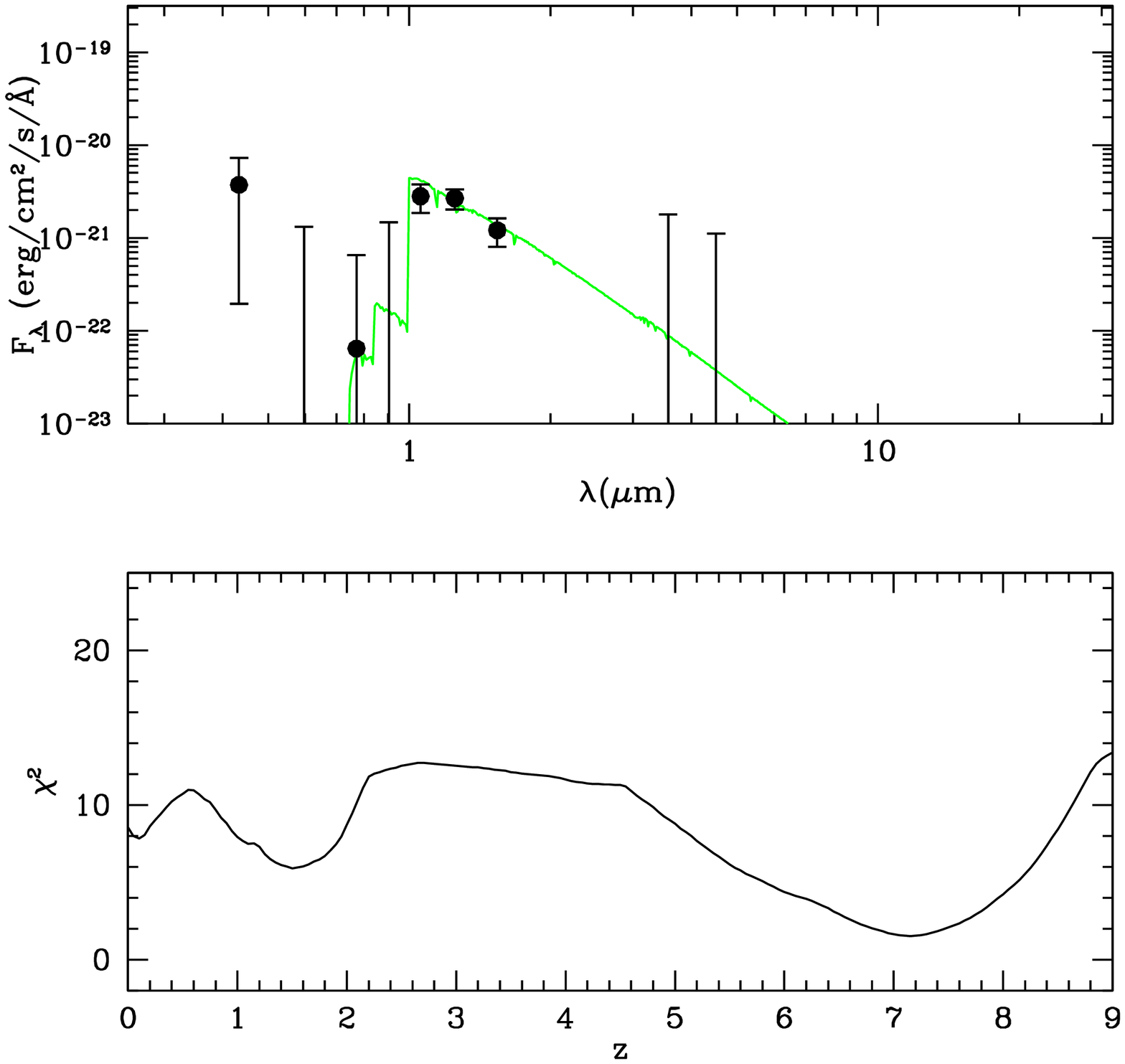}\\
\hspace*{1cm} {\bf 1678:} $z_{\rm est} = 7.05\, (6.60-7.40)$&
\hspace*{1cm}{\bf 2502:} $z_{\rm est} = 7.10\, (6.60-7.65)$\\
\\
\\
\end{tabular}
\addtocounter{figure}{-1}
\caption{continued.}
\end{figure*}

\begin{figure*}
\begin{tabular}{llll}
\hspace*{0.8cm}
\includegraphics[width=0.097\textwidth, angle=270]{1574_z_stamp.ps}
\includegraphics[width=0.097\textwidth, angle=270]{1574_y_stamp.ps}
\includegraphics[width=0.097\textwidth, angle=270]{1574_j_stamp.ps}
\includegraphics[width=0.097\textwidth, angle=270]{1574_h_stamp.ps}&
\hspace*{0.8cm}
\includegraphics[width=0.097\textwidth, angle=270]{835_z_stamp.ps}
\includegraphics[width=0.097\textwidth, angle=270]{835_y_stamp.ps}
\includegraphics[width=0.097\textwidth, angle=270]{835_j_stamp.ps}
\includegraphics[width=0.097\textwidth, angle=270]{835_h_stamp.ps}\\
\includegraphics[width=0.47\textwidth]{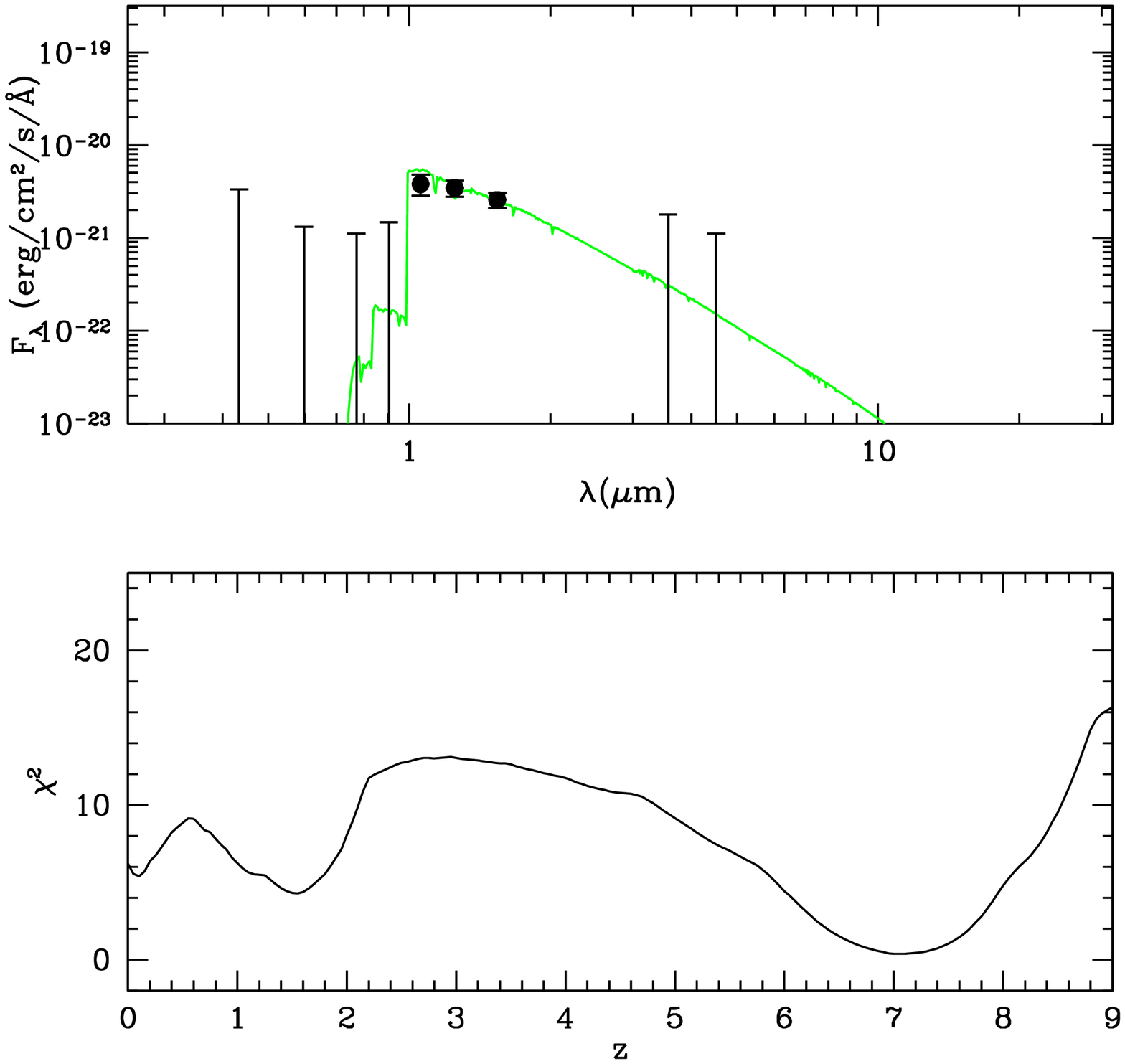}&
\includegraphics[width=0.47\textwidth]{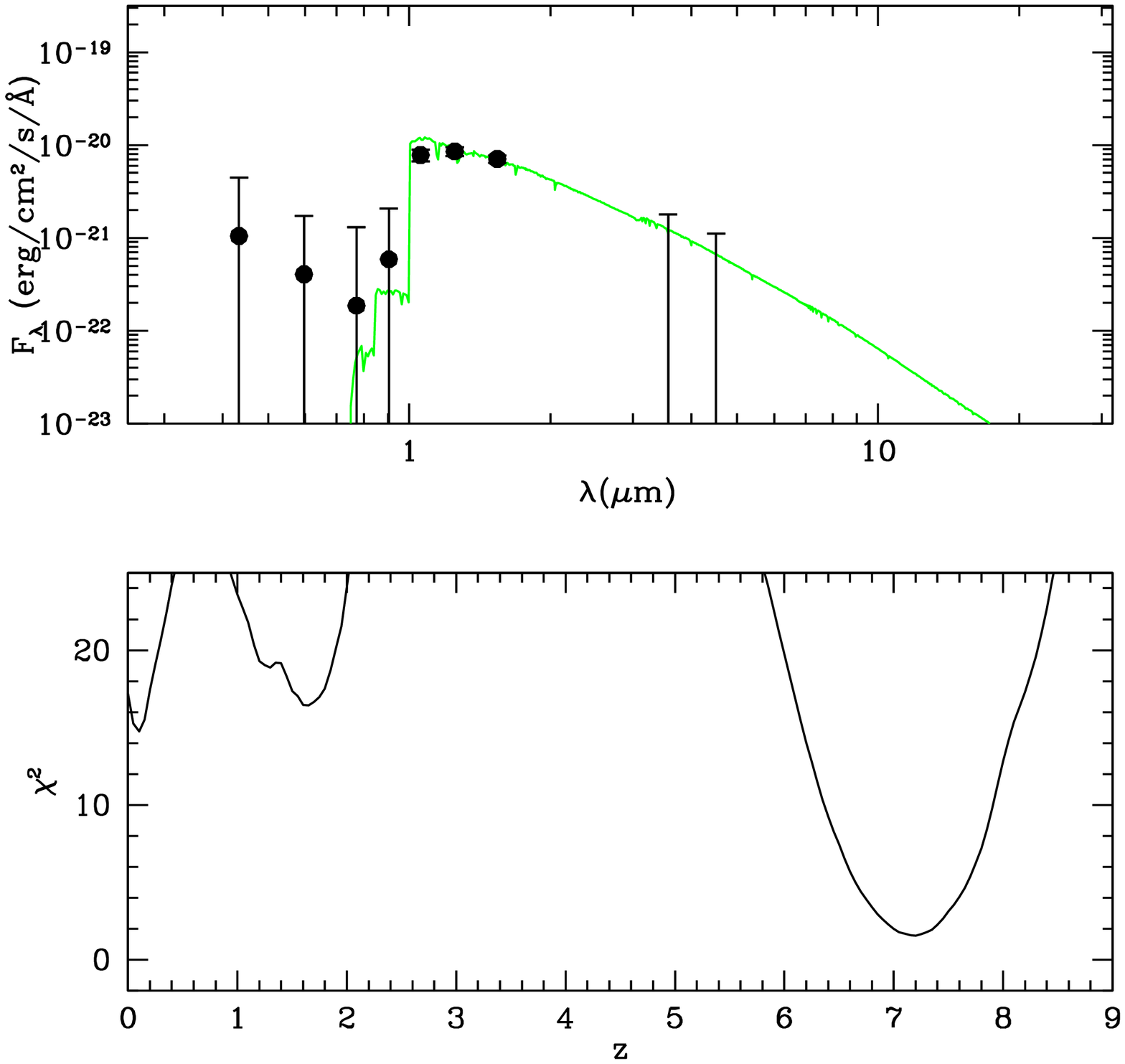}\\
\hspace*{1cm}{\bf 1574:} $z_{\rm est} = 7.20\, (6.55-7.60)$&
\hspace*{1cm} {\bf 835:} $z_{\rm est} = 7.20\, (6.90-7.50)$&\\
\\
\\
\\
\\
\hspace*{0.8cm}
\includegraphics[width=0.097\textwidth, angle=270]{2066_z_stamp.ps}
\includegraphics[width=0.097\textwidth, angle=270]{2066_y_stamp.ps}
\includegraphics[width=0.097\textwidth, angle=270]{2066_j_stamp.ps}
\includegraphics[width=0.097\textwidth, angle=270]{2066_h_stamp.ps}&
\hspace*{0.8cm}
\includegraphics[width=0.097\textwidth, angle=270]{2888_z_stamp.ps}
\includegraphics[width=0.097\textwidth, angle=270]{2888_y_stamp.ps}
\includegraphics[width=0.097\textwidth, angle=270]{2888_j_stamp.ps}
\includegraphics[width=0.097\textwidth, angle=270]{2888_h_stamp.ps}\\
\includegraphics[width=0.47\textwidth]{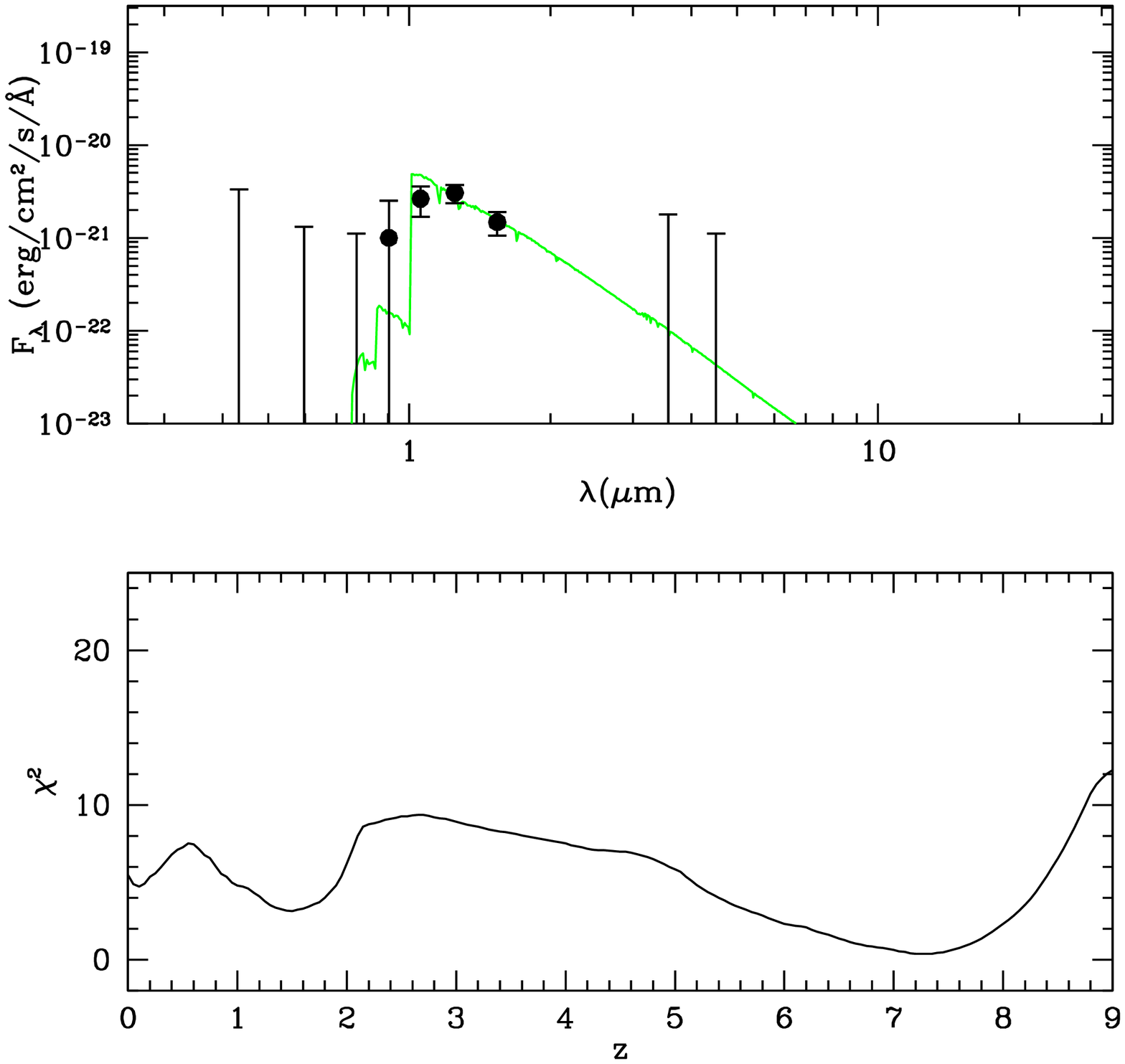}&
\includegraphics[width=0.47\textwidth]{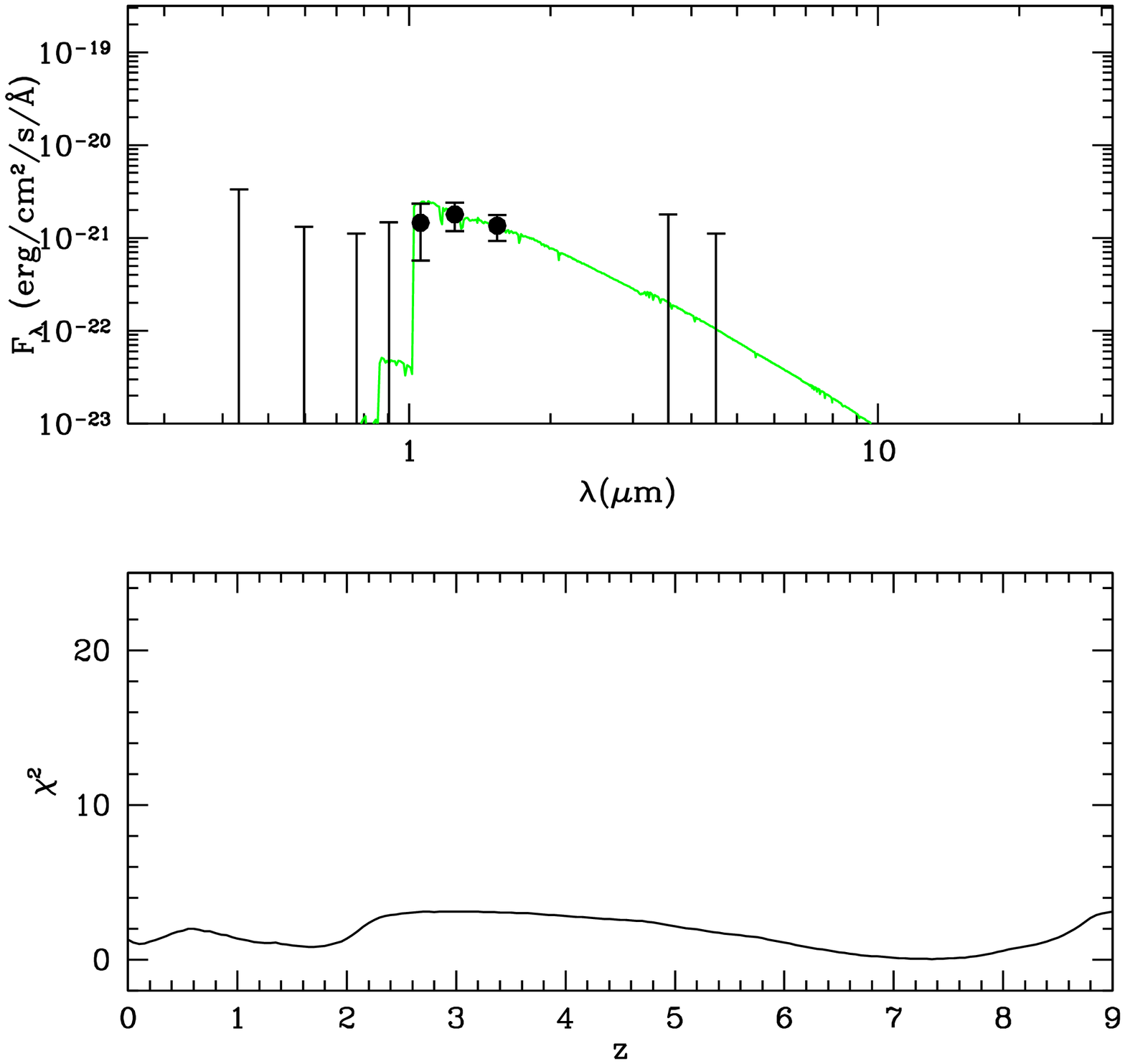}\\
\hspace*{1cm} {\bf 2066:} $z_{\rm est} = 7.20\, (6.50-7.80)$&
\hspace*{1cm}{\bf 2888:} $z_{\rm est} = 7.35\, (6.10-8.35)$\\
\\
\\
\end{tabular}
\addtocounter{figure}{-1}
\caption{continued.}
\end{figure*}

\begin{figure*}
\begin{tabular}{llll}

\hspace*{0.8cm}
\includegraphics[width=0.097\textwidth, angle=270]{2940_z_stamp.ps}
\includegraphics[width=0.097\textwidth, angle=270]{2940_y_stamp.ps}
\includegraphics[width=0.097\textwidth, angle=270]{2940_j_stamp.ps}
\includegraphics[width=0.097\textwidth, angle=270]{2940_h_stamp.ps}&
\hspace*{0.8cm}
\includegraphics[width=0.097\textwidth, angle=270]{2079_z_stamp.ps}
\includegraphics[width=0.097\textwidth, angle=270]{2079_y_stamp.ps}
\includegraphics[width=0.097\textwidth, angle=270]{2079_j_stamp.ps}
\includegraphics[width=0.097\textwidth, angle=270]{2079_h_stamp.ps}\\
\includegraphics[width=0.47\textwidth]{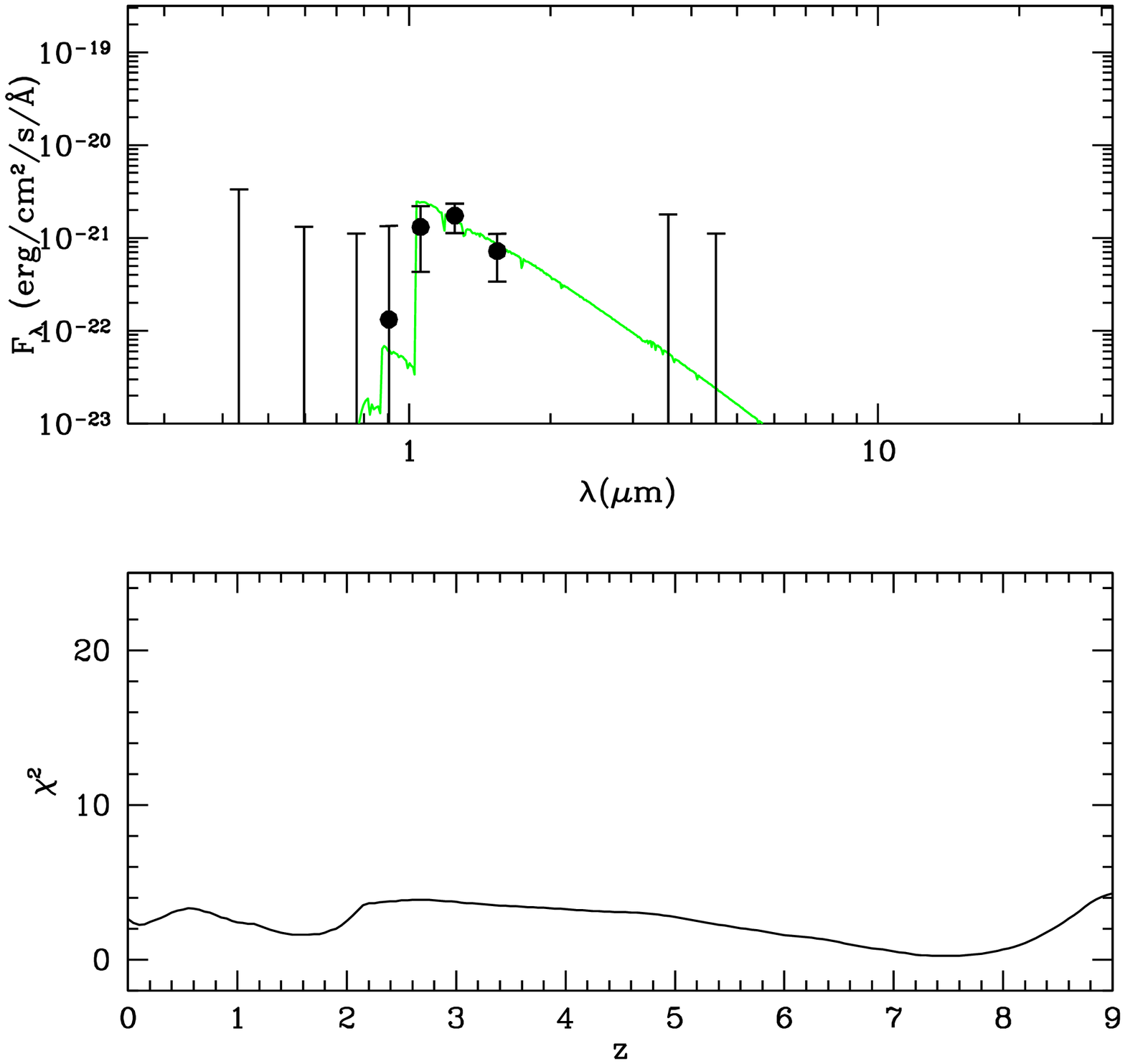}&
\includegraphics[width=0.47\textwidth]{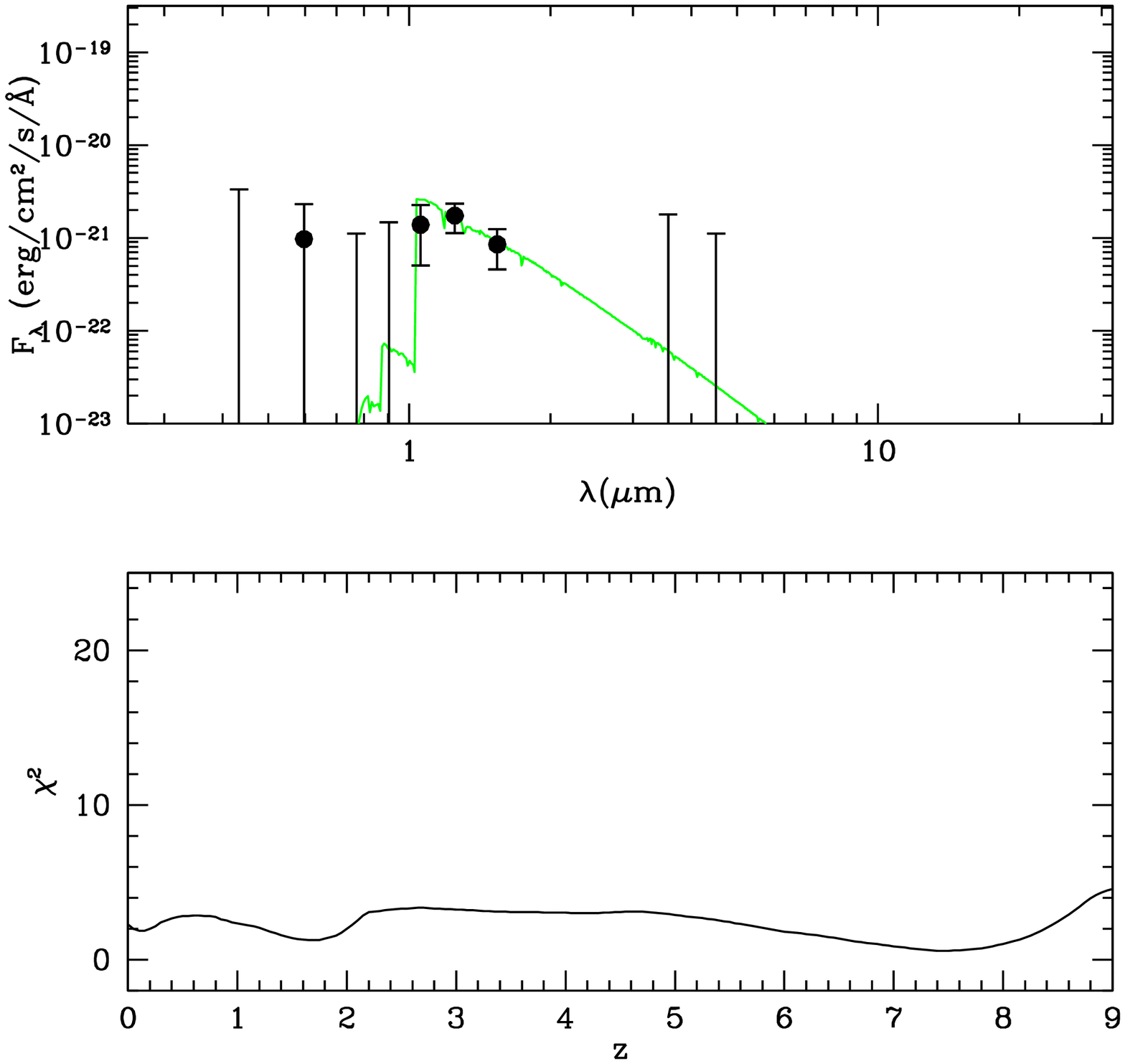}\\
\hspace*{1cm}{\bf 2940:} $z_{\rm est} = 7.40\, (6.40-8.25)$&
\hspace*{1cm} {\bf 2079:} $z_{\rm est} = 7.50\, (6.35-8.25)$\\
\\
\\
\\
\\
\hspace*{0.8cm}
\includegraphics[width=0.097\textwidth, angle=270]{1107_z_stamp.ps}
\includegraphics[width=0.097\textwidth, angle=270]{1107_y_stamp.ps}
\includegraphics[width=0.097\textwidth, angle=270]{1107_j_stamp.ps}
\includegraphics[width=0.097\textwidth, angle=270]{1107_h_stamp.ps}&
\hspace*{0.8cm}
\includegraphics[width=0.097\textwidth, angle=270]{1422_z_stamp.ps}
\includegraphics[width=0.097\textwidth, angle=270]{1422_y_stamp.ps}
\includegraphics[width=0.097\textwidth, angle=270]{1422_j_stamp.ps}
\includegraphics[width=0.097\textwidth, angle=270]{1422_h_stamp.ps}\\
\includegraphics[width=0.47\textwidth]{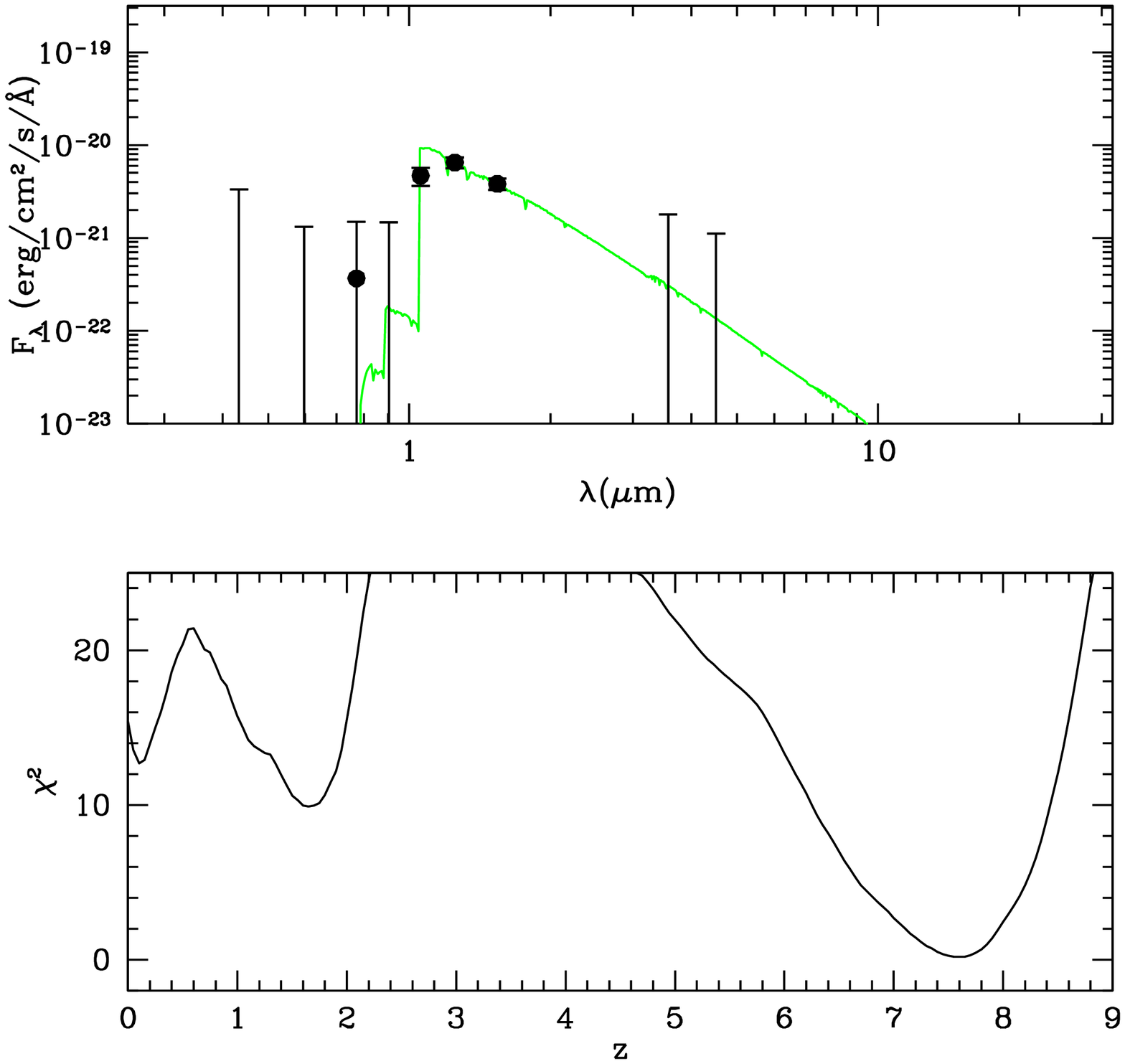}&
\includegraphics[width=0.47\textwidth]{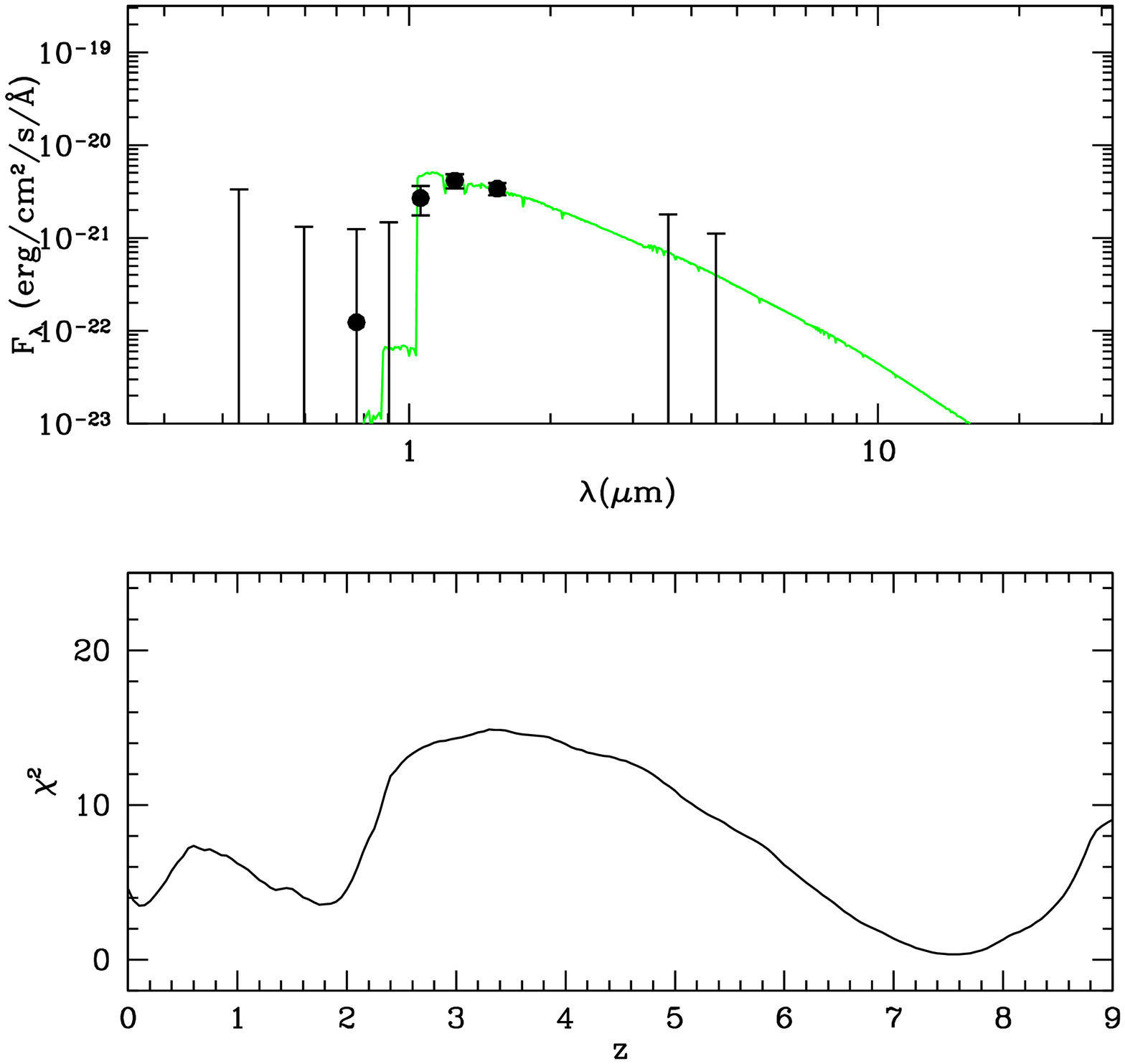}\\
\hspace*{1cm} {\bf 1107:} $z_{\rm est} = 7.60\, (7.30-7.90)$&
\hspace*{1cm}{\bf 1422:} $z_{\rm est} = 7.60\, (7.00-8.05)$\\
\\
\\
\end{tabular}
\addtocounter{figure}{-1}
\caption{continued.}
\end{figure*}

\begin{figure*}
\begin{tabular}{llll}
\hspace*{0.8cm}
\includegraphics[width=0.097\textwidth, angle=270]{2487_z_stamp.ps}
\includegraphics[width=0.097\textwidth, angle=270]{2487_y_stamp.ps}
\includegraphics[width=0.097\textwidth, angle=270]{2487_j_stamp.ps}
\includegraphics[width=0.097\textwidth, angle=270]{2487_h_stamp.ps}&
\hspace*{0.8cm}
\includegraphics[width=0.097\textwidth, angle=270]{1765_z_stamp.ps}
\includegraphics[width=0.097\textwidth, angle=270]{1765_y_stamp.ps}
\includegraphics[width=0.097\textwidth, angle=270]{1765_j_stamp.ps}
\includegraphics[width=0.097\textwidth, angle=270]{1765_h_stamp.ps}\\
\includegraphics[width=0.47\textwidth]{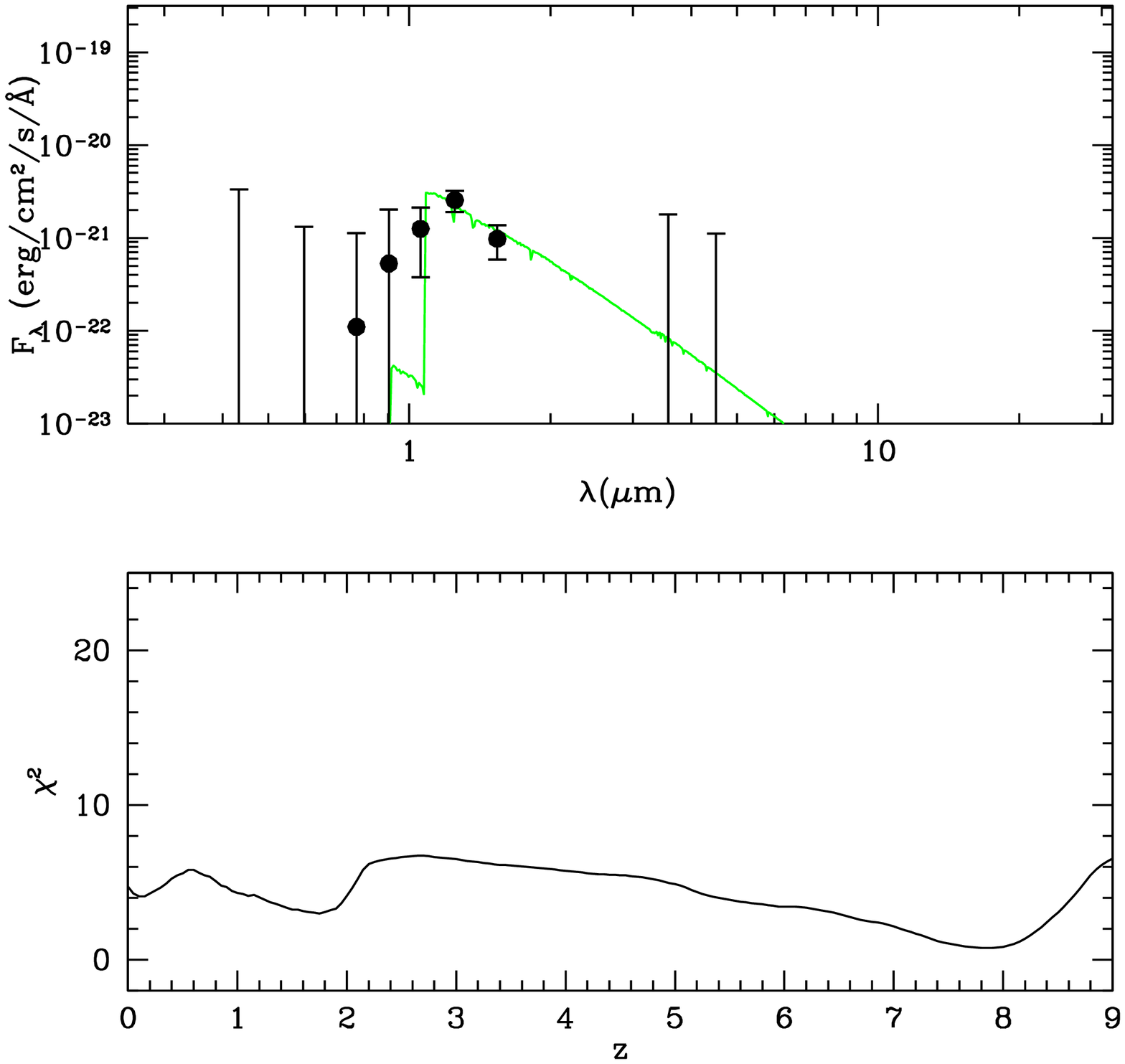}&
\includegraphics[width=0.47\textwidth]{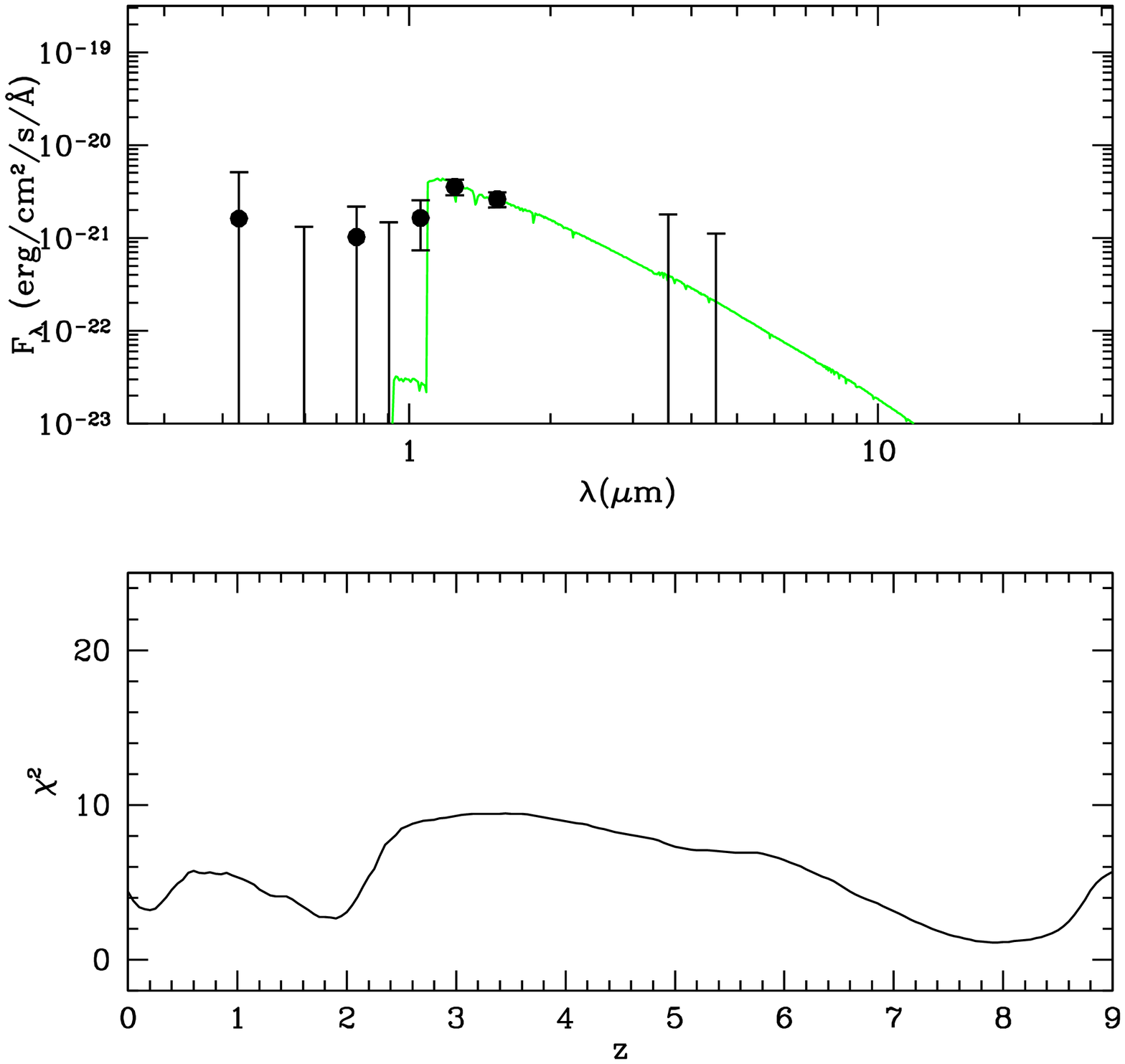}\\
\hspace*{1cm} {\bf 2487:} $z_{\rm est} = 7.80\, (7.10-8.30)$&
\hspace*{1cm}{\bf 1765:} $z_{\rm est} = 7.95\, (7.35-8.50)$\\
\\
\\
\\
\\
\hspace*{0.8cm}
\includegraphics[width=0.097\textwidth, angle=270]{2841_z_stamp.ps}
\includegraphics[width=0.097\textwidth, angle=270]{2841_y_stamp.ps}
\includegraphics[width=0.097\textwidth, angle=270]{2841_j_stamp.ps}
\includegraphics[width=0.097\textwidth, angle=270]{2841_h_stamp.ps}&
\hspace*{0.8cm}
\includegraphics[width=0.097\textwidth, angle=270]{1939_z_stamp.ps}
\includegraphics[width=0.097\textwidth, angle=270]{1939_y_stamp.ps}
\includegraphics[width=0.097\textwidth, angle=270]{1939_j_stamp.ps}
\includegraphics[width=0.097\textwidth, angle=270]{1939_h_stamp.ps}\\
\includegraphics[width=0.47\textwidth]{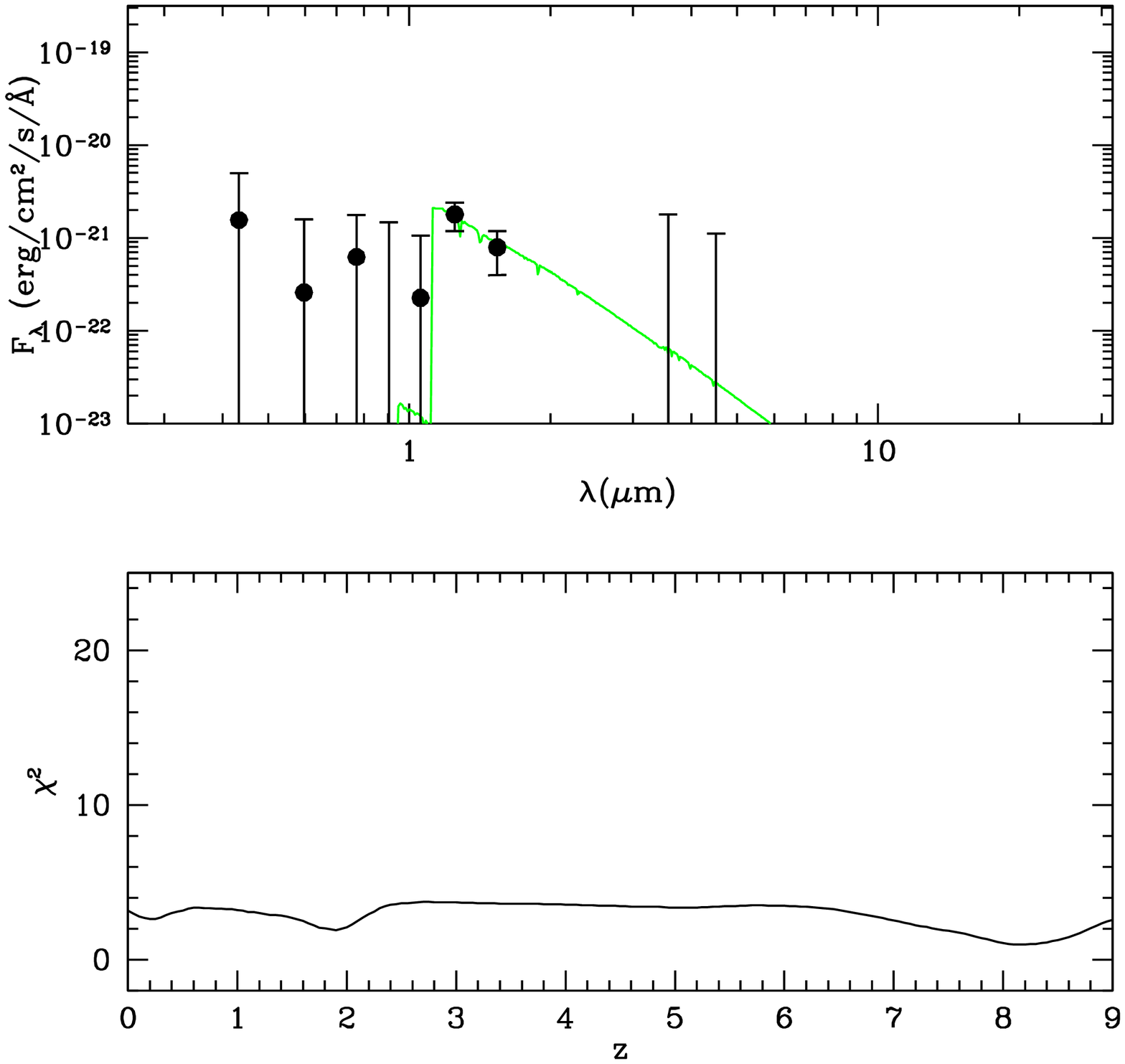}&
\includegraphics[width=0.47\textwidth]{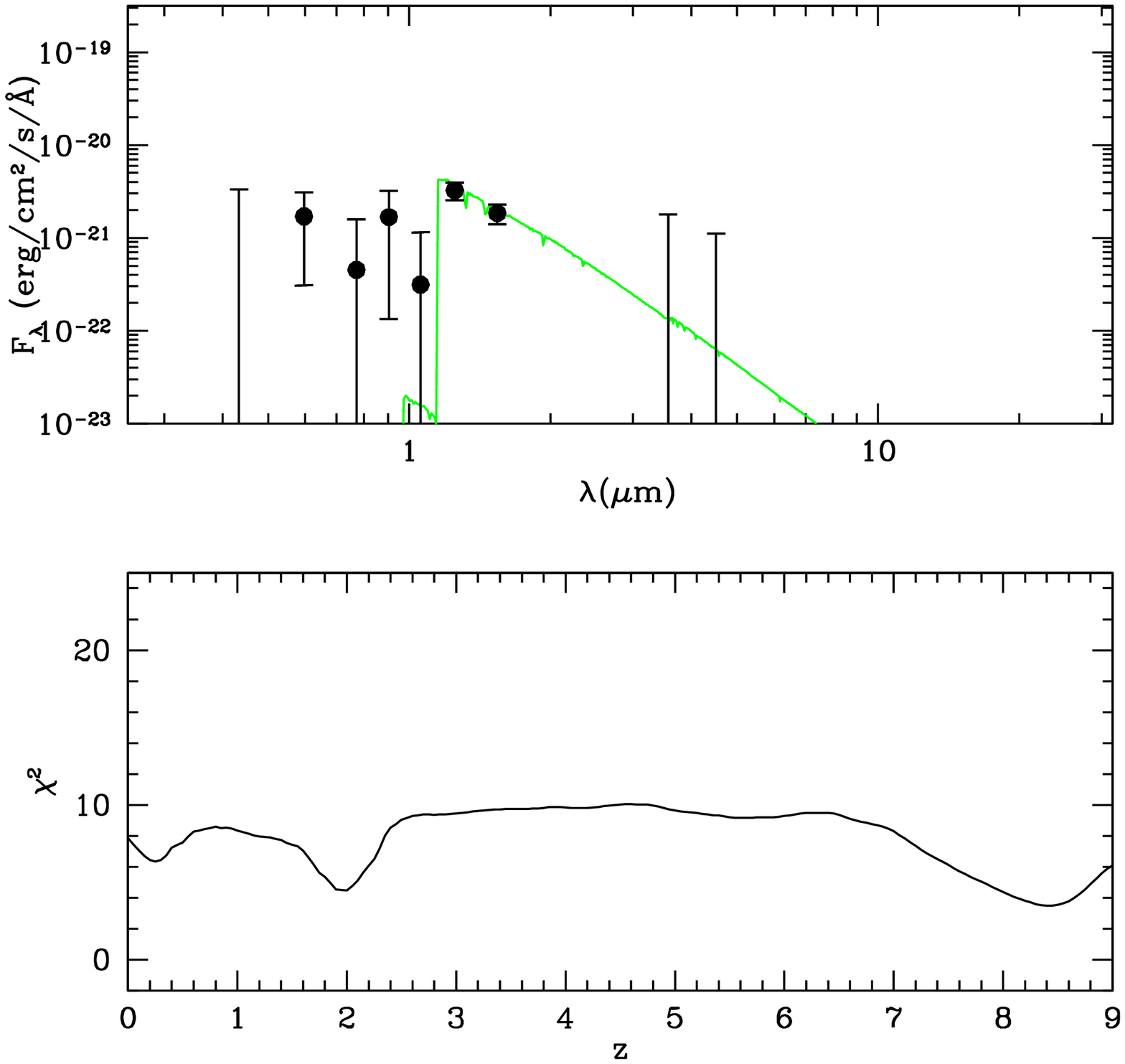}\\
\hspace*{1cm} {\bf 2841:} $z_{\rm est} = 8.10\, (7.40-8.75)$&
\hspace*{1cm} {\bf 1939:} $z_{\rm est} = 8.35\, (7.90-8.70)$\\
\\
\\
\\
\end{tabular}
\addtocounter{figure}{-1}
\caption{continued.}
\end{figure*}

\begin{figure*}
\begin{tabular}{llll}
\hspace*{0.8cm}
\includegraphics[width=0.097\textwidth, angle=270]{1721_z_stamp.ps}
\includegraphics[width=0.097\textwidth, angle=270]{1721_y_stamp.ps}
\includegraphics[width=0.097\textwidth, angle=270]{1721_j_stamp.ps}
\includegraphics[width=0.097\textwidth, angle=270]{1721_h_stamp.ps}\\
\includegraphics[width=0.47\textwidth]{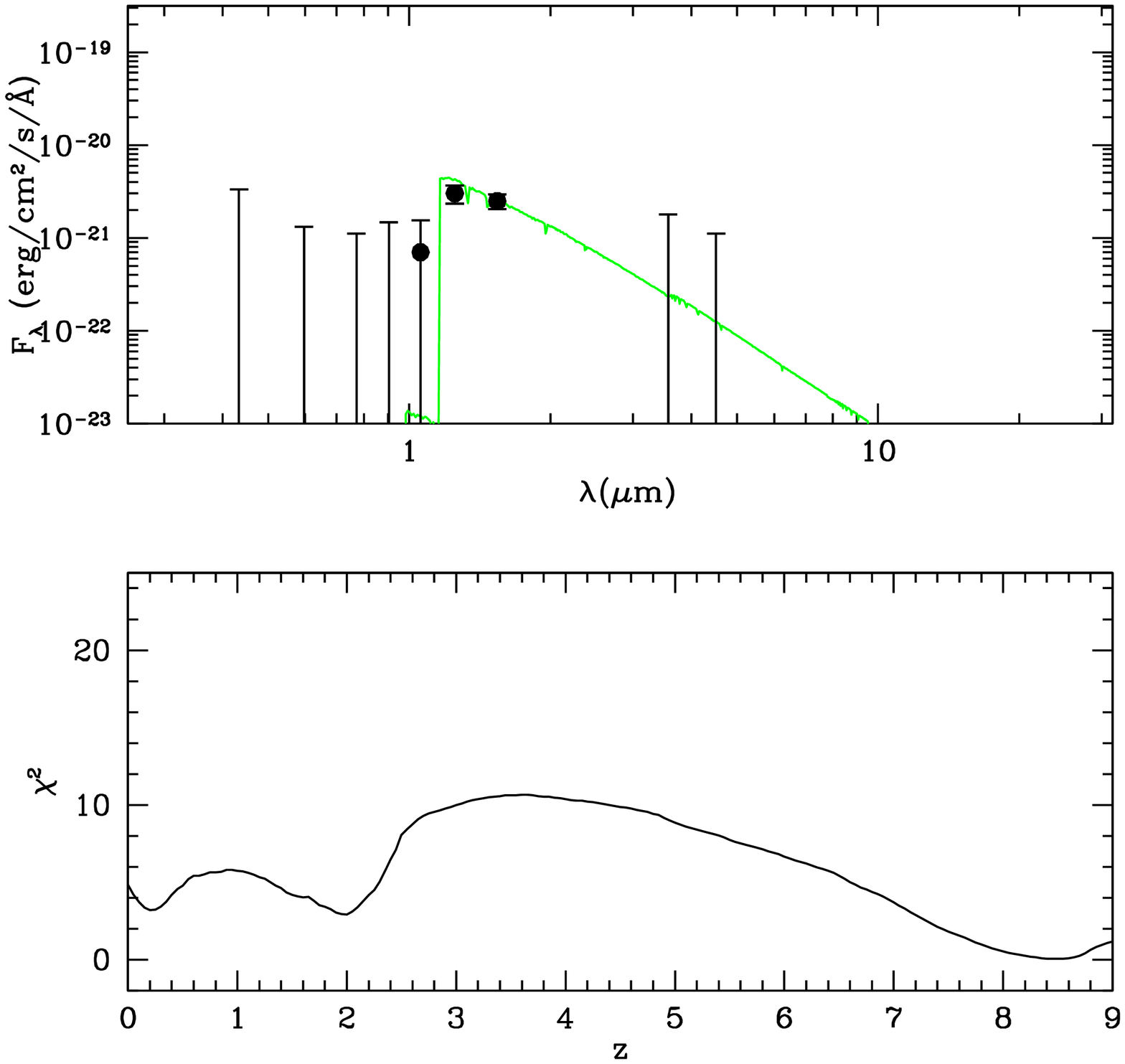}&
\phantom{88888888888888888888888888888888888888888888888888888888888}\\
\hspace*{1cm}{\bf 1721:} $z_{\rm est} = 8.45\, (7.75-8.85)$ \\
\\
\\
\end{tabular}
\addtocounter{figure}{-1}
\caption{continued.}
\end{figure*}

\end{document}